\documentclass[aps,twocolumn,prx,preprintnumbers,amssymb,aps, 
10pt]{revtex4-2}
\usepackage{xpatch}
\makeatletter
\patchcmd{\@ssect@ltx}
    {\addcontentsline{toc}{#1}{\protect\numberline{}#8}}
    {}
    {}
    {}
\makeatother
\pdfoutput=1
\usepackage{color}
\usepackage{enumitem}
\usepackage{comment}
\usepackage{physics}
\usepackage{braket}

\usepackage{ragged2e}

\usepackage[dvipsnames]{xcolor}
\usepackage{hyperref}
\definecolor{amaranth}{rgb}{0.9, 0.17, 0.31}
\definecolor{forestForestGreen(web)}{rgb}{0.13, 0.55, 0.13}
\definecolor{blue(munsell)}{HTML}{005567}
\definecolor{bblue}{rgb}{0.0, 0.58, 0.71}
\hypersetup{
pdfstartview={FitH}, 
pdftitle={}, 
colorlinks=true, 
linkcolor=blue(munsell), 
citecolor=blue(munsell), 
filecolor=magenta, 
urlcolor=blue(munsell)
}
\usepackage{pgfplots}
\pgfplotsset{compat=1.18}
\usepackage{amssymb}
\usepackage{amsfonts}
\usepackage{epstopdf}
\usepackage{dcolumn}
\usepackage{amsmath}
\usepackage{latexsym,bm}
\usepackage{amsthm}
\usepackage{slashed}
\usepackage{color}
\usepackage{rotating}
\usepackage[normalem]{ulem}
\usepackage{tikz-cd}
\usepackage{nicefrac}
\usepackage{mathtools}
\usepackage{yhmath}
\usepackage{graphbox}
\usepackage{mathtools}
\usepackage{enumitem}
\usepackage{scalerel,stackengine}

\usepackage{tikz}
\usepackage[english]{babel}

\usepackage{graphicx}
\graphicspath{{Figures/}}

\allowdisplaybreaks
\usepackage{tikz}
\usepackage{diagbox}
\usetikzlibrary{positioning}

\tikzset{->-/.style={decoration={
  markings,
  mark=at position .5 with {\arrow{>}}},postaction={decorate}}}

\newcommand{\be}{\begin{equation}}
\newcommand{\ee}{\end{equation}}
\newcommand{\ba}{\begin{aligned}}
\newcommand{\ea}{\end{aligned}}
\newcommand{\id}{\text{id}}

\newcommand{\cC}{\mathcal{C}}
\newcommand{\cE}{\mathcal{E}}
\newcommand{\cG}{\mathcal{G}}
\newcommand{\cH}{\mathcal{H}}
\newcommand{\cM}{\mathcal{M}}
\newcommand{\cV}{\mathcal{V}}

\newcommand{\Rep}{\text{Rep}}
\newcommand{\TwoRep}{2\mathsf{Rep}}

\DeclareMathOperator{\Hilb}{Hilb}

\def\bbC{{\mathbb{C}}}
\def\bbZ{{\mathbb{Z}}}

\newcommand\crule[3][black]{\textcolor{#1}{\rule{#2}{#3}}}
 
\newcommand\tub[4][]{\big\langle \unitlength=1pt
 \begin{picture}(20,1)
 \linethickness{0.18mm}
 \put(12,-4.2){\line(0,1){11}}
 \put(-0.5,2.7){\line(1,0){24}}
 \put(12,2.7){\color{gray}\circle*{3}}
 \put(10,8){$\scriptstyle #3$}
 \put(0.3,4.7){$\scriptstyle #2$}
 \put(18.7,4.7){$\scriptstyle #1$}
 \put(13.5,-3){\color{gray} $\scriptstyle #4$}
 \end{picture} \hspace{3.5pt} \big\rangle}

\newcommand\ntub[4][]{\unitlength=1pt
 \begin{picture}(20,1)
 \linethickness{0.18mm}
 \put(12,-4.2){\line(0,1){11}}
 \put(-0.5,2.7){\line(1,0){24}}
 \put(12,2.7){\color{gray}\circle*{3}}
 \put(10,8){$\scriptstyle #3$}
 \put(0.3,4.7){$\scriptstyle #2$}
 \put(18.7,4.7){$\scriptstyle #1$}
 \put(13.5,-3){\color{gray} $\scriptstyle #4$}
 \end{picture} \hspace{3.5pt}}
 
\newcommand\tuub[4][]{\big\langle \unitlength=1pt
 \begin{picture}(20,1)
 \linethickness{0.18mm}
 \put(20,-4.2){\line(0,1){11}}
 \put(-0.5,2.7){\line(1,0){40}}
 \put(20,2.7){\color{gray}\circle*{3}}
 \put(20,8){$\scriptstyle #3$}
 \put(0.3,4.7){$\scriptstyle #2$}
 \put(34,4.7){$\scriptstyle #1$}
 \put(21,-3.4){\color{gray} $\scriptstyle #4$}
 \end{picture} \hspace{18.5pt} \big\rangle}

\newcommand\ntuub[4][]{ \unitlength=1pt
 \begin{picture}(20,1)
 \linethickness{0.18mm}
 \put(20,-4.2){\line(0,1){11}}
 \put(-0.5,2.7){\line(1,0){40}}
 \put(20,2.7){\color{gray}\circle*{3}}
 \put(20,8){$\scriptstyle #3$}
 \put(0.3,4.7){$\scriptstyle #2$}
 \put(34,4.7){$\scriptstyle #1$}
 \put(21,-3.4){\color{gray} $\scriptstyle #4$}
 \end{picture} \hspace{18.5pt} }

\newcommand\nhtub[4][]{\unitlength=1pt
 \begin{picture}(20,12)
 \linethickness{0.18mm}
 \put(9,-4.2){\line(0,1){11}}
 \put(15,4.4){\line(0,1){3.6}}
 \put(-0.5,2.7){\line(1,0){8}}
 \put(12,2.7){\line(1,0){12}}
 \put(12,2.7){\color{gray}\circle*{3}}
 \put(9,9.5){$\scriptstyle #3$}
 \put(0.3,4.7){$\scriptstyle #2$}
 \put(18.7,4.7){$\scriptstyle #1$}
 \put(13,-4){\color{gray} $\scriptstyle #4$}
 \put(9,-4.2){\line(30,6){4}}
 \put(9,6.8){\line(30,6){6}}
 \end{picture} \hspace{3.5pt}}

\newcommand\nhtuub[4][]{ \unitlength=1pt
 \begin{picture}(20,12)
 \linethickness{0.18mm}
 \put(9,-4.2){\line(0,1){11}}
 \put(15,4.4){\line(0,1){3.6}}
 \put(-0.5,2.7){\line(1,0){8}}
 \put(12,2.7){\line(1,0){20}}
 \put(12,2.7){\color{gray}\circle*{3}}
 \put(9,10){$\scriptstyle #3$}
 \put(0.3,4.7){$\scriptstyle #2$}
 \put(26.5,4.7){$\scriptstyle #1$}
 \put(13,-4){\color{gray} $\scriptstyle #4$}
 \put(9,-4.2){\line(30,6){4}}
 \put(9,6.8){\line(30,6){6}}
 \end{picture} \hspace{11.5pt} }

\newcommand\cchtub[3][]{\unitlength=1pt
 \begin{picture}(20,12)
 \linethickness{0.18mm}
 \put(9,-4.2){\line(0,1){11}}
 \put(15,4.4){\line(0,1){3.6}}
 \put(15,-3.02){\line(0,1){4}}
 \put(-0.5,2.7){\line(1,0){8}}
 \put(12,2.7){\line(1,0){12}}
 \put(12,2.7){\color{gray}\circle*{3}}
 \put(9.7,9.5){$\scriptstyle #3$}
 \put(0.3,4.7){$\scriptstyle #2$}
 \put(18.7,4.7){$\scriptstyle #1$}
 \put(9,-4.2){\line(30,6){6}}
 \put(9,6.8){\line(30,6){6}}
 \end{picture} \hspace{3.5pt} 
 }

\newcommand\cchhtub[3][]{\unitlength=1pt
 \big\langle 
 \begin{picture}(20,12)
 \linethickness{0.18mm}
 \put(9,-4.2){\line(0,1){11}}
 \put(15,4.4){\line(0,1){3.6}}
 \put(15,-3.02){\line(0,1){4}}
 \put(-0.5,2.7){\line(1,0){8}}
 \put(12,2.7){\line(1,0){12}}
 \put(12,2.7){\color{gray}\circle*{3}}
 \put(9.7,9.5){$\scriptstyle #3$}
 \put(0.3,4.7){$\scriptstyle #2$}
 \put(18.7,4.7){$\scriptstyle #1$}
 \put(9,-4.2){\line(30,6){6}}
 \put(9,6.8){\line(30,6){6}}
 \end{picture} \hspace{3.5pt} 
 \big\rangle
 }

\newcommand\ncchhtub[3][]{\unitlength=1pt
 \begin{picture}(20,12)
 \linethickness{0.18mm}
 \put(9,-4.2){\line(0,1){11}}
 \put(15,4.4){\line(0,1){3.6}}
 \put(15,-3.02){\line(0,1){4}}
 \put(-0.5,2.7){\line(1,0){8}}
 \put(12,2.7){\line(1,0){12}}
 \put(12,2.7){\color{gray}\circle*{3}}
 \put(9.7,9.5){$\scriptstyle #3$}
 \put(0.3,4.7){$\scriptstyle #2$}
 \put(18.7,4.7){$\scriptstyle #1$}
 \put(9,-4.2){\line(30,6){6}}
 \put(9,6.8){\line(30,6){6}}
 \end{picture} \hspace{3.5pt} 
 }

\newcommand\otub[2][]{\big\langle \unitlength=1pt
 \begin{picture}(20,1)
 \linethickness{0.18mm}
 \put(12,-4.2){\line(0,1){4.6}}
 \put(12,4.9){\line(0,1){2.1}}
 \qbezier(11,4.2)(12,4.2)(12,4.9)
 \qbezier(11,1.2)(12,1.2)(12,0.4)
 \put(-0.5,2.7){\line(1,0){24}}
 \put(-0.5,4.2){\line(1,0){11.5}}
 \put(-0.5,1.2){\line(1,0){11.5}}
 \put(10,8){$\scriptstyle #2$}
 \put(18.7,4.7){$\scriptstyle #1$}
 \end{picture} \hspace{3.5pt} \big\rangle}

\newcommand\notub[2][]{\unitlength=1pt
 \begin{picture}(20,1)
 \linethickness{0.18mm}
 \put(12,-4.2){\line(0,1){4.6}}
 \put(12,4.9){\line(0,1){2.1}}
 \qbezier(11,4.2)(12,4.2)(12,4.9)
 \qbezier(11,1.2)(12,1.2)(12,0.4)
 \put(-0.5,2.7){\line(1,0){24}}
 \put(-0.5,4.2){\line(1,0){11.5}}
 \put(-0.5,1.2){\line(1,0){11.5}}
 \put(10,8){$\scriptstyle #2$}
 \put(18.7,4.7){$\scriptstyle #1$}
 \end{picture} \hspace{3.5pt} }

\newcommand\nstrp[7][]{\hspace{4pt}\unitlength=1pt
 \begin{picture}(20,1)
 \linethickness{0.18mm}
 \put(12,-2){\line(0,1){8.5}}
 \put(3.5,-2){\line(1,0){16}}
 \put(3.5,6.5){\line(1,0){16}}
 \put(12,-2){\color{gray}\circle*{3}}
 \put(12,6.5){\color{gray}\circle*{3}}
 \put(4.5,-0.1){$\scriptstyle #5$}
 \put(20,-4){$\scriptstyle #1$}
 \put(20,4.3){$\scriptstyle #2$}
 \put(-4,-4){$\scriptstyle #3$}
 \put(-4,4.3){$\scriptstyle #4$}
 \put(10,-9.5){\color{gray} $\scriptstyle #6$}
 \put(10,9.5){\color{gray} $\scriptstyle #7$}
 \end{picture}\hspace{7.5pt}}

\newcommand\nsstrp[7][]{\hspace{4pt}\unitlength=1pt
 \begin{picture}(20,1)
 \linethickness{0.18mm}
 \put(15,-2){\line(0,1){8.5}}
 \put(3.5,-2){\line(1,0){16}}
 \put(3.5,6.5){\line(1,0){16}}
 \put(15,-2){\color{gray}\circle*{3}}
 \put(15,6.5){\color{gray}\circle*{3}}
 \put(4.5,-0.1){$\scriptstyle #5$}
 \put(20,-4){$\scriptstyle #1$}
 \put(20,4.3){$\scriptstyle #2$}
 \put(-4,-4){$\scriptstyle #3$}
 \put(-4,4.3){$\scriptstyle #4$}
 \put(13,-9.5){\color{gray} $\scriptstyle #6$}
 \put(13,9.5){\color{gray} $\scriptstyle #7$}
 \end{picture}\hspace{7.5pt}}

\newcommand\nstrrp[7][]{\hspace{4pt}\unitlength=1pt
 \begin{picture}(20,1)
 \linethickness{0.18mm}
 \put(25,-2){\line(0,1){8.5}}
 \put(3.5,-2){\line(1,0){30}}
 \put(3.5,6.5){\line(1,0){30}}
 \put(25,-2){\color{gray}\circle*{3}}
 \put(25,6.5){\color{gray}\circle*{3}}
 \put(4.5,-0.1){$\scriptstyle #5$}
 \put(35,-4){$\scriptstyle #1$}
 \put(35,4.3){$\scriptstyle #2$}
 \put(-4,-4){$\scriptstyle #3$}
 \put(-4,4.3){$\scriptstyle #4$}
 \put(16,-9.5){\color{gray} $\scriptstyle #6$}
 \put(16,9.5){\color{gray} $\scriptstyle #7$}
 \end{picture}\hspace{23pt}}

\usepackage[hang,flushmargin]{footmisc}
\makeatletter
\newcounter{bottomnote}
\newcommand{\bfootnote}[1]{%
    \stepcounter{bottomnote}%
    \edef\thisanchor{bottom.\thebottomnote}%
    \hyperlink{\thisanchor}{\textsuperscript{\thebottomnote}}%
    \insert\footins{%
        \reset@font\footnotesize
        \interlinepenalty\interdisplaylinepenalty
        \splittopskip\footnotesep
        \splitmaxdepth \dp\strutbox \floatingpenalty \@MM
        \hsize\columnwidth \@parboxrestore
        \vskip 0pt 
        \parindent 0pt                
        \leftskip 1.2em               
        \noindent\llap{
            \hypertarget{\thisanchor}{\textsuperscript{\thebottomnote}}\hspace{0.4em}%
        }#1\par%
    }%
}
\makeatother

\makeatletter
\def\l@subsubsection#1#2{}
\makeatother

\makeatletter
\newcommand\xlabel[2][]{\phantomsection\def\@currentlabelname{#1}\label{#2}}
\makeatother

\newtheorem{theorem}{Theorem}

\usetikzlibrary{arrows}
\usetikzlibrary{decorations.pathreplacing,decorations.markings}
\usetikzlibrary{calc}
\tikzset{
  on each segment/.style={
    decorate,
    decoration={
      show path construction,
      moveto code={},
      lineto code={
        \path [#1]
        (\tikzinputsegmentfirst) -- (\tikzinputsegmentlast);
      },
      curveto code={
        \path [#1] (\tikzinputsegmentfirst)
        .. controls
        (\tikzinputsegmentsupporta) and (\tikzinputsegmentsupportb)
        ..
        (\tikzinputsegmentlast);
      },
      closepath code={
        \path [#1]
        (\tikzinputsegmentfirst) -- (\tikzinputsegmentlast);
      },
    },
  },
  mid arrow/.style={postaction={decorate,decoration={
        markings,
        mark=at position .6 with {\arrow[#1]{latex}}
      }}},
}

\begin{document}

\title{Beyond Wigner: Non-Invertible Symmetries Preserve Probabilities}

\author{Thomas Bartsch}
\author{Yuhan Gai}
\author{Sakura Sch\"afer-Nameki}
\affiliation{Mathematical Institute,
University of Oxford, \\
Woodstock Road,
Oxford, OX2 6GG,  
United Kingdom}

\begin{abstract}
\noindent
In recent years, the traditional notion of symmetry in quantum theory was expanded to so-called generalised or categorical symmetries, which, unlike ordinary group symmetries, may be non-invertible. This appears to be at odds with Wigner's theorem, which requires quantum symmetries to be implemented by (anti)unitary -- and hence invertible -- operators in order to preserve probabilities. We resolve this puzzle for (higher) fusion category symmetries $\mathcal{C}$ by proposing that, instead of acting by unitary operators on a fixed Hilbert space, symmetry defects in $\mathcal{C}$ act as isometries between distinct Hilbert spaces constructed from twisted sectors. As a result, we find that non-invertible symmetries naturally act as trace-preserving quantum channels. Crucially, our construction relies on the symmetry category $\mathcal{C}$ being unitary. We illustrate our proposal through several examples that include Tambara-Yamagami, Fibonacci, and Yang-Lee as well as higher categorical symmetries.
\end{abstract}

\maketitle

\smallskip\noindent{\bf Introduction.}
A fundamental feature of quantum symmetries is that they correspond to transformations that preserve transition amplitudes between physical states, leading to probability conservation. For a quantum system with Hilbert space $\mathcal{H}$, the structure of all such transformations is dictated by Wigner's theorem \cite{Wigner1931}: 

\smallskip
\begin{theorem}[Wigner, 1931]
\label{thm-wigner}
Let $T: \mathcal{P}(\mathcal{H}) \to \mathcal{P}(\mathcal{H})$ be a transformation of the ray space \footnote{The \textit{ray space} of a Hilbert space $\mathcal{H}$ is defined to be the set $\mathcal{P}(\mathcal{H}) := (\mathcal{H}\backslash \lbrace 0 \rbrace)/\mathbb{C}^{\times}$ of equivalence classes of non-zero vectors in $\mathcal{H}$, where two such vectors $\Phi,\Psi \in \mathcal{H}\backslash \lbrace 0 \rbrace$ are considered equivalent if there exists a non-zero scalar $\lambda \in \mathbb{C}^{\times}$ such that $\Psi = \lambda \cdot \Phi$.} of a finite-dimensional Hilbert space $\mathcal{H}$ that preserves the ray product
\begin{equation}
\label{eq-ray-product}
[\Phi] \cdot [\Psi] \; := \; \frac{|\!\braket{\Phi,\Psi}\!|}{\norm{\Phi} \!\cdot\! \norm{\Psi}}
\end{equation}
for all $[\Phi],[\Psi] \in \mathcal{P}(\mathcal{H})$. Then, there is an (anti)unitary linear operator $U: \mathcal{H} \to \mathcal{H}$ such that
\begin{equation}
T([\Phi]) \, = \, [U(\Phi)]
\end{equation}
for all $[\Phi] \in \mathcal{P}(\mathcal{H})$ with representative $\Phi \in \mathcal{H}\backslash \lbrace 0 \rbrace$. In particular, the operator $U$ is invertible with inverse $U^{\dagger}$.
\end{theorem}

\noindent
On the other hand, recent years have witnessed the expansion of the traditional notion of symmetry to so-called \textit{generalised} or \textit{categorical symmetries}, which correspond to topological defects of various codimensions \cite{Gaiotto:2014kfa}. In particular, these need not be invertible with respect to the fusion of defects, leading to so-called \textit{non-invertible symmetries} (see \cite{Frohlich:2004ef, Bhardwaj:2017xup} for early works in 2d and \cite{Kaidi:2021xfk, Choi:2021kmx, Bhardwaj:2022yxj} for recent developments in higher dimensions -- for reviews see \cite{Schafer-Nameki:2023jdn, Shao:2023gho}). This raises the following question:
\begin{center}
\emph{Are non-invertible symmetries compatible with Wigner’s theorem and the preservation of transition amplitudes?}
\end{center}
In this letter, we address this question for finite symmetries described by (higher) fusion categories $\mathcal{C}$. {We show that, rather than acting as unitaries on a fixed Hilbert space as in Theorem~\ref{thm-wigner}, symmetry defects in $\cC$ naturally act as \textit{isometries} \footnote{A linear map $U: \mathcal{H} \to \mathcal{H}'$ between Hilbert spaces is called an \emph{isometry} if $\braket{U(\Phi),U(\Psi)} = \braket{\Phi,\Psi}$ for all $\Phi,\Psi \in \mathcal{H}$. Note that every isometry induces a unitary operator upon being restricted to its image $\text{im}(U) \subset \mathcal{H}'$.} between possibly distinct Hilbert spaces that we construct explicitly.} We phrase our result~as~a \textit{Categorical Probability Preservation} (CPP) Theorem~\ref{thm-categorical-wigner}. Crucially, we demonstrate that the latter only holds if the symmetry category $\mathcal{C}$ is \textit{unitary} \cite{Galindo2014, Bartsch2025}.

\noindent
While the tension between non-invertible symmetries and the preservation of probabilities in the sense of Wigner's theorem has been recognised since the outset, it was recently addressed in \cite{Ortiz:2025psr}, where it was argued that, in order to preserve transition amplitudes, non-invertible symmetries need to map physical states into an enlarged Hilbert space. In this work, we provide a systematic construction of the latter in terms of twisted sectors for (higher) fusion category symmetries. Similar ideas appeared in the context of 1+1d lattice models in \cite{Lootens:2022avn, Lootens:2023wnl}, where dualities were realised as isometries using Matrix Product Operators. Moreover, while it was observed in \cite{Bischoff:2016rpu, Bischoff:2022fxf, Okada:2024qmk} that non-invertible symmetries act by \textit{quantum operations} \footnote{A \emph{quantum operation} between Hilbert spaces $\mathcal{H}$ and $\mathcal{H}'$ is a completely positive linear map $\mathcal{E}: B(\mathcal{H}) \to B(\mathcal{H}')$, where $B(\cH)$ is the space of bounded linear maps on $\cH$.}, we show that they in fact act by trace-preserving \textit{quantum channels} \footnote{A \emph{quantum channel} is a quantum operation $\mathcal{E}: B(\mathcal{H}) \to B(\mathcal{H}')$ that is trace-preserving, i.e. $\text{Tr}[\mathcal{E}(\rho)] = \text{Tr}[\hspace{0.5pt}\rho\hspace{0.5pt}]$ for all $\rho \in B(\mathcal{H})$.} upon including all possible twisted sectors and transition channels thereinto. Finally, our results provide a posteriori physical justification for requiring the symmetry category $\mathcal{C}$ to be unitary. While this was motivated in \cite{Bartsch2025} from the perspective of reflection positivity, the CPP Theorem~\ref{thm-categorical-wigner} provides a more direct justification by linking the unitary structure on $\mathcal{C}$ to the preservation of quantum probabilities.

\vspace{4pt}
\smallskip\noindent{\bf Motivation.}
Consider a quantum theory in two spacetime dimensions that has a unitary fusion category symmetry $\mathcal{C}$. Objects $A,B \in \mathcal{C}$ and morphisms $\varphi \in \mathcal{C}(A,B)$ of the latter correspond to topological line defects and junctions between them, respectively, while the monoidal structure $\otimes: \mathcal{C} \times \mathcal{C} \to \mathcal{C}$ captures the fusion of topological defects in the ambient two-dimensional spacetime:
\begin{equation}
\label{eq-fusion-category}
\vspace{-5pt}
\begin{gathered}
\includegraphics[height=1.05cm]{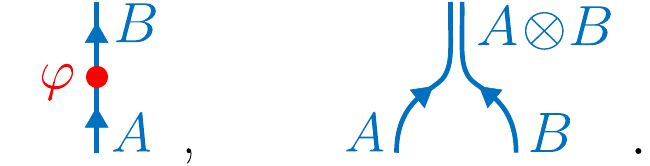}
\end{gathered}
\end{equation}
In this setting, defects $A \in \mathcal{C}$ can act on states $\ket{\Psi}$ in the Hilbert space $\mathcal{H}$ associated to the circle by wrapping the corresponding topological lines around the cylinder:
\begin{equation}
\label{eq-fusion-ring-action}
\vspace{-5pt}
\begin{gathered}
\includegraphics[height=2.3cm]{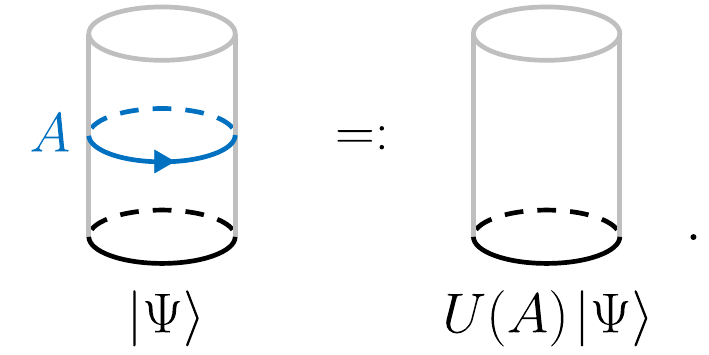}
\end{gathered}
\end{equation}
Compatibility with the fusion of symmetry defects then requires the operators $U(A)$ to satisfy
\begin{equation}
    U(A) \hspace{1pt} \circ \hspace{1pt} U(B) \; = \; U(A \otimes B) \; ,
\end{equation}
which turns $U$ into a representation of the so-called \textit{fusion algebra} \footnote{Given a fusion category $\mathcal{C}$, its \textit{fusion algebra} is the finite-dimensional associative algebra that is spanned by the simple objects $S_i$ of $\mathcal{C}$ with algebra multiplication given by $S_i \cdot S_j = \raisebox{-.48ex}{\large \text{$\Sigma$}}_{k}\hspace{1pt} N_{ij}^k \hspace{1pt} S_k$ (where $N_{ij}^k \in \mathbb{N}$ denote the so-called \textit{fusion coefficients} of $\mathcal{C}$).} associated to $\mathcal{C}$.

As an example, consider the two-dimensional critical Ising CFT \cite{Frohlich:2004ef, Aasen:2016dop, Chang2019,Thorngren2024}, which, in addition to an invertible $\bbZ_2$-symmetry generated by a line defect $\eta$, hosts the non-invertible Kramers-Wannier duality line $D$ with
\begin{equation}
\label{eq-ising-fusion-rule}
    D \otimes D \; = \; 1 \oplus \eta \; .
\end{equation}
Let $\ket{\sigma} \in \mathcal{H}$ be the primary state associated to the spin field $\sigma$, on which $\eta$ acts by $U(\eta) = -1$. Using the fusion rule (\ref{eq-ising-fusion-rule}), we find that  $D$ acts on $\ket{\sigma}$ by $U(D)=0$, which clearly does not preserve inner products. However, upon vertically attaching the invertible line $\eta$ to the wrapped defect $D$, the spin field $\sigma$ is mapped to (a non-zero multiple of) the disorder operator $\mu$ \cite{Frohlich:2004ef}:
\begin{equation}
\vspace{-5pt}
\begin{gathered}
\includegraphics[height=2.3cm]{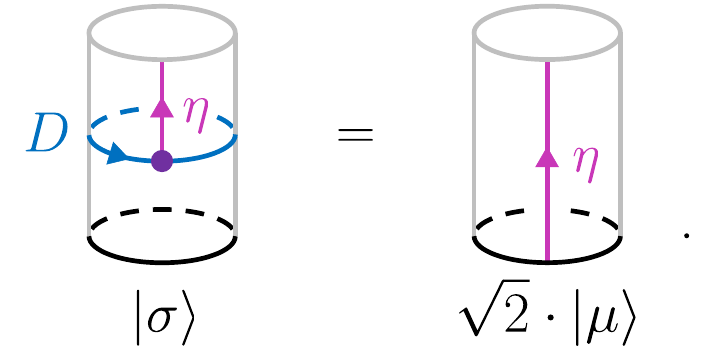}
\end{gathered}
\end{equation}
Importantly, the resulting state now lives in the \textit{twisted sector} Hilbert space $\mathcal{H}_\eta$ that is punctured {by the vertical}  line $\eta$. Hence, upon setting $\mathcal{H}' := \mathcal{H} \oplus \mathcal{H}_\eta$, we can embed $U(D)$ into the $\mathcal{H}'$-valued operator $U'(D) = (0,\sqrt{2})^{\text{T}}$ that preserves inner products up to a scalar factor. Suitably rescaling $U'(D)$ then yields a linear isometry associated to the non-invertible defect $D$, which maps the original Hilbert space $\mathcal{H}$ into the enlarged Hilbert space $\mathcal{H}'$.

In general, we would like to emulate the above discussion for any given unitary fusion category $\mathcal{C}$. To this end, we denote by $\mathcal{H}_X$ the \textit{$X$-twisted Hilbert space}, i.e. the Hilbert space on $S^1$ that is punctured by a vertical line defect $X \in \mathcal{C}$. As before, symmetry defects $A \in \mathcal{C}$ can act on twisted sector states $\ket{\Psi} \in \mathcal{H}_X$ by wrapping the corresponding topological lines around the cylinder:
\begin{equation}
\label{eq-tube-action}
\vspace{-5pt}
\begin{gathered}
\includegraphics[height=2.3cm]{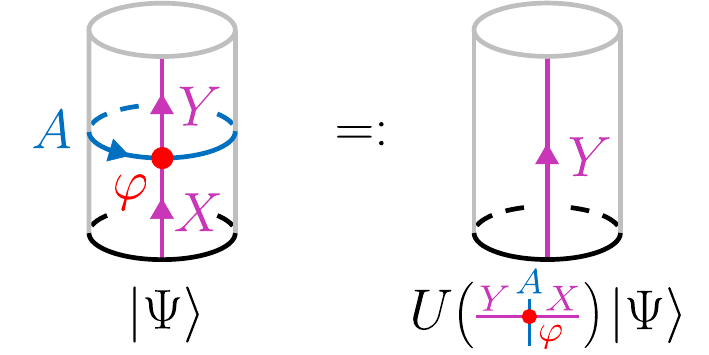}
\end{gathered}
\end{equation}
(note the rotation by 90° on the LHS for notational convenience).
However, the presence of twisted sector lines now requires the choice of a topological local operator
\begin{equation}
\label{eq-intersection-morphism}
    \varphi : \; A \otimes X \, \to \, Y \otimes A
\end{equation}
that sits at the junction between $A$, the incoming twisted sector $X$, and the outgoing twisted sector $Y$. We refer to $\varphi$ as a \textit{transition channel} between $X$ and $Y$ and denote the space of all transition channels for $A$ by 
\begin{equation}
\label{eq-transition-channels}
    \mathcal{C}^{\hspace{0.5pt}A\hspace{-0.5pt}}(X,Y) \; := \; \mathcal{C}(A \hspace{-1pt}\otimes\hspace{-1pt} X \hspace{1pt}, Y \hspace{-1pt}\otimes\hspace{-1pt} A) \; .
\end{equation}
This is naturally a vector space of dimension
\begin{equation}
\label{eq-transition-channel-dimension}
    D_{XY}^A \; := \; \text{dim}(\hspace{0.5pt} \mathcal{C}^{\hspace{0.5pt}A\hspace{-0.5pt}}(X,Y) ) \; .
\end{equation}
In particular, if $D_{XY}^A > 1$, the action of $A$ admits more than one (linearly independent) transition channel from the $X$-twisted into the $Y$-twisted sector. Upon choosing a particular channel $\varphi \in \mathcal{C}^{\hspace{0.5pt}A\hspace{-0.5pt}}(X,Y)$, we denote by
\begin{equation}
\label{eq-tube-wrapping-action}
    U\Big( \ntub[\hspace{-2.4pt}X]{\hspace{0.5pt}Y}{\hspace{-1pt}A}{\scriptstyle\hspace{1pt} \varphi} \Big): \;\;  \mathcal{H}_X \; \to \; \mathcal{H}_Y
\end{equation}
the corresponding linear map that captures the wrapping action of $A$ on twisted sectors as illustrated in (\ref{eq-tube-action}). Compatibility with the fusion of symmetry defects then requires the above maps to satisfy
\begin{equation}
\label{eq-composition-rule}
U\Big(\ntub[\hspace{-1.5pt}Y]{\hspace{0.5pt}Z}{\hspace{-1.5pt}A}{\hspace{1pt} \varphi}\Big) \, \circ \, U\Big(\ntub[\hspace{-2.4pt}X]{\hspace{0.5pt}Y}{\hspace{-1pt}B}{\raisebox{-0.4pt}{\hspace{1pt}$\scriptstyle \psi$}}\Big) \;\, = \,\;  U\Big(\ntuub[\hspace{-2.4pt}X]{\hspace{0.4pt}Z}{\raisebox{0.7pt}{\hspace{-10pt}$\scriptstyle A \hspace{0.5pt} \otimes B$}}{\hspace{1pt} \varphi \hspace{0.6pt}\circ\hspace{0.6pt} \psi}\Big) \; .
\end{equation}
Furthermore, we require $U$ to be compatible with spacetime reflections in the sense that
\begin{equation}
\label{eq-dagger-condition}
    U\Big( \ntub[\hspace{-2.4pt}X]{\hspace{0.5pt}Y}{\hspace{-1pt}A}{\scriptstyle\hspace{1pt} \varphi} \Big)^{\dagger} \; = \; U\Big( \ntub[\hspace{-1.2pt}Y]{\hspace{0pt}X}{\hspace{-2pt}A^{\hspace{-0.8pt}\vee}}{\raisebox{-2.2pt}{\hspace{1pt}$\scriptstyle \varphi^{\dagger}$}} \Big) \; ,
\end{equation}
where $A^{\vee}$ denotes the orientation reversal of the symmetry defect $A$ and $\varphi^{\dagger}$ denotes the adjoint of the intersection channel $\varphi$. Mathematically, this turns $U$ into a representation of the so-called \textit{tube algebra} of $\mathcal{C}$ \cite{Ocneanu2016ChiralityFO,Izumi:2000qa, Mueger:2001crc, Lin:2022dhv,Bartsch2023a}. We will refer to representations $U$ of this type as \textit{generalised charges} for the symmetry $\mathcal{C}$ in what follows \cite{Bhardwaj:2023wzd,Bhardwaj:2023ayw}. Physically, these capture the full action of symmetry defects in $\mathcal{C}$ on all possible twisted sectors.

Given a generalised charge $U$, the operator $U\big( \raisebox{0.5pt}{$\scaleobj{0.8}{\ntub[\hspace{-2.4pt}X]{\hspace{0.5pt}Y}{\hspace{-1pt}A}{\scriptstyle\hspace{1pt} \varphi}}$} \big)$ associated to a generic transition channel will typically \textit{not} preserve inner products (cf. the critical Ising CFT discussed above). However, we propose that there exist preferred choices of outgoing twisted sector $Y$ and transition channel $\varphi$ such that the resulting operator yields an inner-product-preserving isometry. We summarise our main proposal below.

\vspace{10pt}
\noindent
{\bf Main Result.} Consider a quantum theory in two spacetime dimensions whose generalised symmetries are described by a unitary fusion category $\mathcal{C}$ (for more mathematical background on the notion of unitary fusion categories we refer the reader to section~\ref{ssec-fusion-cats} in the Supplementary Material (SM)  -- see e.g. \cite{Etingof2017} for a textbook review). Then, the action of $\mathcal{C}$ on twisted sectors is compatible with the preservation of quantum transition amplitudes in the sense of the following \textit{Categorical Probability Preservation} (CPP) Theorem:

\begin{theorem}[CPP]
\label{thm-categorical-wigner}
    Let $A \in \mathcal{C}$ be a symmetry defect and let $X \in \mathcal{C}$ be a fixed incoming twisted sector. Then for each simple outgoing sector $S \in \mathcal{C}$ there exists a basis
    \begin{equation}
    \label{eqn:transition-channel-basis}
        \lbrace e_i^{\raisebox{-2pt}{$\scriptstyle S$}} \rbrace_i \, \subset \, \mathcal{C}^{\hspace{0.5pt}A\hspace{-0.5pt}}(X,S)
    \end{equation}
    of transition channels such that in any given generalised charge $U$ it holds that
    \begin{equation}
    \label{eq-generalised-preservation}
        \sum_{S\hspace{0.5pt},\hspace{0.5pt}i} \hspace{1pt} \Big\langle U\Big( \ntub[\hspace{-2.4pt}X]{\hspace{0.5pt}S}{\hspace{-1pt}A}{\scriptstyle\hspace{1pt} \raisebox{-2pt}{$\scriptstyle e_i^{\raisebox{-2pt}{$\scriptscriptstyle S$}}$}} \Big) \hspace{1pt} \Phi \hspace{1pt}, \, U\Big( \ntub[\hspace{-2.4pt}X]{\hspace{0.5pt}S}{\hspace{-1pt}A}{\scriptstyle\hspace{1pt} \raisebox{-2pt}{$\scriptstyle e_i^{\raisebox{-2pt}{$\scriptscriptstyle S$}}$}} \Big) \hspace{1pt} \Psi \Big\rangle \,\; = \,\; \braket{\Phi,\Psi}
    \end{equation}
    for all $\ket{\Phi}, \ket{\Psi} \in \mathcal{H}_X$, where sum runs over all simple objects $S$ and basis indices $i$. In other words 
    \begin{equation}
    \label{eq-isometry}
        U(A)_X \; := \; \bigoplus_{S\hspace{0.5pt},\hspace{1pt}i} \, U\Big( \ntub[\hspace{-2.4pt}X]{\hspace{0.5pt}S}{\hspace{-1pt}A}{\scriptstyle\hspace{1pt} \raisebox{-2pt}{$\scriptstyle e_i^{\raisebox{-2pt}{$\scriptscriptstyle S$}}$}} \Big)
    \end{equation}
defines a linear isometry from the $X$-twisted Hilbert space $\mathcal{H}_X$ into the enlarged Hilbert space
\be \label{eqn:enlargedHilb}
\mathcal{H}_X^A \; := \; \bigoplus_S \, D_{XS}^A \cdot \mathcal{H}_S \,,
\ee    
where $D_{XS}^A \in \mathbb{N}$ is as in (\ref{eq-transition-channel-dimension}). Moreover, the basis $\lbrace e_i^{\raisebox{-2pt}{$\scriptstyle S$}} \rbrace_i$ is unique up to transformations $\vec{e}^{\hspace{1.5pt}S} \mapsto M \cdot \vec{e}^{\hspace{1.5pt}S}$ by unitary matrices $M \in \mathcal{U}(D_{XS}^A)$.
\end{theorem}

\noindent
A constructive proof of this theorem is given in SM \ref{ssec-proof}. Physically, the above means that in order for the action of a symmetry defect $A$ on $X$-twisted sector states to preserve inner products, we have to consider \textit{all} possible outgoing twisted sectors together with \textit{all} possible (linearly independent) transition channels thereinto. Given a non-zero state $\ket{\Psi} \in \mathcal{H}_X \!\setminus\! \lbrace 0 \rbrace$, we set
\begin{equation}
\label{eq-transition-probability-1}
    p\Big( \ntub[\hspace{-2.4pt}X]{\hspace{0.5pt}S}{\hspace{-1pt}A}{\scriptstyle\hspace{1pt} \raisebox{-2pt}{$\scriptstyle e_i^{\raisebox{-2pt}{$\scriptscriptstyle S$}}$}} \Big)_{\!\Psi} \; := \; \norm{ \hspace{1pt} U\Big( \ntub[\hspace{-2.4pt}X]{\hspace{0.5pt}S}{\hspace{-1pt}A}{\scriptstyle\hspace{1pt} \raisebox{-2pt}{$\scriptstyle e_i^{\raisebox{-2pt}{$\scriptscriptstyle S$}}$}} \Big) \hspace{1pt} \Psi \hspace{0.5pt}}^2 \hspace{-1pt}\Big/ \norm{\Psi}^2 \; ,
\end{equation}
which as a consequence of (\ref{eq-generalised-preservation}) satisfies
\begin{equation}
    \textstyle \sum_{S\hspace{0.5pt},\hspace{0.5pt}i} \hspace{2pt} p\Big( \ntub[\hspace{-2.4pt}X]{\hspace{0.5pt}S}{\hspace{-1pt}A}{\scriptstyle\hspace{1pt} \raisebox{-2pt}{$\scriptstyle e_i^{\raisebox{-2pt}{$\scriptscriptstyle S$}}$}} \Big)_{\!\Psi} \; = \; 1 \; .
\end{equation}
Thus, we can interpret $p\big( \scaleobj{0.8}{\ntub[\hspace{-2.4pt}X]{\hspace{0.5pt}S}{\hspace{-1pt}A}{\scriptstyle\hspace{1pt} \raisebox{-2pt}{$\scriptstyle e_i^{\raisebox{-2pt}{$\scriptscriptstyle S$}}$}}} \big)_{\hspace{-1pt}\Psi} \in [0,1]$ as the probability that the symmetry defect $A$ maps $\ket{\Psi}$ to an $S$-twisted sector state via the transition channel $e_i^{\raisebox{-2pt}{$\scriptstyle S$}}$.

More generally, we can use the transition channels $e_i^{\raisebox{-2pt}{$\scriptstyle S$}}$ to construct \textit{quantum operations} 
\begin{equation}
    \mathcal{E}\Big( \ntub[\hspace{-2.4pt}X]{\hspace{0.5pt}S}{\hspace{-1pt}A}{\scriptstyle\hspace{1pt} \raisebox{-2pt}{$\scriptstyle e_i^{\raisebox{-2pt}{$\scriptscriptstyle S$}}$}} \Big): \; B(\mathcal{H}_X) \; \to \; B(\mathcal{H}_S)
\end{equation}
that act on \textit{density matrices} $\rho \in B(\mathcal{H}_X)$ \footnote{A \emph{density matrix} on a Hilbert space $\mathcal{H}$ is a positive-semidefinite operator $\rho \in B(\mathcal{H})$ s.t. $\text{Tr}(\rho) = 1$.} via their so-called \textit{Stinespring representation} \cite{Stinespring1955} \footnote{Given a quantum operation $\mathcal{E}: B(\mathcal{H}) \to B (\mathcal{H}')$, a \emph{Stinespring representation} for $\mathcal{E}$ consists of a Hilbert space $\mathcal{K}$ and a linear map $V: \mathcal{H} \to \mathcal{H}' \otimes \mathcal{K}$, such that $\mathcal{E}(\rho) = \text{Tr}_{\hspace{1pt}\mathcal{K}}\big(V \hspace{-1pt}\circ \hspace{-1pt} \rho \hspace{0pt} \circ \hspace{-1pt} V^{\dagger}\big)$ for all $\rho \in B(\mathcal{H})$. Every quantum operation $\cE$ admits a Stinespring representation. If $\mathcal{E}$ is moreover a quantum channel (i.e. trace-preserving), the linear map $V$ is an isometry, i.e. $V^{\dagger} \circ V = \text{Id}_{\mathcal{H}}$.}
\begin{equation}
\label{eqn:Stinespring}
    \cE\Big( \ntub[\hspace{-2.4pt}X]{\hspace{0.5pt}S}{\hspace{-1pt}A}{\scriptstyle\hspace{1pt} \raisebox{-2pt}{$\scriptstyle e_i^{\raisebox{-2pt}{$\scriptscriptstyle S$}}$}} \Big) (\rho) \;\, := \;\, U\Big( \ntub[\hspace{-2.4pt}X]{\hspace{0.5pt}S}{\hspace{-1pt}A}{\scriptstyle\hspace{1pt} \raisebox{-2pt}{$\scriptstyle e_i^{\raisebox{-2pt}{$\scriptscriptstyle S$}}$}} \Big) \hspace{0.7pt}\circ\hspace{0.7pt} \rho \hspace{1.2pt}\circ\hspace{1.2pt} U\Big( \ntub[\hspace{-2.4pt}X]{\hspace{0.5pt}S}{\hspace{-1pt}A}{\scriptstyle\hspace{1pt} \raisebox{-2pt}{$\scriptstyle e_i^{\raisebox{-2pt}{$\scriptscriptstyle S$}}$}} \Big)^{\dagger} 
\end{equation}
with associated \emph{Kraus operators} $U\big( \scaleobj{0.8}{\ntub[\hspace{-2.4pt}X]{\hspace{0.5pt}S}{\hspace{-1pt}A}{\scriptstyle\hspace{1pt} \raisebox{-2pt}{$\scriptstyle e_i^{\raisebox{-2pt}{$\scriptscriptstyle S$}}$}}}\big)$ \cite{kraus_states_1983} \footnote{Given a quantum operation $\cE: B(\cH) \rightarrow B(\cH')$, a \textit{Kraus representation} for $\mathcal{E}$ consists of a collection $\{ K_i \}_i$ of so-called \textit{Kraus operators} $K_i: \mathcal{H} \to \mathcal{H}'$ such that $\mathcal{E}(\rho) = \raisebox{-1.5pt}{\text{\large $\Sigma$}}_i \hspace{1pt} K_i \hspace{-1pt} \circ \hspace{-1pt} \rho \circ \hspace{-1pt} K_i^{\dagger}$ for all $\rho \in B(\cH)$. Every quantum operation admits a Kraus representation. If $\mathcal{E}$ is moreover a quantum channel (i.e. trace-preserving), the Kraus operators $K_i$ satisfy the Kraus completeness relation $\raisebox{-.48ex}{\large \text{$\Sigma$}}_i \hspace{1pt} K_i^{\dagger} \circ K_i = \text{Id}_{\mathcal{H}}$.}. Due to (\ref{eq-generalised-preservation}), the latter obey the Kraus completeness relation
\be
\sum_{S\hspace{0.5pt},\hspace{0.5pt} i} \; U\Big( \ntub[\hspace{-2.4pt}X]{\hspace{0.5pt}S}{\hspace{-1pt}A}{\scriptstyle\hspace{1pt} \raisebox{-2pt}{$\scriptstyle e_i^{\raisebox{-2pt}{$\scriptscriptstyle S$}}$}} \Big)^\dagger \! \circ \hspace{1.4pt} U\Big( \ntub[\hspace{-2.4pt}X]{\hspace{0.5pt}S}{\hspace{-1pt}A}{\scriptstyle\hspace{1pt} \raisebox{-2pt}{$\scriptstyle e_i^{\raisebox{-2pt}{$\scriptscriptstyle S$}}$}} \Big) \; = \; \text{Id}_{\hspace{0.5pt}\mathcal{H}_X} \; ,
\ee
which implies that setting 
\begin{equation}
    \mathcal{E}(A)_X \; := \; \bigoplus_{S\hspace{0.5pt},\hspace{1pt}i} \hspace{2pt} \mathcal{E}\Big( \ntub[\hspace{-2.4pt}X]{\hspace{0.5pt}S}{\hspace{-1pt}A}{\scriptstyle\hspace{1pt} \raisebox{-2pt}{$\scriptstyle e_i^{\raisebox{-2pt}{$\scriptscriptstyle S$}}$}} \Big) 
\end{equation}
yields a quantum operation $\mathcal{E}(A)_X: B(\mathcal{H}_X) \to B(\mathcal{H}_X^A)$ that satisfies the trace-preservation condition
\be
\text{Tr}\big[\hspace{1pt}\mathcal{E}(A)_X(\rho)\hspace{0.5pt}\big] \; = \; \text{Tr}[\hspace{0.5pt}\rho\hspace{0.5pt}]
\ee 
for all $\rho \in B(\mathcal{H}_X)$. As a result, we naturally obtain a \textit{quantum channel} associated to the action of the defect $A$ on $X$-twisted sector states.

\vspace{3pt}
\smallskip\noindent{\bf Example -- $\Rep (S_3)$.}
Let $\mathcal{C} = \Rep(S_3)$ be the unitary fusion category of finite-dimensional representations of the permutation group $S_3$. The simple objects are the trivial irrep $1$, the one-dimensional sign irrep $\psi$, and the two-dimensional irrep $\pi$, which fuse according to
\begin{equation}
\label{eq-reps3-fusion-rules} 
\pi \otimes \pi \; = \; 1 \oplus \psi \oplus \pi \, .
\end{equation}
The following then yield bases of transition channels for the non-invertible defect $\pi$ that obey Theorem~\ref{thm-categorical-wigner}:
\begin{equation}
\label{eqn:RepS3basis}
\begin{split}
    \left\{ \tfrac{1}{2} \hspace{0.5pt}\tub[1]{1}{\pi}{\pi} \right\} \, & \subset \, \cC^{\hspace{1pt}\pi}(1,1) \, , \hspace{18.5pt} \left\{\tfrac{1}{2} \tub[1]{\raisebox{0.9pt}{$\scriptstyle \psi$}}{\pi}{\pi} \right\} \, \subset \, \cC^{\hspace{1pt}\pi}(1,\psi) \, , \\[3pt]
    \left\{\tfrac{1}{\sqrt{2}} \tub[1]{\pi}{\pi}{\pi} \right\} \, & \subset \, \cC^{\hspace{1pt}\pi}(1,\pi) \, , \hspace{16.95pt} \left\{ \tfrac{1}{2} \hspace{0.5pt}\tub[\hspace{-1pt}\raisebox{0.9pt}{$\scriptstyle \psi$}]{1}{\pi}{\pi} \right\} \, \subset \, \cC^{\hspace{1pt}\pi}(\psi,1) \, , \\[3pt]
    \left\{ \tfrac{1}{2} \hspace{0.5pt} \tub[\hspace{-1pt}\raisebox{0.9pt}{$\scriptstyle \psi$}]{\raisebox{0.9pt}{$\scriptstyle \psi$}}{\pi}{\pi} \right\} \, & \subset \, \cC^{\hspace{1pt}\pi}(\psi,\psi) \, , \hspace{7pt} \left\{ \tfrac{1}{\sqrt{2}} \hspace{0.5pt} \tub[\hspace{-1pt}\raisebox{0.9pt}{$\scriptstyle \psi$}]{\pi}{\pi}{\pi} \right\} \, \subset \, \cC^{\hspace{1pt}\pi}(\psi,\pi) \, , \\[3pt]
    \left\{ \tfrac{1}{2} \hspace{0.5pt}\tub[\pi]{1}{\pi}{\pi} \right\} \, & \subset \, \cC^{\hspace{1pt}\pi}(\pi,1) \, , \hspace{16.95pt} \left\{ \tfrac{1}{2} \hspace{0.5pt} \tub[\pi]{\raisebox{0.9pt}{$\scriptstyle \psi$}}{\pi}{\pi} \right\} \, \subset \, \cC^{\hspace{1pt}\pi}(\pi,\psi) \, , \\[3pt]
    &\hspace{-55.4pt}\left\{ \tub[\pi]{\pi}{\pi}{1} \hspace{1pt} , \tub[\pi]{\pi}{\pi}{\psi} \hspace{1pt} , \tfrac{1}{\sqrt{2}} \hspace{0.5pt} \tub[\pi]{\pi}{\pi}{\pi} \right\} \, \subset \, \cC^{\hspace{1pt}\pi}(\pi,\pi)\,. 
\end{split}
\raisetag{15pt}
\end{equation}

\noindent
Upon choosing a generalised charge $U$, the above induce isometries $U(\pi)_X$ for the action of $\pi$ on $X$-twisted sectors (cf. \cite{Bartsch2025} and {SM \ref{app-RepS3}} for a list of all generalised charges): 
\begin{itemize}[leftmargin=3ex]
    \item For the generalised charge $U = U_{1,\psi}$ supported on the twisted sectors $\cH_1 \cong \cH_\psi \cong \mathbb{C}$, we have that
    \begin{equation}
    \begin{aligned}
    \quad U_{1,\psi}(\pi)_1 \, = \,  \tfrac{1}{2} \begin{bmatrix} -1 ,\, \sqrt{3} \end{bmatrix}^\text{T} : \;\; &\mathcal{H}_1 \; \to \; \mathcal{H}_1 \oplus \mathcal{H}_{\psi} \; , \\[2pt]
    \quad U_{1,\psi}(\pi)_{\psi} \, = \,  \tfrac{1}{2} \begin{bmatrix} \sqrt{3} ,\, 1 \end{bmatrix}^\text{T} : \;\; &\mathcal{H}_{\psi} \; \to \; \mathcal{H}_1 \oplus \mathcal{H}_{\psi} \; .
    \end{aligned}
    \end{equation}
The corresponding transition probabilities are
    \begin{equation}
    \begin{alignedat}{2}
        \;\quad p\Big( \tfrac{1}{2} \hspace{0.5pt}\tub[1]{1}{\pi}{\pi} \Big) \, &= \, 1/4 \; , \;\quad p\Big( \tfrac{1}{2} \hspace{0.5pt}\tub[\hspace{-1pt}\raisebox{0.9pt}{$\scriptstyle \psi$}]{1}{\pi}{\pi} \Big) \, &&= \, 3/4 \; , \\
        \;\quad p\Big( \tfrac{1}{2} \hspace{0.5pt} \tub[1]{\raisebox{0.9pt}{$\scriptstyle \psi$}}{\pi}{\pi} \Big) \, &= \, 3/4 \; , \;\quad p\Big( \tfrac{1}{2} \hspace{0.5pt} \tub[\hspace{-1pt}\raisebox{0.9pt}{$\scriptstyle \psi$}]{\raisebox{0.9pt}{$\scriptstyle \psi$}}{\pi}{\pi} \Big) \, &&= \, 1/4 \; . 
    \end{alignedat}
    \end{equation}
    \item For $U = U_{1,\pi}$ supported on  $\cH_1 \cong \cH_\pi \cong \mathbb{C}$, we find
\begin{equation}
\begin{aligned}
    \hspace{5pt} U_{1,\pi}(\pi)_1 \, = \,  \begin{bmatrix} 0,\hspace{1pt}1 \end{bmatrix}^\text{T} \!\!: \;\; &\mathcal{H}_1 \; \to \; \mathcal{H}_1 \oplus \mathcal{H}_{\pi} , \\[0pt]
    \hspace{5pt} U_{1,\pi}(\pi)_{\pi} \, = \,  \tfrac{1}{2} \begin{bmatrix} \sqrt{2},\hspace{1pt} 1 ,\hspace{1pt} 1 ,\hspace{1pt} 0 \end{bmatrix}^\text{T} \!\!: \;\; &\mathcal{H}_{\pi} \; \to \; \mathcal{H}_1 \oplus \mathcal{H}_{\pi}^{\hspace{1pt}\oplus \hspace{1pt} 3} . \\[0pt]
\end{aligned}
\end{equation}
The corresponding transition probabilities are
\begin{equation}
\hspace{10pt}
     \begin{aligned}
         p\Big( \tfrac{1}{2} \hspace{0.5pt}\tub[1]{1}{\pi}{\pi} \Big) \; &= \; 0 \; , \\[-1pt]
         p\Big( \tfrac{1}{\sqrt{2}} \tub[1]{\pi}{\pi}{\pi} \Big) \; &= \; 1 \; , 
     \end{aligned}
     \hspace{6pt}
     \begin{aligned}
        p\Big( \tfrac{1}{2} \hspace{0.5pt}\tub[\pi]{1}{\pi}{\pi} \Big) \; &= \; 1/2 \; , \\[-1pt]
         p\Big( \tub[\pi]{\pi}{\pi}{1} \Big) \; &= \; 1/4 \; , \\[-1pt]
         p\Big( \tub[\pi]{\pi}{\pi}{\psi} \Big) \; &= \; 1/4 \; , \\[-1pt]
         p\Big( \tfrac{1}{\sqrt{2}} \hspace{0.5pt} \tub[\pi]{\pi}{\pi}{\pi} \Big) \; &= \; 0 \; .
    \end{aligned}
    \end{equation}
\end{itemize}

\smallskip\noindent{\bf Examples -- Fibonacci and Yang-Lee.}
Consider the unitary Fibonacci fusion category $\mathcal{C} = \text{Fib}$, which has simple objects $1$ and $\tau$ that fuse according to
\begin{equation}
\label{eq-fib-fusion-rules}
    \tau \otimes \tau \; = \; 1\oplus \tau \, .
\end{equation}
The quantum dimension of $\tau$ is given by the golden ratio $d_{\tau} = \phi \equiv \tfrac{1}{2}(1 + \sqrt{5})$. The following then yield bases of transition channels for $\tau$ that obey Theorem~\ref{thm-categorical-wigner}:
\begin{equation}
\label{eqn:basis_Fib}
\ba
    \left\{ \phi^{-1} \tub[1]{1}{\tau}{\tau} \right\} \, &\subset \, \cC^{\hspace{1pt}\tau}(1,1)\,, \\[1pt]
    \left\{\phi^{-1} \tub[1]{\tau}{\tau}{\tau} \right\} \,& \subset \, \cC^{\hspace{1pt}\tau}(1,\tau) \; , \\[1pt]
    \left\{\phi^{-1} \tub[\tau]{1}{\tau}{\tau} \right\} \, &\subset \, \cC^{\hspace{1pt}\tau}(\tau,1) \,, \\[1pt]
    \left\{\phi^{{\nicefrac{1}{2}}} \hspace{2pt} \tub[\tau]{\tau}{\tau}{1}\,,\, \phi^{-1} \tub[\tau]{\tau}{\tau}{\tau} \right\} \, &\subset \, \cC^{\hspace{1pt}\tau}(\tau,\tau) \; . 
\ea
\end{equation}
Upon choosing a generalised charge $U$, the above induce isometries $U(\tau)_X$ for the action of $\tau$ on $X$-twisted sectors (cf. \cite{Bartsch2025} and {SM \ref{app-fib}} for a list of all generalised charges):

\begin{itemize}[leftmargin=3ex]
    \item For the generalized charges $U = U_{\tau}^{\pm}$ supported on the twisted sector $\cH_\tau \cong \mathbb{C}$, we have that
\begin{equation}
    \qquad U_{\tau}^{\pm}(\tau)_{\tau} \, = \,  \begin{bmatrix} \phi^{\nicefrac{1}{2}} \hspace{1pt} x_{\pm} \\ -(1+\phi \hspace{1pt} x_{\pm})\end{bmatrix} : \;\, \mathcal{H}_{\tau} \; \to \; \mathcal{H}_{\tau}^{\hspace{1pt}\oplus \hspace{1pt} 2} \; ,
\end{equation}
where $x_{\pm} \in \mathbb{C}$ solve  $x^2+x+\phi^{-2}=0\hspace{1pt}$. The corresponding transition probabilities are given by 
\be
    \quad p\Big( \phi^{{\nicefrac{1}{2}}} \tub[\tau]{\tau}{\tau}{1} \Big) = \phi - 1 \, , \quad
    p\Big( \tfrac{1}{\phi} \tub[\tau]{\tau}{\tau}{\tau} \Big) = 2 - \phi \, .
\ee

\item For the generalised charges $U = U_{1,\tau}$ supported on the twisted sectors $\cH_1 \cong \cH_\tau \cong \mathbb{C}$, we obtain
    \be
    \begin{split}
    \quad U_{1,\tau}(\tau)_1 =  \begin{bmatrix} -\phi^{-2} \\ \phi^{-\nicefrac{1}{2}}\hspace{1pt}(1-i\phi^{-1}) \end{bmatrix} : \;\, &\mathcal{H}_1 \; \to \; \mathcal{H}_1 \oplus \mathcal{H}_{\tau} \, , \\[2pt]
    \quad U_{1,\tau}(\tau)_{\tau} =  \begin{bmatrix} \phi^{-\nicefrac{3}{2}}\hspace{1pt}(1+i\phi^{-1}) \\ \phi^{-\nicefrac{1}{2}} \\ \phi^{-3} \end{bmatrix} : \;\, &\mathcal{H}_{\tau} \; \to \; \mathcal{H}_1 \oplus \mathcal{H}_{\tau}^{\hspace{1pt}\oplus \hspace{1pt} 2} \, . \\[-15pt]
    \end{split}
    \raisetag{-7pt}
    \ee
The corresponding transition probabilities are
    \be
    \begin{split}
        \quad\; p\Big( \phi^{{\nicefrac{1}{2}}} \tub[\tau]{\tau}{\tau}{1} \Big) &= \phi^{-1} , \hspace{4ex}
        p\Big( \tfrac{1}{\phi} \tub[1]{1}{\tau}{\tau} \Big) = \phi^{-4} , \\
        p\Big( \tfrac{1}{\phi} \hspace{1pt}\tub[\tau]{\tau}{\tau}{\tau} \Big) & = \phi^{-6} , \hspace{4ex}
        p\Big( \tfrac{1}{\phi} \hspace{1pt}\tub[1]{\tau}{\tau}{\tau} \Big) = 1 - \phi^{-4} , \\
        p\Big( \tfrac{1}{\phi} \tub[\tau]{1}{\tau}{\tau} \Big) &= 1-\phi^{-1}-\phi^{-6} . 
    \end{split}
    \raisetag{15.8pt}
    \ee
\end{itemize}

The Fibonacci category is to be contrasted with its non-unitary counterpart, the Yang-Lee category $\mathcal{C} = \text{YL}$, which has the same simple objects $1$ and $\tau$ and fusion rules (\ref{eq-fib-fusion-rules}) as $\text{Fib}$ but is crucially \textit{non-unitary}, as evidenced by the fact that the quantum dimension of $\tau$ is $d_\tau=-1/\phi<0$ in $\text{YL}$. In SM~\ref{sec:Yang-Lee}, we show that the CPP Theorem~\ref{thm-categorical-wigner} does not hold in this case, highlighting the importance of imposing unitarity on the symmetry category $\cC$ in order for the action of symmetry defects to preserve quantum probabilities.

\smallskip\noindent{\bf Generalisations.}
The CPP Theorem~\ref{thm-categorical-wigner} admits a direct generalisation to theories in arbitrary spacetime dimension $d \geq 2$, whose generalised symmetries are described by a unitary fusion $(d\!-\!1)$-category $\mathcal{C}$. Similarly to before, objects and ($p\!-\!1$)-morphisms of the latter correspond to topological symmetry defects of codimension $p=1,...,d$. In particular, codimension-one defects $A \in \mathcal{C}$ can act on states $\ket{\mathcal{O}}$ on the $(d\!-\!1)$-sphere (here depicted as local operators $\mathcal{O}$ using a conformal mapping) via linking:
\begin{equation}
\label{eq-higher-d-tube-action-main}
\vspace{-5pt}
\begin{gathered}
\includegraphics[height=2cm]{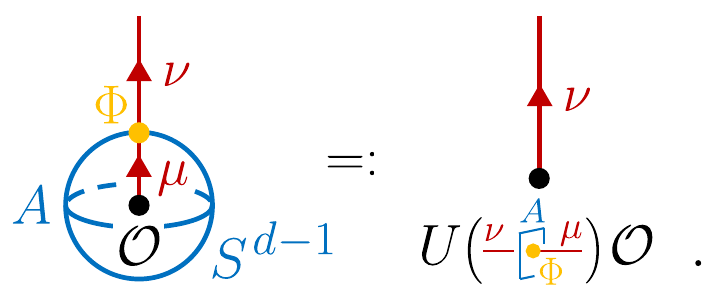}
\end{gathered}
\end{equation}
Since states on the sphere may be twisted by genuine line defects $\mu$ and $\nu$, this again requires the choice of a \textit{transition channel} $\Phi \in \mathcal{C}^{\hspace{1pt}A}(\mu,\nu)$. The corresponding linear map that captures the linking action of $A$ on the Hilbert space $\mathcal{H}_{\mu}$ of $\mu$-twisted operators is denoted by
\begin{equation}
    U\Big( \nhtub[\mu]{\nu}{A}{\Phi} \Big): \; \mathcal{H}_{\mu} \; \to \; \mathcal{H}_{\nu} \; .
\end{equation}
Mathematically, this defines a representation of the \textit{tube algebra} associated to $\mathcal{C}$ as introduced in \cite{Bartsch2023a}, which we will again refer to as a \textit{generalised charge} \cite{Bhardwaj:2023wzd,Bhardwaj:2023ayw}.

The analogue of the CPP Theorem \ref{thm-categorical-wigner} for higher unitary fusion categories $\mathcal{C}$ can then be formulated as follows:

\begin{theorem}[Higher CPP]
\label{thm-higher_cat_wigner-main}
Let $A$ be a codimension-one symmetry defect and let $\mu$ be a genuine line defect in $\mathcal{C}$. Then for each simple genuine line $\sigma$ there exists a basis $\lbrace e_i^{\raisebox{-2pt}{$\scriptstyle \sigma$}} \rbrace_i \, \subset \, \mathcal{C}^{\hspace{0.5pt}A\hspace{-0.5pt}}(\mu,\sigma)$ of transition channels such that in any given generalised charge $U$ it holds that
\begin{equation}
\label{eq-generalised-preservation-higher}
    \sum_{\sigma\hspace{0.5pt},\hspace{1pt}i} \hspace{1pt} \Big\langle U\Big( \nhtub[\mu]{\sigma}{A}{\scriptstyle\hspace{1pt} \raisebox{-0pt}{$\scriptstyle e_i^{\raisebox{-2pt}{$\scriptscriptstyle \sigma$}}$}} \Big) \hspace{1pt} \Phi \hspace{1pt}, \, U\Big( \nhtub[\mu]{\sigma}{A}{\scriptstyle\hspace{1pt} \raisebox{-0pt}{$\scriptstyle e_i^{\raisebox{-2pt}{$\scriptscriptstyle \sigma$}}$}} \Big) \hspace{1pt} \Psi \Big\rangle \,\; = \,\; \braket{\Phi,\Psi}
\end{equation}
for all $\ket{\Phi}, \ket{\Psi} \in \mathcal{H}_{\mu}$. The basis $\lbrace e_i^{\raisebox{-2pt}{$\scriptstyle \sigma$}} \rbrace_i$ is unique up to transformations $\vec{e}^{\hspace{1.5pt}\sigma} \mapsto M \hspace{-2pt}\cdot\hspace{-1pt} \vec{e}^{\hspace{1.5pt}\sigma}$ by unitary matrices $M$. 
\end{theorem}

\noindent
A more detailed discussion of the above can be found in SM \ref{ssec-higher-d}. In SM \ref{ssec-boundaries}, we discuss further generalisations of the CPP Theorem to quantum systems with boundary.

\vspace{4pt}
\smallskip
\noindent
{\bf Conclusion and Outlook.} In this work, we resolved the apparent tension between probability preservation in the sense of Wigner's theorem and the existence of non-invertible symmetries by proving the CPP Theorem~\ref{thm-categorical-wigner} (and higher-dimensional versions thereof). Our results shed light on the physical relevance of including twisted sectors and imposing unitarity on the symmetry category.

We expect our results to have direct applications to scattering, in particular off non-invertible defects, {where the different transition channels encode the possible scattering outcomes of a given operator.} This should elucidate the observations in \cite{vanBeest:2023dbu}. {Moreover, as generalised non-invertible symmetries act not only on states/local operators but also on extended operators, it would be interesting to generalise our results to this case.}

\smallskip
\noindent{\bf Acknowledgements.}
We thank Yuji Tachikawa for comments on the draft and Andrea Antinucci, Christian Copetti, Mark Mezei and Alison Warman for discussions. This work is supported by the UKRI Frontier Research Grant, underwriting the ERC Advanced Grant ``Generalized Symmetries in Quantum Field Theory and Quantum Gravity''. Authors are ordered by contribution.

\let\oldaddcontentsline\addcontentsline
\renewcommand{\addcontentsline}[3]{}
\setlength{\bibsep}{4pt}
\AtBeginEnvironment{thebibliography}{\justifying}
\bibliographystyle{JHEP.bst}
\small
\bibliography{references}
\let\addcontentsline\oldaddcontentsline
\onecolumngrid

\newpage
\clearpage

\widetext
\begin{center}
\textit{\large Supplementary Material} \\[4.5pt]
\textbf{\large Beyond Wigner: Non-Invertible Symmetries Preserve Probabilities}\\

\bigskip

{Thomas Bartsch, Yuhan Gai, and Sakura Sch\"afer-Nameki}\\[2pt]
{\it Mathematical Institute, University
of Oxford, Woodstock Road, Oxford, OX2 6GG, United Kingdom}
\bigskip 
\bigskip
\end{center}

\setcounter{equation}{0}
\setcounter{figure}{0}
\setcounter{table}{0}

\makeatletter
\renewcommand{\theequation}{\thesection.\arabic{equation}}
\@addtoreset{equation}{section}
\makeatother

\twocolumngrid

\tableofcontents

\section{Mathematical Background}
\label{sec-mathematical-background}

\noindent
In this section, we provide the mathematical background underlying the Categorical Probability Preservation (CPP) Theorem presented in the main text. We begin by reviewing (unitary) fusion categories and their associated tube categories, and proceed by giving a constructive proof of Theorem \ref{thm-categorical-wigner}.

\subsection{Fusion Categories}
\label{ssec-fusion-cats}

\noindent
Throughout this letter, we assume the finite bosonic generalised symmetries of a two-dimensional quantum theory to be described by a fusion category $\mathcal{C}$, whose objects and morphisms correspond to topological line defects and their local junctions, respectively. Below, we summarise the most salient features of this mathematical notion (for more extensive reviews we refer the reader to e.g. \cite{Etingof2017,Etingof2002,Bakalov2001}):
\begin{itemize}[leftmargin=2.5ex]
    \item \textbf{Linearity:} As a category, $\mathcal{C}$ is enriched over $\text{Vect}$, meaning that the morphism space $\mathcal{C}(A,B)$ is a finite-dimensional vector space for all objects $A,B \in \mathcal{C}$ such that composition of morphisms is a linear operation.
    \item \textbf{Finite semisimplicity:} Every object $A \in \mathcal{C}$ can be decomposed into a finite direct sum
    \begin{equation}
    A \;\, \cong \;\, \bigoplus_{i \hspace{1pt} = \hspace{1pt} 1}^n \; A_i \cdot S_i
    \end{equation} 
    (with multiplicities $A_i \in \mathbb{N}$) of finitely many \textit{simple objects} $S_i \in \mathcal{C}$, which are such that 
    \begin{equation}
        \mathcal{C}(S_i,S_j) \; = \; \delta_{ij} \, \mathbb{C} \; .
    \end{equation}
    We denote by $\text{Irr}(\mathcal{C})$ the finite set of (representatives of isomorphism classes of) simple objects in $\mathcal{C}$.
    
    \item \textbf{Monoidal structure:} The category $\mathcal{C}$ is a \textit{monoidal} in the sense that it is equipped with a linear functor 
    \begin{equation}
        \otimes: \; \mathcal{C} \hspace{0.5pt}\boxtimes\hspace{0.5pt} \mathcal{C} \; \to \; \mathcal{C}
    \end{equation}
    that captures the fusion of of topological defects as illustrated on the RHS of (\ref{eq-fusion-category}). The associativity of fusion is controlled by a natural isomorphism called the \textit{associator}, which mediates between the two possible ways of bringing three parallel lines $A,B,C \in \mathcal{C}$ together: 
    \begin{equation}
    \vspace{-5pt}
    \begin{gathered}
    \includegraphics[height=1.05cm]{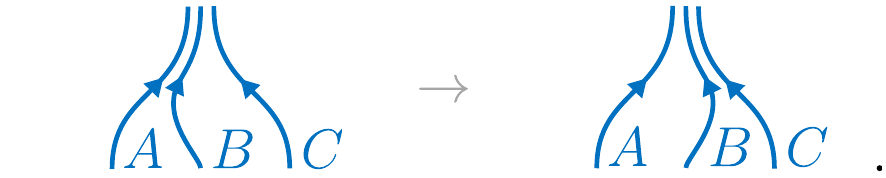}
    \end{gathered}
    \end{equation}
    Compatibility with bringing four lines together then requires the associator to obey the \textit{pentagon relation} \cite{Etingof2017}.
    
    \item \textbf{Dual structure:} For each object $A \in \mathcal{C}$ there exists a \textit{dual object} $A^{\vee} \in \mathcal{C}$ (the orientation reversal of $A$) together with \textit{evaluation} and \textit{coevaluation morphisms}
    \begin{equation}
    \vspace{-5pt}
    \begin{gathered}
    \includegraphics[height=1.05cm]{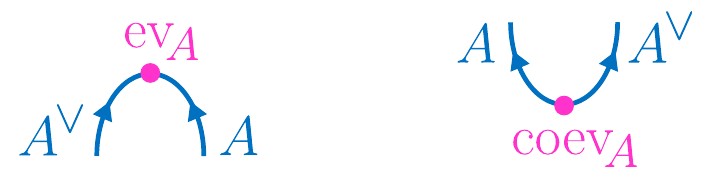}
    \end{gathered}
    \end{equation}
    that satisfy suitable \textit{zig-zag relations} \cite{Etingof2017}. The assignment $A \mapsto A^{\vee}$ then yields a contravariant functor $\vee: \mathcal{C} \to \mathcal{C}$ that acts on morphisms via
    \begin{equation}
    \vspace{-5pt}
    \begin{gathered}
    \includegraphics[height=1.6cm]{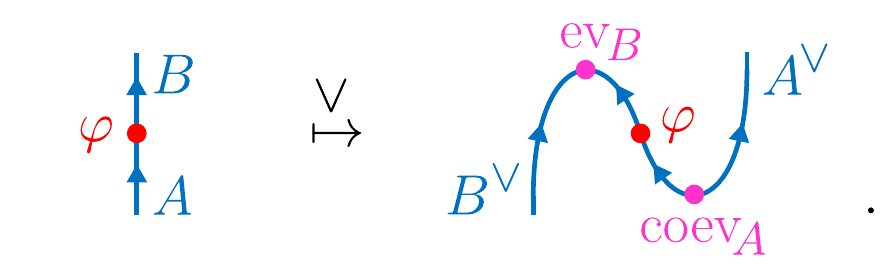}
    \end{gathered}
    \end{equation}
    A \textit{pivotal structure} on $\mathcal{C}$ is a choice of natural isomorphism $\xi: \vee \circ \vee \Rightarrow \text{Id}_{\mathcal{C}}$. We say that $\xi$ is \textit{spherical} if its associated left and right traces agree, i.e. 
    \vspace{-2pt}
    \begin{equation}
    \vspace{-5pt}
    \begin{gathered}
    \quad\includegraphics[height=1.7cm]{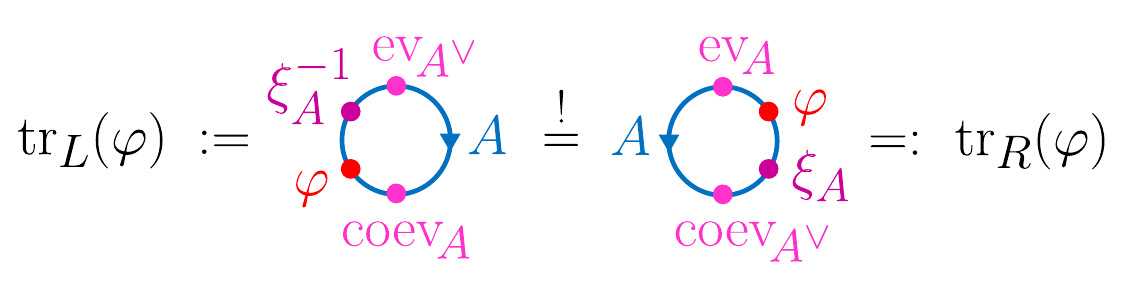}
    \vspace{-8pt}
    \end{gathered}
    \vspace{8pt}
    \end{equation}
    for all endomorphisms $\varphi \in \mathcal{C}(A,A)$. In this case, we define the \textit{quantum dimension} of an object $A \in \mathcal{C}$ by 
    \begin{equation}
    \label{eq-quantum-dimension}
        d_A \; := \; \text{tr}_{L/R}(\text{id}_A) \, \in \, \mathbb{C} \; .
    \end{equation}
\end{itemize}
For the purposes of this letter, of particular importance are those fusion categories $\mathcal{C}$ that are \textit{unitary}, meaning that they are equipped with the following additional structure:
\begin{itemize}[leftmargin=2.5ex]
    \item \textbf{Unitarity:} We require $\mathcal{C}$ to be a \textit{$\dagger$-category} in the sense that it is equipped with a contravariant functor 
    \begin{equation}
        \dagger: \; \mathcal{C} \, \to \, \mathcal{C}
    \end{equation}
    that acts as the identity on objects and antilinearly on morphisms such that $\dagger \circ \dagger = \text{Id}_{\mathcal{C}}$. Physically, $\dagger$ implements spacetime reflections of topological defects and their junctions about a fixed hyperplane:
    \begin{equation}
    \vspace{-5pt}
    \begin{gathered}
    \includegraphics[height=1.05cm]{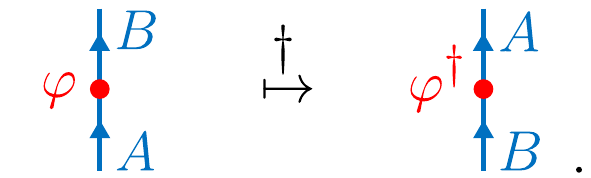}
    \end{gathered}
    \end{equation}
    We demand this $\dagger$-structure to be compatible with the remaining structures on $\mathcal{C}$ in the follwoing ways:
    \begin{itemize}[leftmargin=2.5ex,label=$\circ$]
        \item The monoidal structure is required to be compatible with the $\dagger$-structure in the sense that $\otimes$ is a $\dagger$-functor\bfootnote{A functor $F: \mathcal{C} \to \mathcal{C}'$ between two $\dagger$-categories is said to be a \textit{$\dagger$-functor} if $F \circ \dagger = \dagger' \circ F$.} and the components of the are associator unitary\bfootnote{A morphism $\varphi: A \to B$ in $\mathcal{C}$ is called \textit{unitary} if $\varphi^{\dagger} \circ \varphi = \text{id}_A$ and $\varphi \circ \varphi^{\dagger} = \text{id}_B$.}. If such a $\dagger$-structure exists, it is unique up to equivalence \cite{Reutter2019}.

        \item We require the dual structure to be \textit{unitary} in the sense that $\vee$ is a $\dagger$-functor with unitary coherence isomorphisms. In this case, the unitary isomorphisms
        \vspace{-2pt}
        \begin{equation}
        \label{eq-unitary-pivotal}
        \vspace{-5pt}
        \begin{gathered}
        \qquad\quad \includegraphics[height=1.45cm]{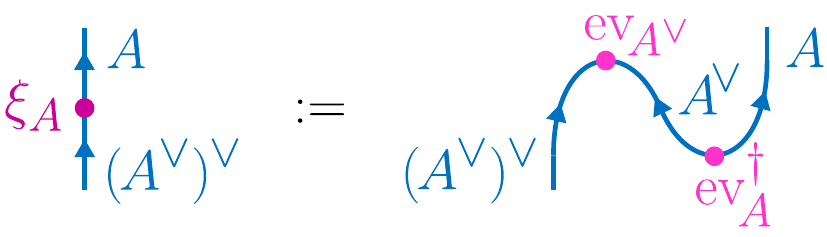}
        \vspace{-8pt}
        \end{gathered}
        \vspace{8pt}
        \end{equation}
        yield a canonical pivotal structure on $\mathcal{C}$. There exists a unique unitary dual structure on $\mathcal{C}$ such that the associated pivotal structure (\ref{eq-unitary-pivotal}) is spherical \cite{Penneys2021}. We will assume that $\mathcal{C}$ is equipped with this unique unitary dual structure in what follows. As a result, we have that the quantum dimensions of objects $A \in \mathcal{C}$ satisfy $d_A > 0$.
    \end{itemize}
\end{itemize}

\subsection{Tube Categories}

\noindent
Given a fusion category $\mathcal{C}$, we can associate to it its so-called \textit{tube category}\bfootnote{The notion of the tube category is closely related to the notion of the \textit{tube algebra} associated to $\mathcal{C}$ \cite{Ocneanu2016ChiralityFO}, which is given by
\begin{equation}
    \text{Tube}(\mathcal{C}) \; = \; \text{End}_{\hspace{1pt} \text{T}\mathcal{C}}\Big( \bigoplus\nolimits_i S_i \Big) \; .
\end{equation}
The tube algebra is `Morita equivalent' to the tube category in the sense that there is an equivalence
\begin{equation}
    \text{Rep}(\text{Tube}(\mathcal{C})) \; \cong \; [\text{T}\mathcal{C},\text{Hilb}] \; ,
\end{equation}
where the LHS denotes the category of representations of the tube algebra and the RHS denotes the category of linear functors from $\text{T}\mathcal{C}$ into the category of Hilbert spaces.} \cite{HARDIMAN2020,Bartsch2023a}, which is the linear category $\text{T}\mathcal{C}$ that has the same objects $X$,$Y$ as $\mathcal{C}$ but enlarged morphism spaces given by the quotient vector spaces
\begin{equation}
\label{eq-tube-morphism-space}
\text{T}\mathcal{C}(X,Y) \;\, := \,\; \bigoplus_{A \hspace{1pt} \in \hspace{1pt} \mathcal{C}} \, \mathcal{C}^{\hspace{0.5pt}A\hspace{-0.5pt}}(X,Y) \; \Big/ \sim
\end{equation}
(cf. equation (\ref{eq-transition-channels}) for notation) of transition channels
\begin{equation}
\label{eq-transition-channel}
\vspace{-5pt}
\begin{gathered}
\includegraphics[height=1.05cm]{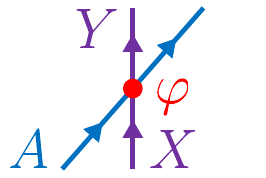}
\end{gathered}
\end{equation}
subjected to the equivalence relation that is generated by
\begin{equation}
\label{eq-equivalence-relation}
\vspace{-5pt}
\begin{gathered}
\includegraphics[height=1.39cm]{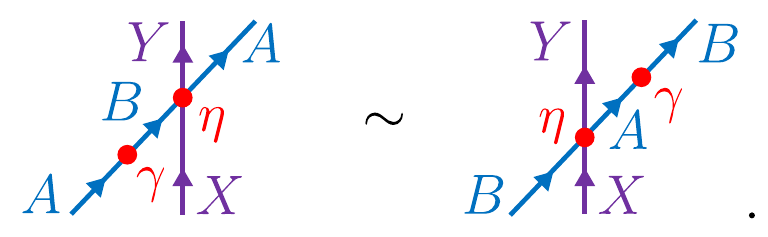}
\end{gathered}
\end{equation}
We will denote the equivalence class of a channel (\ref{eq-transition-channel}) under the equivalence relation (\ref{eq-equivalence-relation}) by
\begin{equation}
\tub[\hspace{-2.4pt}X]{\hspace{0.5pt}Y}{\hspace{-1pt}A}{\scriptstyle\hspace{1pt} \varphi} \; \in \; \text{T}\mathcal{C}(X,Y) \;.
\end{equation}
As a vector space, (\ref{eq-tube-morphism-space}) is isomorphic to
\begin{equation}
\text{T}\mathcal{C}(X,Y) \;\, \cong \,\; \bigoplus_{i \hspace{1pt} = \hspace{1pt} 1}^n \; \mathcal{C}^{\hspace{0.8pt}S_i\hspace{-0.5pt}}(X,Y)
\end{equation}
and is hence finite-dimensional. The composition of morphisms in $\text{T}\mathcal{C}$ is induced by the stacking
\begin{equation}
\vspace{-5pt}
\begin{gathered}
\includegraphics[height=1.77cm]{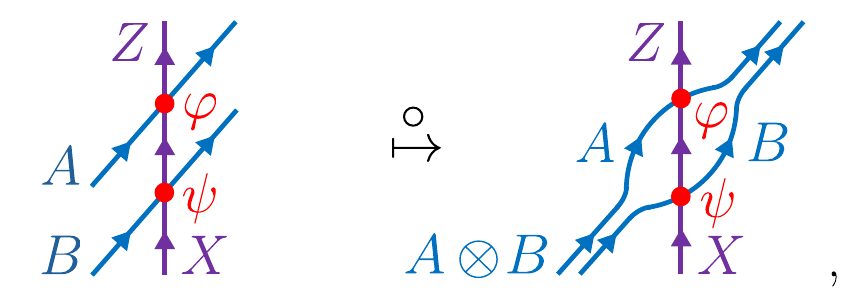}
\end{gathered}
\end{equation}
which we denote schematically by 
\begin{equation}
\tub[\hspace{-1.5pt}Y]{\hspace{0.5pt}Z}{\hspace{-1.5pt}A}{\hspace{1pt} \varphi} \; \circ \; \tub[\hspace{-2.4pt}X]{\hspace{0.5pt}Y}{\hspace{-1pt}B}{\raisebox{-0.4pt}{\hspace{1pt}$\scriptstyle \psi$}} \;\; = \;\;  \tuub[\hspace{-2.4pt}X]{\hspace{0.4pt}Z}{\raisebox{0.7pt}{\hspace{-10pt}$\scriptstyle A \hspace{0.5pt} \otimes B$}}{\hspace{1pt} \varphi \hspace{0.6pt}\circ\hspace{0.6pt} \psi} \; .
\end{equation}
If the fusion category $\mathcal{C}$ is unitary, there exists a canonical $\dagger$-structure on $\text{T}\mathcal{C}$ induced by
\begin{equation}
\vspace{-5pt}
\begin{gathered}
\includegraphics[height=1.22cm]{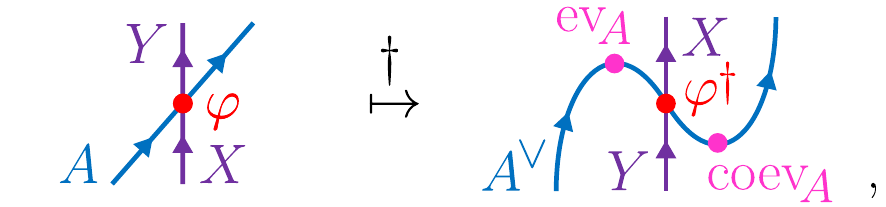}
\end{gathered}
\end{equation}
which we denote schematically by
\begin{equation}
\label{eq-tube-dagger}
\tub[\hspace{-2.4pt}X]{\hspace{0.5pt}Y}{\hspace{-1pt}A}{\hspace{1pt} \varphi}^{\dagger} \; = \;\, \tub[\hspace{-1.2pt}Y]{\hspace{0pt}X}{\hspace{-2pt}A^{\hspace{-0.8pt}\vee}}{\raisebox{-2.2pt}{\hspace{1pt}$\scriptstyle \varphi^{\dagger}$}} \; .
\end{equation}

A \textit{generalised charge} for $\mathcal{C}$ is defined to be a linear functor 
\begin{equation}
    U: \; \text{T}\mathcal{C} \, \to \, \text{Hilb}
\end{equation}
from the tube category into the category of Hilbert spaces, which assigns to each object $X \in \mathcal{C}$ a Hilbert space $\mathcal{H}_X$ and to each morphism $\scaleobj{0.85}{\tub[\hspace{-2.4pt}X]{\hspace{0.5pt}Y}{\hspace{-1pt}A}{\scriptstyle\hspace{1pt} \varphi}} \in \text{T}\mathcal{C}(X,Y)$ a linear map as in (\ref{eq-tube-wrapping-action}), such that the composition rule (\ref{eq-composition-rule}) is satisfied. If the fusion category $\mathcal{C}$ is unitary, we further require $U$ to be a $\dagger$-functor\bfootnote{For a unitary fusion category $\mathcal{C}$, it can be shown that every linear functor $U: \text{T}\mathcal{C} \to \text{Hilb}$ is equivalent to a $\dagger$-functor \cite{Galindo2014,Bartsch2025}. This can be seen as an analogue of the fact that every representation of a finite group is equivalent to a unitary one.}, meaning that it obeys equation (\ref{eq-dagger-condition}). As discussed in the main text of this letter, generalised charges $U$ capture the possible actions of the symmetry category $\mathcal{C}$ on twisted sector states of a two-dimensional quantum theory \cite{Bartsch:2023pzl,Bartsch2023a,Bhardwaj:2023wzd,Bhardwaj:2023ayw}.

\subsection{Proof of Main Theorem}
\label{ssec-proof}

\noindent
Having reviewed the notion of unitary fusion categories $\mathcal{C}$ and their associated tube categories, we are now in the position to prove the CPP Theorem~\ref{thm-categorical-wigner}. Our strategy is to give a constructive proof of the existence of a basis obeying equation~\eqref{eq-generalised-preservation}. To this end, we fix objects $A, X \in \mathcal{C}$ and let
\begin{equation}
    {}^{A\!}X \, := \, (A \hspace{-0.5pt}\otimes\hspace{-0.5pt} X) \hspace{-0.5pt}\otimes\hspace{-0.5pt} A^{\vee} \, \in \, \mathcal{C}
\end{equation}
denote the `conjugation of $X$ by $A$'. We then consider the direct sum decomposition
\begin{equation}
    {}^{A\!}X \; \cong \; \textstyle \bigoplus_S \hspace{1pt} D_{XS}^A \cdot S \; ,
\end{equation}
where $S \in \text{Irr}(\mathcal{C})$ runs over the simple objects of $\mathcal{C}$ and 
\begin{equation}
\label{eqn:multiplicity_NAXS}
\begin{aligned}
    D_{XS}^A \; &\equiv \; \text{dim}(\mathcal{C}({}^{A\!}X, S))  \; .
\end{aligned}
\end{equation}
For a fixed simple object $S$, we denote by $\pi_i^S: {}^{A\!}X \twoheadrightarrow S$ and $\imath_i^S: S \hookrightarrow {}^{A\!}X$ (where $i=1,...,D_{XS}^A$) the corresponding projection and inclusion morphisms satisfying
\begin{equation}
\label{eq-projection-inclusion}
    \pi_i^S \circ \imath_j^S \; = \; \delta_{ij} \cdot \text{id}_S \;\quad\; \text{and} \;\quad\; (\pi_i^S)^{\dagger} \; = \; \imath_i^S \; .
\end{equation}
Note that redefining $\vec{\pi}^{\hspace{1pt}S} \to M \hspace{-0.5pt}\cdot\hspace{0pt} \vec{\pi}^{\hspace{1pt}S}$ and $(\vec{\imath}^{\hspace{2pt}S})^{\text{T}} \to (\vec{\imath}^{\hspace{2pt}S})^{\text{T}} \hspace{-0.5pt}\cdot\hspace{0pt} M^{\dagger}$ for some unitary matrix $M \in \mathcal{U}(D_{XS}^A)$ gives an equivalent set of projections and inclusions obeying (\ref{eq-projection-inclusion}). For each index $i=1,...,D_{XS}^A$, we then define the transition channel 
\begin{equation}
\label{eqn:string_diagram_channel}
\vspace{-5pt}
\begin{gathered}
\includegraphics[height=1.9cm]{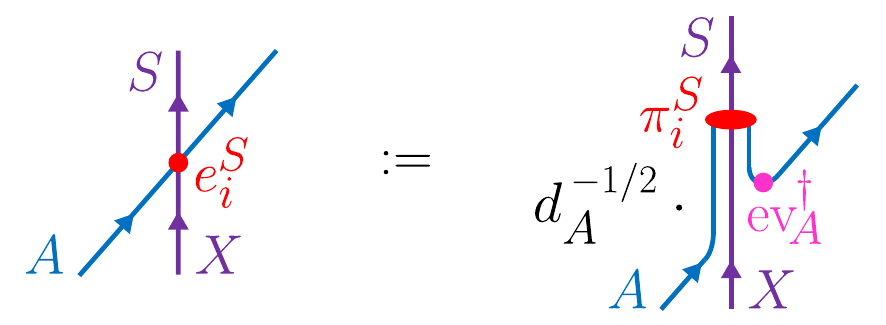}
\end{gathered}
\end{equation}
(where $d_A > 0$ is the quantum dimension of $A$ as in (\ref{eq-quantum-dimension})), which yields a basis $\lbrace e_i^S \rbrace_i$ of $\mathcal{C}^{\hspace{0.5pt}A\hspace{-0.5pt}}(X,S)$. Indeed, suppose that $\varphi \in \mathcal{C}^{\hspace{0.5pt}A\hspace{-0.5pt}}(X,S)$ is an arbitrary transition channel and let
\vspace{-2pt}
\begin{equation}
\vspace{-5pt}
\begin{gathered}
\includegraphics[height=2.3cm]{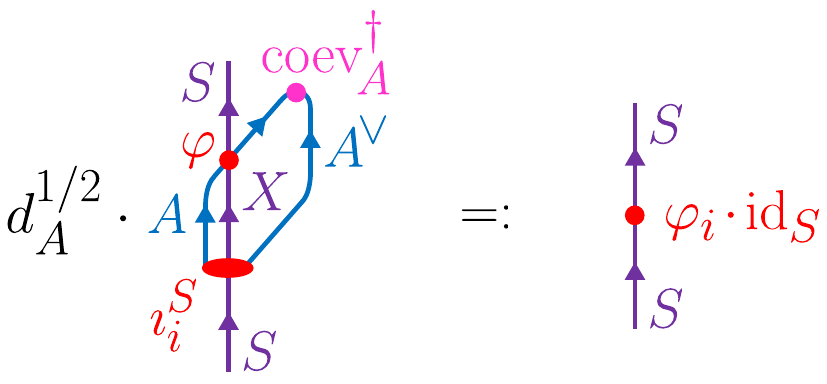}
\end{gathered}
\end{equation}
for all $i=1,...,D_{XS}^A$. Then, we have that 
\vspace{-5pt}
\begin{equation*}
\vspace{-5pt}
\begin{gathered}
\includegraphics[height=3.6cm]{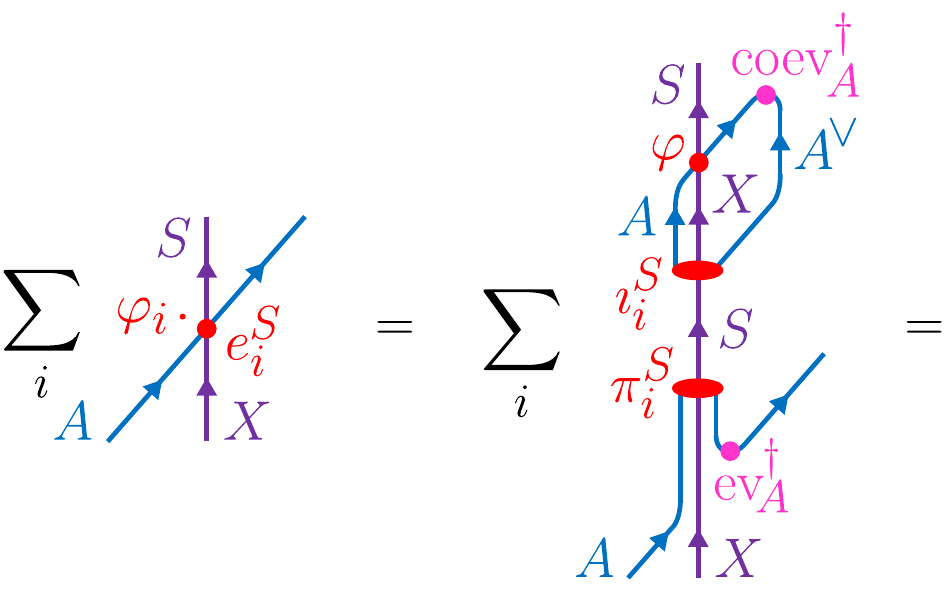} 
\end{gathered}
\end{equation*}
\vspace{-20pt}
\begin{equation}
\vspace{3pt}
\begin{gathered}
\includegraphics[height=3.6cm]{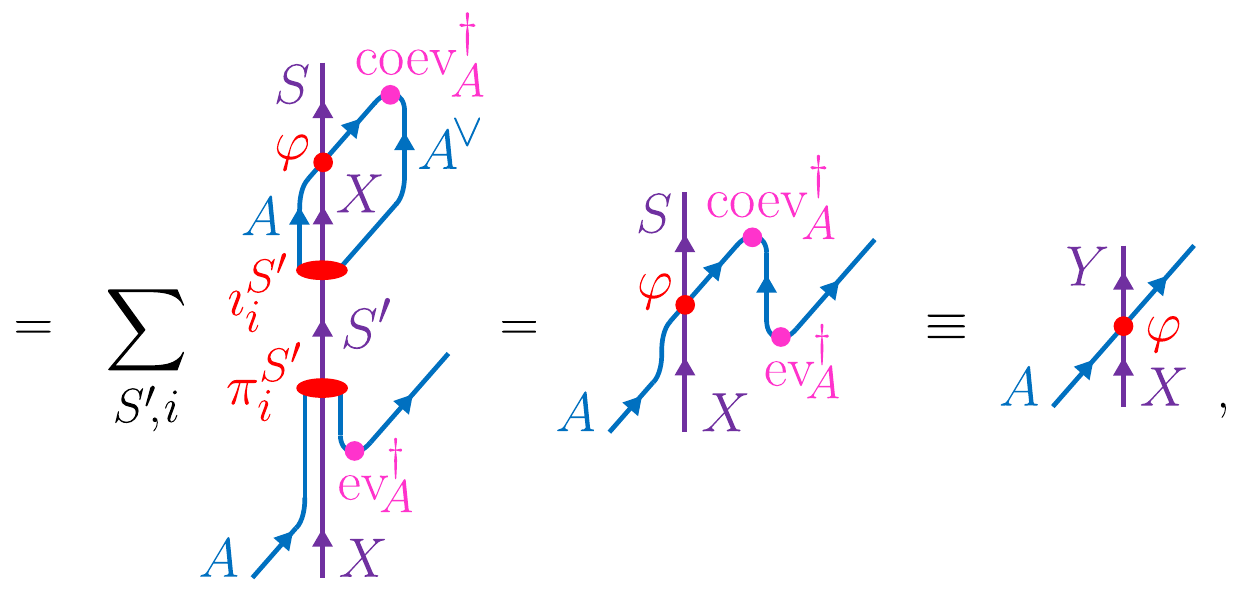}
\end{gathered}
\end{equation}
where we used the fact that $\mathcal{C}(S',S) = \delta_{S,S'} \,\mathbb{C}$ in the first step, the completeness relation
\begin{equation}
\label{eq-completeness-relation}
    \sum_{S',i} \hspace{0.8pt} \imath_i^{S'} \circ \pi_i^{S'} \; = \; \text{id}_{({}^{A\!}X)}
\end{equation}
in the second step, and the (daggered) zig-zag relation in the last step. For each $e^S_i \in \mathcal{C}^{\hspace{0.5pt}A\hspace{-0.5pt}}(X,S)$ from \eqref{eqn:string_diagram_channel}, the corresponding morphism in the tube category
\begin{equation}
    \tub[\hspace{-2.4pt}X]{\hspace{0.5pt}S}{\hspace{-1pt}A}{\scriptstyle\hspace{1pt} \raisebox{-2pt}{$\scriptstyle e_i^{\raisebox{-2pt}{$\scriptscriptstyle S$}}$}} \, \in \, \text{T}\mathcal{C}(X,S)
\end{equation}
then has the property that its adjoint is given by
\begin{equation}
\vspace{-5pt}
\begin{gathered}
\includegraphics[height=3.3cm]{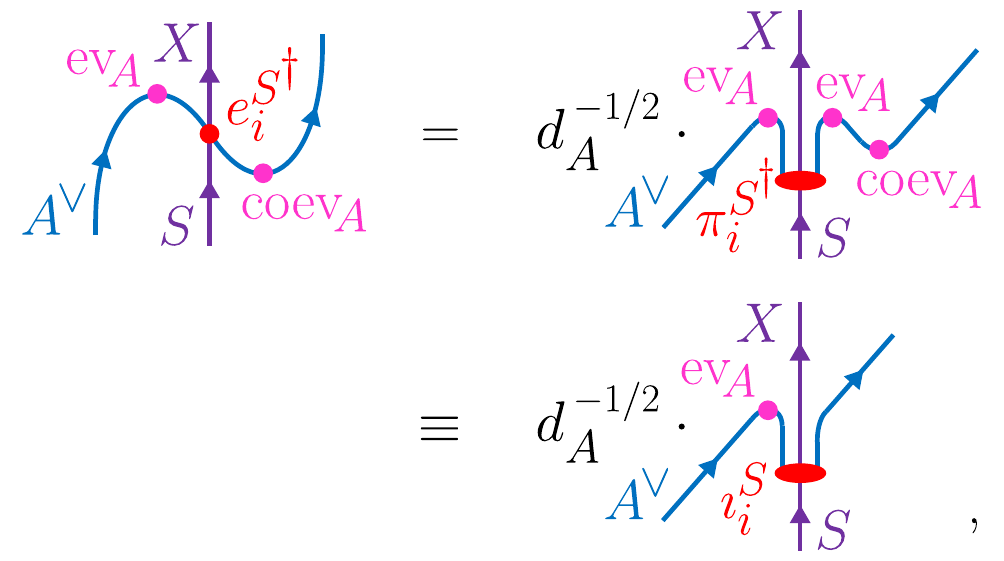}
\end{gathered}
\end{equation}
where we used the zig-zag relation as well as (\ref{eq-projection-inclusion}). In particular, composing $e_i^{\raisebox{-2pt}{$\scriptstyle S$}}$ with its adjoint and summing over all simple objects $S$ and basis indices $i$ gives $X$
\begin{equation}
\label{eqn:string_diagram_isometry}
\vspace{-5pt}
\begin{gathered}
\includegraphics[height=5.25cm]{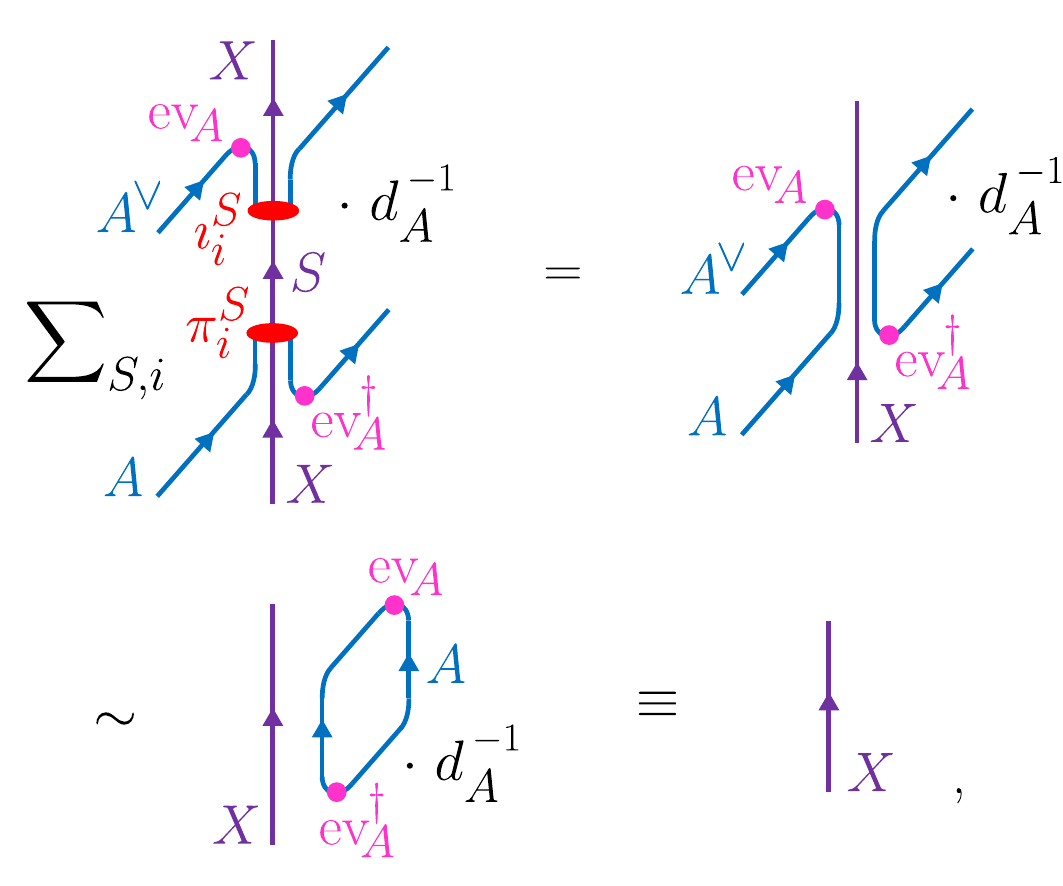}
\end{gathered}
\end{equation}
where we used the completeness relation (\ref{eq-completeness-relation}) in the first step and the equivalence relation (\ref{eq-equivalence-relation}) in the second step. As a result, the morphism
\begin{equation}
\label{eqn:notation_triple_line}
    \otub[\hspace{-2.4pt}X]{\hspace{-1pt}A} \; := \; \bigoplus_{S,i} \hspace{1pt} \tub[\hspace{-2.4pt}X]{\hspace{0.5pt}S}{\hspace{-1pt}A}{\scriptstyle\hspace{1pt} \raisebox{-2pt}{$\scriptstyle e_i^{\raisebox{-2pt}{$\scriptscriptstyle S$}}$}}
\end{equation}
from $X$ to the composite object $\bigoplus_S D_{XS}^A \hspace{1pt} S$ in $\text{T}\mathcal{C}$ obeys
\begin{equation}
    \otub[\hspace{-2.4pt}X]{\hspace{-1pt}A}^{\dagger} \circ \, \otub[\hspace{-2.4pt}X]{\hspace{-1pt}A} \; = \; \tub[\hspace{-2.4pt}X]{X}{\hspace{-0.2pt}1}{} \; .
\end{equation}
Consequently, given any generalised charge $U: \text{T}\mathcal{C} \to \text{Hilb}$ (which for unitary $\mathcal{C}$ can always be taken to be a $\dagger$-functor), the associated linear operator
\begin{equation}
\label{eq-induced-isometry}
    U(A)_X \; := \; U\Big( \notub[\hspace{-2.4pt}X]{\hspace{-1pt}A} \Big) \; \equiv \; \bigoplus_{S\hspace{0.5pt},\hspace{1pt}i} \, U\Big( \ntub[\hspace{-2.4pt}X]{\hspace{0.5pt}S}{\hspace{-1pt}A}{\scriptstyle\hspace{1pt} \raisebox{-2pt}{$\scriptstyle e_i^{\raisebox{-2pt}{$\scriptscriptstyle S$}}$}} \Big)
\end{equation}
defines an isometry from $\mathcal{H}_X$ to the enlarged Hilbert space $\bigoplus_S D_{XS}^A \hspace{1pt} \mathcal{H}_S$ that obeys equation (\ref{eq-generalised-preservation}) as claimed. This concludes the proof of Theorem \ref{thm-categorical-wigner}.

\section{Examples}
\label{sec-examples}

\noindent
In this section, we provide examples of unitary fusion categories, their corresponding tube categories and generalised charges, and the resulting isometries associated to the action of symmetry defects that preserve transition amplitudes in the sense of the CPP Theorem \ref{thm-categorical-wigner}. We discuss invertible and group-theoretical symmetries as well as symmetries of Tambara-Yamagami and Fibonacci type.

\subsection{Group Symmetry}
\label{app-pointed}

\noindent
Given a finite group $G$ and a 3-cocycle $\omega \in Z^3(G,U(1))$, the unitary pointed fusion category $\mathcal{C} = \Hilb_G^{\omega}$ has simple objects labelled by group elements $g \in G$ that fuse according to the group law of $G$. The associator is given by 
\be
\quad (g\otimes h) \otimes k \; \xrightarrow{\displaystyle \,\omega(g,h,k)\cdot \id_{ghk}\,} \; g \otimes (h \otimes k) \, .
\ee

\vspace{5pt}
\noindent\textbf{Tube Category.}
The associated tube category (whose corresponding tube algebra is given by the twisted Drinfeld double of $G$ \cite{Roche1990, Willerton2008}) has morphism spaces
\begin{equation}
    \text{T}\mathcal{C}(x,y) \, = \; \mathbb{C}\hspace{0.5pt}\raisebox{-1pt}{$\scaleobj{1.5}{[}$} \hspace{0.5pt} \tub[x]{y}{\raisebox{1pt}{$\scriptstyle g$}}{} \hspace{1.5pt} \raisebox{-1pt}{$\scaleobj{1.5}{|}$} \hspace{2pt} g \hspace{-1pt}\in\hspace{-1pt} G \;\; \text{s.t.} \;\, {}^gx = y \hspace{1pt}\raisebox{-1pt}{$\scaleobj{1.5}{]}$}
\end{equation}
(where $x,y \in G$ and ${}^gx:=gxg^{-1}$) and composition rules
\begin{equation}
\tub[\hspace{-3.3pt}{}^{h}\hspace{-0.95pt}x]{\hspace{-6pt}{}^{\mathrlap{\raisebox{-2.5pt}{\hspace{-1pt}\crule[white]{8pt}{8pt}}}g\hspace{-0.5pt}h}\hspace{-0.95pt}x}{\hspace{-0.2pt}\raisebox{1.5pt}{$\scriptstyle g$}}{} \, \circ \, \tub[x]{\scriptstyle \hspace{-2.7pt}{}^{\mathrlap{\raisebox{-1.5pt}{\hspace{-5pt}\crule[white]{8pt}{8pt}}}h}\hspace{-0.95pt}x}{\hspace{-0.2pt}h}{} \; = \; \tau_x(\omega)(g,h) \hspace{1pt} \cdot \hspace{1pt} \tub[x]{\hspace{-6pt}{}^{\mathrlap{\raisebox{-2.5pt}{\hspace{-1pt}\crule[white]{8pt}{8pt}}}g\hspace{-0.5pt}h}\hspace{-0.95pt}x}{\hspace{-2.4pt} gh}{}  \; ,
\end{equation}
where we defined the multiplicative phases 
\begin{equation}
\tau_x(\omega)(g,h) \; := \; \frac{\omega(g,h,x) \cdot \omega({}^{gh}x,g,h)}{\omega(g,{}^hx,h)} \; .
\end{equation}
Mathematically, this defines a 2-cocycle 
\begin{equation}
    \tau(\omega) \, \in \, Z^2(G/\!/G,U(1))
\end{equation}
on the action groupoid $G/\!/G$ for the conjugation action of $G$ on itself, which is called the \textit{transgression} of the 3-cocycle $\omega$. The $\dagger$-structure acts on morphisms via 
\begin{equation}
\tub[x]{\hspace{-2.7pt}{}^{\mathrlap{\raisebox{-2.5pt}{\hspace{-5pt}\crule[white]{8pt}{8pt}}}g}\hspace{-0.95pt}x}{\raisebox{1pt}{$\scriptstyle g$}}{}^{\dagger} \; = \; \tau^{\ast}_x(\omega)(g^{-1},g) \hspace{1pt} \cdot \hspace{1pt} \tub[\hspace{-3pt}{}^g\hspace{-0.95pt}x]{x}{\raisebox{1pt}{$\scriptstyle g^{-1}$}}{} \; .
\end{equation}

\vspace{5pt}
\noindent\textbf{Generalised Charges.}
As shown in \cite{Roche1990}, the irreducible generalised charges are labelled by pairs $(x,\rho)$ consisting of 
\begin{enumerate}[leftmargin=4ex]
    \item a representative $x$ of a conjugacy class $[x]\in\text{Cl}(G)$,
    \item an irreducible unitary projective representation $\rho$ of the centraliser $G_x:=\{g\in G \,|\, {}^gx=x\}$ of $x$ on a Hilbert space $\mathcal{V}$ with projective 2-cocycle $\tau_x(\omega) \in Z^2(G_x,U(1))$.
\end{enumerate}
Concretely, the associated generalised charge $U_{(x,\rho)}$ can be constructed by fixing for each $y \in [x]$ a representative $r_y \in G$ such that ${}^{(r_y)}y = x$ (with $r_x \equiv 1$) and setting
\begin{equation}
\label{eqn:gy_centraliser}
g_y \; := \; r_{({}^gy)} \cdot g \cdot r_y^{-1} \; \in \; G_x
\end{equation}
for $g \in G$ and $y \in [x]$. Then, $U_{(x,\rho)}$ acts on the twisted sectors
\begin{equation}
\mathcal{H}_y \; = \; \begin{cases} \mathcal{V} &\text{if} \; y \in [x] \\ 0 &\text{otherwise} \end{cases}
\end{equation}
via the linear operators
\begin{equation}
U_{(x,\rho)}\Big( \hspace{2pt}\ntub[\raisebox{1pt}{$\scriptstyle y$}]{\hspace{1pt}\raisebox{1pt}{$\scriptstyle \hspace{-2.7pt}{}^{g}\hspace{-0.95pt}y$}}{\raisebox{1pt}{$\scriptstyle g$}}{} \hspace{0.7pt}\Big) \; := \; \kappa_y(g) \cdot \rho(g_y ) \; ,
\end{equation}
where we defined the multiplicative phases
\begin{equation}
\label{eqn:kappaphase}
\kappa_y(g) \; := \; \frac{\tau_y(\omega)(r_{({}^gy)},g)}{\tau_y(\omega)(g_y,r_y)} \; .
\end{equation}
This construction depends on the choice of the representatives $r_y$ only up to isomorphism.

\vspace{5pt}
\noindent\textbf{Isometry Actions.}
For a given $g \in G$, the following yield bases of transition channels that obey Theorem~\ref{thm-categorical-wigner}:
\begin{equation}
    \Big\{ \tub[y]{\hspace{-2.7pt}{}^{\mathrlap{\raisebox{-2.5pt}{\hspace{0pt}\crule[white]{4pt}{8pt}}}g}\hspace{-0.95pt}y}{\raisebox{1pt}{$\scriptstyle{g}$}}{} \Big\} \; \subset \; \cC^{\hspace{1pt}g}(y, {}^gy) \; , 
\end{equation}
where $y\in G$. Upon choosing a generalised charge $U = U_{(x,\rho)}$, we then obtain the following linear isometries associated to the action of $g$ on $y$-twisted sectors (which in the present invertible case are in fact unitary as expected):
\be
U_{(x, \rho)}(g)_y \; = \; \delta_{\hspace{0.5pt}y \hspace{0.5pt} \in \hspace{0.5pt} [x]} \hspace{-0.5pt} \cdot \hspace{-0.5pt} \kappa_y(g) \hspace{-0.5pt} \cdot \hspace{-0.5pt} \rho(g_y ): \; \cH_{y} \; \rightarrow \; \cH_{({}^{g\hspace{-0.5pt}}y)}\, ,
\ee
In particular, this yields the following transition probability for the action of $g$ on $y$-twisted sector states:
\be
    \qquad p\Big( \hspace{2pt} \ntub[\raisebox{1pt}{$\scriptstyle y$}]{\hspace{-2pt}\raisebox{1pt}{$\scriptstyle {}^{g}\hspace{-0.95pt}y$}}{\raisebox{1pt}{$\scriptstyle g$}}{} \Big) \; = \; 1 \; .
\ee

\subsection{Tambara-Yamagami Symmetry}
\label{app-ty}
\noindent Given a finite abelian group $A$, a Tambara-Yamagami fusion category based on $A$ has simple objects given by group elements $a \in A$ together with a distinguished non-invertible object $m$ of dimension $d_m = |A|^{1/2}$, which fuse according to
\begin{equation}
\label{eq-ty-fusion-rules}
\begin{aligned}
    a \otimes b \; &= \; a \cdot b \; , \\
    a \otimes m \; &= \; m \otimes a \; = \; m \; , \\
    m \otimes m \; &= \; \bigoplus_{a \hspace{1pt} \in \hspace{1pt} A} \hspace{1pt} a \; .
\end{aligned}
\end{equation}
The different solutions to the pentagon equation for the associator are then classified by a choice of \cite{Tambara1998}
\begin{enumerate}[leftmargin=4ex]
    \item a non-deg. symmetric bicharacter $\chi: A \times A \to U(1)$,
    \item a square-root $s$ of $1/|A|$.
\end{enumerate}
We denote the corresponding fusion category by $\mathcal{C} = \text{TY}_A^{\chi,s}$ henceforth, which is unitary for all choices of $\chi$ and $s$. The non-trivial components of the associator are given by
\begin{equation*}
\vspace{-5pt}
\begin{gathered}
\includegraphics[height=4.45cm]{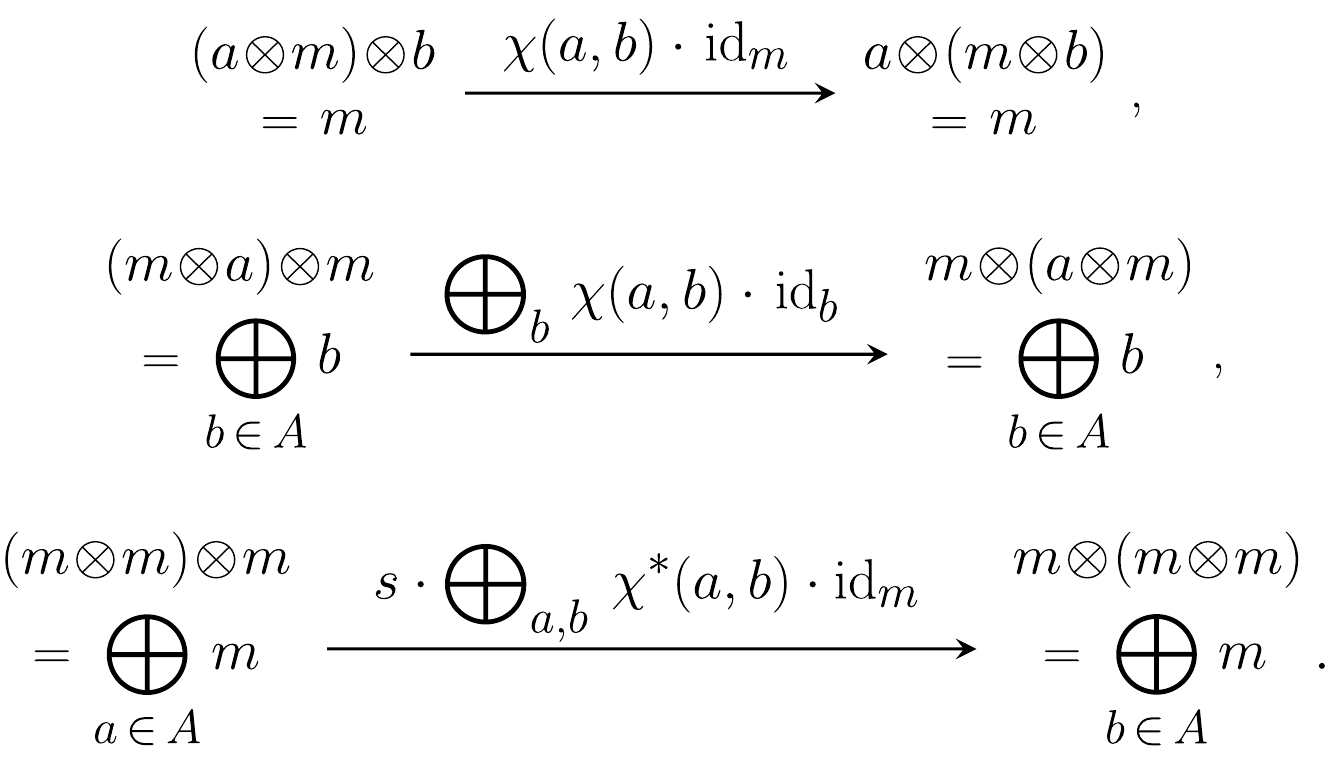}
\vspace{-2pt}
\end{gathered}
\vspace{4pt}
\end{equation*}

\vspace{5pt}
\noindent\textbf{Tube Category.}
The associated tube category was computed in \cite{Bartsch2025} and has morphism spaces
\begin{alignat*}{2}
    &\text{T}\mathcal{C}(x,y) \, &&= \; \delta_{x,y} \cdot \mathbb{C}\hspace{0.5pt}\raisebox{-1pt}{$\scaleobj{1.5}{[}$} \hspace{0.5pt} \tub[x]{x}{a}{}\hspace{1.5pt} \raisebox{-1pt}{$\scaleobj{1.5}{|}$} \hspace{1.5pt} a \hspace{-1pt}\in\hspace{-1pt} A \hspace{1pt}\raisebox{-1pt}{$\scaleobj{1.5}{]}$} \, \oplus \, \mathbb{C}\hspace{0.5pt}\raisebox{-1pt}{$\scaleobj{1.5}{[}$} \hspace{0.5pt} \tub[x]{y}{\!m}{} \hspace{0.5pt}\raisebox{-1pt}{$\scaleobj{1.5}{]}$} \; , \\[5pt]
    &\text{T}\mathcal{C}(m,m) \, &&= \; \mathbb{C}\hspace{0.5pt}\raisebox{-1pt}{$\scaleobj{1.5}{[}$} \hspace{0.5pt} \tub[\hspace{-2pt}m]{m}{a}{}\hspace{1pt}, \hspace{1pt}\tub[\hspace{-2pt}m]{m}{\!m}{b} \hspace{1.5pt} \raisebox{-1pt}{$\scaleobj{1.5}{|}$} \hspace{1.5pt} a,b \hspace{-1pt}\in\hspace{-1pt} A \hspace{1pt}\raisebox{-1pt}{$\scaleobj{1.5}{]}$} 
\end{alignat*}
(where $x,y \in A$), whose composition rules are given by
\begin{equation}
\begin{aligned}
\tub[x]{x}{a}{} \, \circ \, \tub[x]{x}{b}{} \; &= \; \tub[x]{x}{\hspace{-2.2pt} ab}{} \, , \\[3pt]
\tub[x]{y}{\!m}{} \, \circ \, \tub[x]{x}{a}{} \; &= \; \chi(a,y) \cdot \tub[x]{y}{\!m}{} \, , \\[3pt]
\tub[y]{y}{a}{} \, \circ \, \tub[x]{y}{\!m}{} \; &= \; \chi(a,x) \cdot \tub[x]{y}{\!m}{} \, , \\[3pt]
\tub[y]{z}{\!m}{} \, \circ \, \tub[x]{y}{\!m}{} \; &= \; \delta_{x,z} \cdot \chi^{\ast}(x,y) \hspace{1pt} \\[3pt]
&\hphantom{= \;} \cdot \hspace{1pt}\sum\nolimits_a \chi^{\ast}(a,y) \cdot \tub[x]{x}{a}{} \, , \\[4pt]
\tub[\hspace{-2.5pt} m]{m}{a}{} \, \circ \, \tub[\hspace{-2.5pt} m]{m}{b}{} \; &= \; \chi^{\ast}(a,b) \cdot \tub[\hspace{-2.5pt} m]{m}{\hspace{-2.2pt} ab}{} \, , \\[3pt]
\tub[\hspace{-2.5pt} m]{m}{a}{} \, \circ \, \tub[\hspace{-2.5pt} m]{m}{\hspace{-2pt}m}{b} \; &= \; \tub[\hspace{-2.5pt} m]{m}{\hspace{-2pt}m}{b} \, \circ \, \tub[\hspace{-2.5pt} m]{m}{a}{} \\[3pt] 
&= \; \chi(a,ab) \cdot \tub[\hspace{-2.5pt} m]{m}{\hspace{-2pt}m}{ab} \, , \\[3pt]
\tub[\hspace{-2.5pt} m]{m}{\hspace{-2pt}m}{a} \, \circ \, \tub[\hspace{-2.5pt} m]{m}{\hspace{-2pt}m}{b} \; &= \; \frac{s}{|A|} \cdot \chi(a,b) \\[3pt]
&\hphantom{= \;} \cdot \hspace{1pt} \sum\nolimits_c \chi^{\ast}(ab,c) \cdot \tub[\hspace{-2.5pt} m]{m}{c}{} \, .
\end{aligned}
\end{equation}

The $\dagger$-structure acts on morphisms via
\begin{equation}
\begin{aligned}
\tub[x]{x}{a}{}^{\dagger} \; &= \; \tub[x]{x}{\! a^{\scalebox{0.4}[0.4]{$-1$}}}{} \; , \\[3pt]
\tub[x]{y}{\!m}{}^{\dagger} \; &= \; \chi(x,y) \cdot \tub[y]{x}{\!m}{} \; , \\[3pt]
\tub[\hspace{-2.5pt} m]{m}{a}{}^{\dagger} \; &= \; \chi^{\ast}(a,a) \cdot \tub[\hspace{-2.5pt} m]{m}{\! a^{\scalebox{0.4}[0.4]{$-1$}}}{} \; , \\[3pt]
\tub[\hspace{-2.5pt} m]{m}{\hspace{-2pt}m}{a}^{\dagger} \; &= \; s \cdot \sum_b \, \chi^{\ast}(a,b) \cdot \tub[\hspace{-2.5pt} m]{m}{\hspace{-2pt}m}{b} \; .
\end{aligned}
\end{equation}

\vspace{5pt}
\noindent\textbf{Generalised Charges.}
There is a total of $\frac{1}{2} \hspace{1pt}|A| \hspace{-1pt}\cdot\hspace{-1pt}(|A| + 7)$ irreducible generalised charges \cite{Bartsch2023a,Bartsch2025}, which fall into the following three categories:
\begin{itemize}[leftmargin=3ex]
\item There are $2 \hspace{-1pt} \cdot \hspace{-1pt} |A|$ one-dimensional charges $U^{\Delta}_{\hspace{-1pt}x}$ labelled by group elements $x \in A$ and a choice of square-root $\Delta$ of $\chi^{\ast}(x,x) \in U(1)$. These act on $\mathcal{H}_x \cong \mathbb{C}$ via
\begin{equation}
\begin{aligned}
U^{\Delta}_{\hspace{-1pt}x}\Big(\ntub[x]{x}{a}{} \Big) \; &= \; \chi(a,x) \; , \\[3pt]
U^{\Delta}_{\hspace{-1pt}x}\Big( \ntub[x]{x}{\!m}{} \Big) \; &= \; \frac{1}{s} \cdot \Delta \; .
\end{aligned}
\end{equation}

\item There are $2 \! \cdot \! |A|$ one-dimensional charges $U^{\Delta}_{\hspace{-1pt}\rho}$ labelled by antiderivatives\bfootnote{An antiderivative of $\chi$ is a map $\rho: A \to U(1)$ such that $(d\rho)(a,b) := \rho(a)\cdot \rho(b) \cdot \rho^{\ast}(ab) \equiv \chi(a,b)$ for all $a,b \in A$. Since $\chi$ is symmetric, such a $\rho$ always exists and the set of all antiderivatives forms a torsor over $A^{\vee} = \text{Hom}(A,U(1))$. This shows that there are $|A^{\vee}| = |A|$ antiderivatives of $\chi$.} $\rho$ of $\chi$ and a choice of square-root $\Delta$ of $s \cdot \sum_c \rho^{\ast}(c) \in U(1)$. These act on $\mathcal{H}_m \cong \mathbb{C}$ via
\begin{equation}
\begin{aligned}
U^{\Delta}_{\hspace{-1pt}\rho}\Big( \ntub[\hspace{-2.5pt} m]{m}{a}{} \Big) \; &= \; \rho^{\ast}(a) \; , \\[3pt]
U^{\Delta}_{\hspace{-1pt}\rho}\Big( \ntub[\hspace{-2.5pt} m]{m}{\hspace{-2pt}m}{a} \Big) \; &= \; s \cdot \Delta \cdot \rho(a^{-1}) \; .
\end{aligned}
\end{equation}

\item There are $\frac{1}{2}|A|\hspace{-1pt}\cdot \hspace{-1pt}(|A|-1)$ two-dimensional charges $U_{x,y}$ labelled by distinct group elements $x \neq y \in A$. These act on the twisted sectors $\mathcal{H}_x \cong \mathcal{H}_y \cong \mathbb{C}$ via
\begin{equation}
\begin{aligned}
U_{x,y}\Big( \ntub[x]{x}{a}{} \Big) \; &= \; \chi(a,y) \; , \\[3pt]
U_{x,y}\Big( \ntub[y]{y}{a}{} \Big) \; &= \; \chi(a,x) \; , \\[3pt]
U_{x,y}\Big( \ntub[x]{y}{\!m}{} \Big) \; &= \; \frac{1}{s} \cdot \chi^{\ast}(x,y) \; , \\[3pt]
U_{x,y}\Big( \ntub[y]{x}{\!m}{} \Big) \; &= \; \frac{1}{s} \; .
\end{aligned}
\end{equation}
\end{itemize}

\vspace{5pt}
\noindent\textbf{Isometry Actions.}
The following yield bases of transition channels for the non-invertible defect $m$ that obey Theorem~\ref{thm-categorical-wigner}:
\begin{equation}
\begin{aligned}
    \left\{ s \cdot \tub[x]{y}{\!m}{} \right\} & \; \subset \; \cC^{\hspace{0.5pt}m}(x,y) \; , \\[3pt]
    \left\{ \tub[\hspace{-2.5pt} m]{m}{\hspace{-2pt}m}{a} \,|\, a\in A \right\} & \; \subset \; \cC^{\hspace{0.5pt}m}(m,m) \; ,
\end{aligned}
\end{equation}
where $x,y \in A$. Upon choosing a generalised charge $U$, the above then induce linear isometries $U(m)_X$ associated to the action of $m$ on $X$-twisted sector states:
\begin{itemize}[leftmargin=3ex]
    \item For the generalised charges $U = U_{\rho}^{\Delta}$ supported on the twisted sector $\cH_{m}\cong \bbC$, we obtain 
    \begin{equation}
    \begin{gathered}
    \qquad U_{\rho}^{\Delta}(m)_{m} \, = \, \bigoplus\nolimits_a \hspace{-1pt}s \hspace{-1pt}\cdot\hspace{-1pt} \Delta \hspace{-1pt}\cdot\hspace{-1pt} \rho(a^{-1}) : \; \cH_{m} \;\, \to \,\; \bigoplus\nolimits_a \hspace{-1pt}\cH_{m} \, , \\[5pt]
    \end{gathered}
    \vspace{3pt}
    \end{equation}
    As a result, we obtain the following transition probabilities for the action of $m$ on $m$-twisted sector states:
    \begin{equation}
        \qquad p\Big( \hspace{0.5pt}\ntub[\hspace{-2.5pt} m]{m}{\hspace{-2pt}m}{a} \Big) \; = \; \frac{1}{|A|} \; .
    \end{equation}
    
    \item For the generalised charge $U = U_{x,y}$ supported on the twisted sectors $\cH_x \cong \cH_y \cong \bbC$, we have that
    \begin{equation}
    \begin{aligned}
    \qquad U_{x,y}(m)_x \, = \,  \begin{bmatrix} 0 \\ \chi(x,y)^* \end{bmatrix} : \;\; &\mathcal{H}_x \;\, \to \,\; \mathcal{H}_x \oplus \mathcal{H}_y \; , \\[2pt]
    \qquad U_{x,y}(m)_y \, = \,  \begin{bmatrix} \;1\;\, \\ \;0\;\, \end{bmatrix} : \;\; &\mathcal{H}_y \;\, \to \,\; \mathcal{H}_x \oplus \mathcal{H}_y \; .
    \end{aligned}
    \vspace{4pt}
    \end{equation}
    As a result, we obtain the following transition probabilities for the action of $m$ on twisted sector states:
    \begin{equation}
    \begin{alignedat}{2}
        \qquad p\Big( s \hspace{1pt}\tub[x]{x}{\!m}{} \Big) \; &= \; 0 \; , \qquad p\Big( s \hspace{1pt}\tub[y]{x}{\!m}{} \Big) \; &&= \; 1 \; , \\
        \qquad p\Big( s \hspace{1pt}\tub[x]{y}{\!m}{} \Big) \; &= \; 1 \; , \qquad p\Big( s \hspace{1pt}\tub[y]{y}{\!m}{} \Big) \; &&= \; 0 \; . \\[3pt]
    \end{alignedat}
    \end{equation}
\end{itemize}

\subsection{\texorpdfstring{$\Rep(S_3)$}{Rep(S3)} Symmetry}
\label{app-RepS3}
\noindent The unitary fusion category $\mathcal{C} = \Rep(S_3)$ of finite-dimensional representations of the permutation group $S_3$ has simple objects $1$, $\psi$, and $\pi$ of dimensions $d_1 = d_{\psi} = 1$ and $d_{\pi}=2$, respectively, which fuse according to 
\begin{equation}
\ba
\psi \otimes \psi \, &= \, 1 \, , \\ 
\psi \otimes \pi \, &= \, \pi \otimes \psi \, = \, \pi \, , \\ 
\pi \otimes \pi \, &= \, 1 \oplus \psi \oplus \pi \, .
\ea
\end{equation}
The non-trivial components of the associator are given by
\begin{equation*}
\vspace{-5pt}
\begin{gathered}
\includegraphics[height=3.8cm]{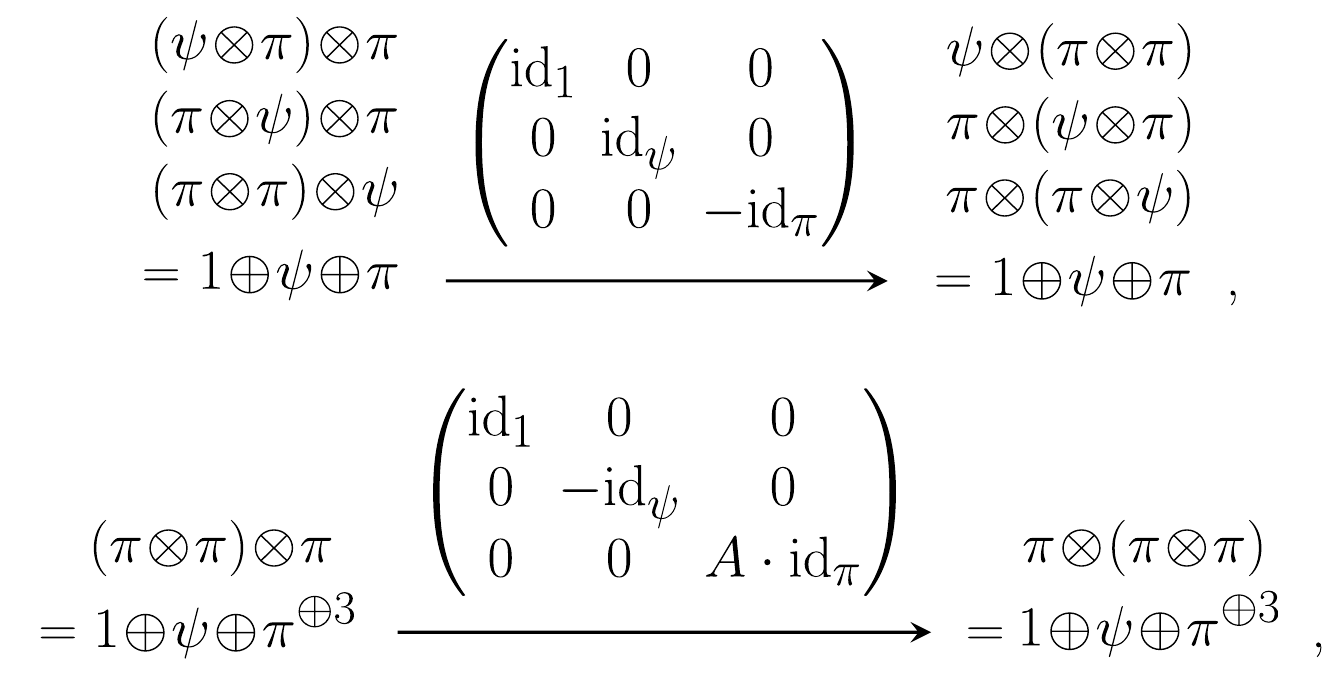}
\vspace{-2pt}
\end{gathered}
\vspace{4pt}
\end{equation*}
where the $(3\!\times\!3)$-matrix $A$ is given by 
\begin{equation}
A \; = \; \frac{1}{2}
\begin{pmatrix}
1 \; & 1 \; & \sqrt{2} \\
1 \; & 1 \; & -\sqrt{2} \\
\sqrt{2} \; & -\sqrt{2} \; & 0
\end{pmatrix} \; .
\end{equation}

\vspace{5pt}
\noindent\textbf{Tube Category.}
The associated tube category (which to our knowledge has not yet been computed in the literature) has morphism spaces
\begin{alignat*}{2}
    &\text{T}\mathcal{C}(1,1) \, &&= \; \mathbb{C}\hspace{0.5pt}\raisebox{-1pt}{$\scaleobj{1.5}{[}$} \hspace{0.5pt} \tub[1]{1}{1}{}\hspace{1pt}, \hspace{1pt}\tub[1]{1}{\raisebox{0.9pt}{$\scriptstyle \psi$}}{}\hspace{1pt}, \hspace{1pt}\tub[1]{1}{\pi}{} \hspace{0.5pt}\raisebox{-1pt}{$\scaleobj{1.5}{]}$} \; , \\[2pt]
    &\text{T}\mathcal{C}(1,\psi) \, &&= \; \mathbb{C}\hspace{0.5pt}\raisebox{-1pt}{$\scaleobj{1.5}{[}$} \hspace{0.5pt} \tub[1]{\raisebox{0.9pt}{$\scriptstyle \psi$}}{\pi}{} \hspace{0.5pt}\raisebox{-1pt}{$\scaleobj{1.5}{]}$} \; , \\[2pt]
    &\text{T}\mathcal{C}(1,\pi) \, &&= \; \mathbb{C}\hspace{0.5pt}\raisebox{-1pt}{$\scaleobj{1.5}{[}$} \hspace{0.5pt} \tub[1]{\pi}{\pi}{} \hspace{0.5pt}\raisebox{-1pt}{$\scaleobj{1.5}{]}$} \; , \\[2pt]
    &\text{T}\mathcal{C}(\psi,1) \, &&= \; \mathbb{C}\hspace{0.5pt}\raisebox{-1pt}{$\scaleobj{1.5}{[}$} \hspace{0.5pt} \tub[\hspace{-1pt}\raisebox{0.9pt}{$\scriptstyle \psi$}]{1}{\pi}{} \hspace{0.5pt}\raisebox{-1pt}{$\scaleobj{1.5}{]}$} \; , \\[2pt]
    &\text{T}\mathcal{C}(\psi,\psi) \, &&= \; \mathbb{C}\hspace{0.5pt}\raisebox{-1pt}{$\scaleobj{1.5}{[}$} \hspace{0.5pt} \tub[\hspace{-1pt}\raisebox{0.9pt}{$\scriptstyle \psi$}]{\raisebox{0.9pt}{$\scriptstyle \psi$}}{1}{}\hspace{1pt}, \hspace{1pt}\tub[\hspace{-1pt}\raisebox{0.9pt}{$\scriptstyle \psi$}]{\raisebox{0.9pt}{$\scriptstyle \psi$}}{\raisebox{0.9pt}{$\scriptstyle \psi$}}{}\hspace{1pt}, \hspace{1pt}\tub[\hspace{-1pt}\raisebox{0.9pt}{$\scriptstyle \psi$}]{\raisebox{0.9pt}{$\scriptstyle \psi$}}{\pi}{} \hspace{0.5pt}\raisebox{-1pt}{$\scaleobj{1.5}{]}$} \; , \\[2pt]
    &\text{T}\mathcal{C}(\psi,\pi) \, &&= \; \mathbb{C}\hspace{0.5pt}\raisebox{-1pt}{$\scaleobj{1.5}{[}$} \hspace{0.5pt} \tub[\hspace{-1pt}\raisebox{0.9pt}{$\scriptstyle \psi$}]{\pi}{\pi}{} \hspace{0.5pt}\raisebox{-1pt}{$\scaleobj{1.5}{]}$} \; , \\[2pt]
    &\text{T}\mathcal{C}(\pi,1) \, &&= \; \mathbb{C}\hspace{0.5pt}\raisebox{-1pt}{$\scaleobj{1.5}{[}$} \hspace{0.5pt} \tub[\pi]{1}{\pi}{} \hspace{0.5pt}\raisebox{-1pt}{$\scaleobj{1.5}{]}$} \; , \\[2pt]
    &\text{T}\mathcal{C}(\pi,\psi) \, &&= \; \mathbb{C}\hspace{0.5pt}\raisebox{-1pt}{$\scaleobj{1.5}{[}$} \hspace{0.5pt} \tub[\pi]{\raisebox{0.9pt}{$\scriptstyle \psi$}}{\pi}{} \hspace{0.5pt}\raisebox{-1pt}{$\scaleobj{1.5}{]}$} \; , \\[2pt]
    &\text{T}\mathcal{C}(\pi,\pi) \, &&= \; \mathbb{C}\hspace{0.5pt}\raisebox{-1pt}{$\scaleobj{1.5}{[}$} \hspace{0.5pt} \tub[\pi]{\pi}{1}{}\hspace{1pt}, \hspace{1pt}\tub[\pi]{\pi}{\raisebox{0.9pt}{$\scriptstyle \psi$}}{}\hspace{1pt},  \\[2pt]
    & && \hspace{22pt} \tub[\pi]{\pi}{\pi}{1}, \hspace{1pt} \tub[\pi]{\pi}{\pi}{\psi}, \hspace{1pt} \tub[\pi]{\pi}{\pi}{\pi} \hspace{0.5pt}\raisebox{-1pt}{$\scaleobj{1.5}{]}$} \; ,
\end{alignat*}
whose composition rules are given by
\begin{align*}
\tub[1]{1}{\raisebox{0.9pt}{$\scriptstyle \psi$}}{} \, \circ \, \tub[1]{1}{\raisebox{0.9pt}{$\scriptstyle \psi$}}{} \; &= \; \tub[1]{1}{1}{} \; , \\[3pt]
\tub[1]{1}{\pi}{} \, \circ \, \tub[1]{1}{\raisebox{0.9pt}{$\scriptstyle \psi$}}{} \; &= \; \tub[1]{1}{\raisebox{0.9pt}{$\scriptstyle \psi$}}{} \, \circ \, \tub[1]{1}{\pi}{} \; = \; \tub[1]{1}{\pi}{} \; , \\[3pt]
\tub[1]{1}{\pi}{} \, \circ \, \tub[1]{1}{\pi}{} \; &= \; \tub[1]{1}{1}{} \, + \, \tub[1]{1}{\raisebox{0.9pt}{$\scriptstyle \psi$}}{} \, + \, \tub[1]{1}{\pi}{} \; , \\[3pt]
\tub[1]{\raisebox{0.9pt}{$\scriptstyle \psi$}}{\pi}{} \, \circ \, \tub[1]{1}{\raisebox{0.9pt}{$\scriptstyle \psi$}}{} \; &= \; \tub[1]{\raisebox{0.9pt}{$\scriptstyle \psi$}}{\pi}{} \; , \\[3pt]
\tub[1]{\pi}{\pi}{} \, \circ \, \tub[1]{1}{\raisebox{0.9pt}{$\scriptstyle \psi$}}{} \; &= \; - \tub[1]{\pi}{\pi}{} \; , \\[3pt]
\tub[1]{\raisebox{0.9pt}{$\scriptstyle \psi$}}{\pi}{} \, \circ \, \tub[1]{1}{\pi}{} \; &= \; -\tub[1]{\raisebox{0.9pt}{$\scriptstyle \psi$}}{\pi}{} \; , \\[3pt]
\tub[1]{\pi}{\pi}{} \, \circ \, \tub[1]{1}{\pi}{} \; &= \; 0 \; , \\[3pt]
\tub[\hspace{-1pt}\raisebox{0.9pt}{$\scriptstyle \psi$}]{1}{\pi}{} \, \circ \, \tub[1]{\raisebox{0.9pt}{$\scriptstyle \psi$}}{\pi}{} \; &= \; \tub[1]{1}{1}{} \, + \, \tub[1]{1}{\raisebox{0.9pt}{$\scriptstyle \psi$}}{} \, - \, \tub[1]{1}{\pi}{} \; , \\[3pt]
\tub[\hspace{-1pt}\raisebox{0.9pt}{$\scriptstyle \psi$}]{\raisebox{0.9pt}{$\scriptstyle \psi$}}{\raisebox{0.9pt}{$\scriptstyle \psi$}}{} \, \circ \, \tub[1]{\raisebox{0.9pt}{$\scriptstyle \psi$}}{\pi}{} \; &= \; \tub[1]{\raisebox{0.9pt}{$\scriptstyle \psi$}}{\pi}{} \; , \\[3pt]
\tub[\hspace{-1pt}\raisebox{0.9pt}{$\scriptstyle \psi$}]{\raisebox{0.9pt}{$\scriptstyle \psi$}}{\pi}{} \, \circ \, \tub[1]{\raisebox{0.9pt}{$\scriptstyle \psi$}}{\pi}{} \; &= \; \tub[1]{\raisebox{0.9pt}{$\scriptstyle \psi$}}{\pi}{} \; , \\[3pt]
\tub[\hspace{-1pt}\raisebox{0.9pt}{$\scriptstyle \psi$}]{\pi}{\pi}{} \, \circ \, \tub[1]{\raisebox{0.9pt}{$\scriptstyle \psi$}}{\pi}{} \; &= \; 0 \; , \\[3pt]
\tub[\pi]{1}{\pi}{} \, \circ \, \tub[1]{\pi}{\pi}{} \; &= \; \tub[1]{1}{1}{} \, - \, \tub[1]{1}{\raisebox{0.9pt}{$\scriptstyle \psi$}}{} \; , \\[3pt]
\tub[\pi]{\raisebox{0.9pt}{$\scriptstyle \psi$}}{\pi}{} \, \circ \, \tub[1]{\pi}{\pi}{} \; &= \; 0 \; , \\[3pt]
\tub[\pi]{\pi}{\raisebox{0.9pt}{$\scriptstyle \psi$}}{} \, \circ \, \tub[1]{\pi}{\pi}{} \; &= \; \tub[1]{\pi}{\pi}{} \; , \\[3pt]
\tub[\pi]{\pi}{\pi}{1} \, \circ \, \tub[1]{\pi}{\pi}{} \; &= \; \tfrac{1}{2} \hspace{1pt}\tub[1]{\pi}{\pi}{} \; , \\[3pt]
\tub[\pi]{\pi}{\pi}{\psi} \, \circ \, \tub[1]{\pi}{\pi}{} \; &= \; \tfrac{1}{2} \hspace{1pt}\tub[1]{\pi}{\pi}{} \; , \\[3pt]
\tub[\pi]{\pi}{\pi}{\pi} \, \circ \, \tub[1]{\pi}{\pi}{} \; &= \; 0 \; , \\[3pt]
\tub[1]{1}{\raisebox{0.9pt}{$\scriptstyle \psi$}}{}  \, \circ \, \tub[\hspace{-1pt}\raisebox{0.9pt}{$\scriptstyle \psi$}]{1}{\pi}{} \; &= \; \tub[\hspace{-1pt}\raisebox{0.9pt}{$\scriptstyle \psi$}]{1}{\pi}{} \; , \\[3pt]
\tub[1]{1}{\pi}{}  \, \circ \, \tub[\hspace{-1pt}\raisebox{0.9pt}{$\scriptstyle \psi$}]{1}{\pi}{} \; &= \; -\tub[\hspace{-1pt}\raisebox{0.9pt}{$\scriptstyle \psi$}]{1}{\pi}{} \; , \\[3pt]
\tub[1]{\raisebox{0.9pt}{$\scriptstyle \psi$}}{\pi}{}  \, \circ \, \tub[\hspace{-1pt}\raisebox{0.9pt}{$\scriptstyle \psi$}]{1}{\pi}{} \; &= \; \tub[\hspace{-1pt}\raisebox{0.9pt}{$\scriptstyle \psi$}]{\raisebox{0.9pt}{$\scriptstyle \psi$}}{1}{} \, + \, \tub[\hspace{-1pt}\raisebox{0.9pt}{$\scriptstyle \psi$}]{\raisebox{0.9pt}{$\scriptstyle \psi$}}{\raisebox{0.9pt}{$\scriptstyle \psi$}}{} \, + \, \tub[\hspace{-1pt}\raisebox{0.9pt}{$\scriptstyle \psi$}]{\raisebox{0.9pt}{$\scriptstyle \psi$}}{\pi}{} \; , \\[3pt]
\tub[1]{\pi}{\pi}{}  \, \circ \, \tub[\hspace{-1pt}\raisebox{0.9pt}{$\scriptstyle \psi$}]{1}{\pi}{} \; &= \; 0 \; , \\[3pt]
\tub[\hspace{-1pt}\raisebox{0.9pt}{$\scriptstyle \psi$}]{1}{\pi}{} \, \circ \, \tub[\hspace{-1pt}\raisebox{0.9pt}{$\scriptstyle \psi$}]{\raisebox{0.9pt}{$\scriptstyle \psi$}}{\raisebox{0.9pt}{$\scriptstyle \psi$}}{} \; &= \; \tub[\hspace{-1pt}\raisebox{0.9pt}{$\scriptstyle \psi$}]{1}{\pi}{} \; , \\[3pt]
\tub[\hspace{-1pt}\raisebox{0.9pt}{$\scriptstyle \psi$}]{\raisebox{0.9pt}{$\scriptstyle \psi$}}{\raisebox{0.9pt}{$\scriptstyle \psi$}}{}  \, \circ \, \tub[\hspace{-1pt}\raisebox{0.9pt}{$\scriptstyle \psi$}]{\raisebox{0.9pt}{$\scriptstyle \psi$}}{\raisebox{0.9pt}{$\scriptstyle \psi$}}{} \; &= \; \tub[\hspace{-1pt}\raisebox{0.9pt}{$\scriptstyle \psi$}]{\raisebox{0.9pt}{$\scriptstyle \psi$}}{1}{} \; , \\[3pt]
\tub[\hspace{-1pt}\raisebox{0.9pt}{$\scriptstyle \psi$}]{\raisebox{0.9pt}{$\scriptstyle \psi$}}{\pi}{}  \, \circ \, \tub[\hspace{-1pt}\raisebox{0.9pt}{$\scriptstyle \psi$}]{\raisebox{0.9pt}{$\scriptstyle \psi$}}{\raisebox{0.9pt}{$\scriptstyle \psi$}}{} \; &= \; \tub[\hspace{-1pt}\raisebox{0.9pt}{$\scriptstyle \psi$}]{\raisebox{0.9pt}{$\scriptstyle \psi$}}{\pi}{} \; , \\[3pt]
\tub[\hspace{-1pt}\raisebox{0.9pt}{$\scriptstyle \psi$}]{\pi}{\pi}{}  \, \circ \, \tub[\hspace{-1pt}\raisebox{0.9pt}{$\scriptstyle \psi$}]{\raisebox{0.9pt}{$\scriptstyle \psi$}}{\raisebox{0.9pt}{$\scriptstyle \psi$}}{} \; &= \; - \tub[\hspace{-1pt}\raisebox{0.9pt}{$\scriptstyle \psi$}]{\pi}{\pi}{} \; , \\[3pt]
\tub[\hspace{-1pt}\raisebox{0.9pt}{$\scriptstyle \psi$}]{1}{\pi}{}  \, \circ \, \tub[\hspace{-1pt}\raisebox{0.9pt}{$\scriptstyle \psi$}]{\raisebox{0.9pt}{$\scriptstyle \psi$}}{\pi}{} \; &= \; \tub[\hspace{-1pt}\raisebox{0.9pt}{$\scriptstyle \psi$}]{1}{\pi}{} \; , \\[3pt]
\tub[\hspace{-1pt}\raisebox{0.9pt}{$\scriptstyle \psi$}]{\raisebox{0.9pt}{$\scriptstyle \psi$}}{\raisebox{0.9pt}{$\scriptstyle \psi$}}{} \, \circ \, \tub[\hspace{-1pt}\raisebox{0.9pt}{$\scriptstyle \psi$}]{\raisebox{0.9pt}{$\scriptstyle \psi$}}{\pi}{} \; &= \; \tub[\hspace{-1pt}\raisebox{0.9pt}{$\scriptstyle \psi$}]{\raisebox{0.9pt}{$\scriptstyle \psi$}}{\pi}{} \; , \\[3pt]
\tub[\hspace{-1pt}\raisebox{0.9pt}{$\scriptstyle \psi$}]{\raisebox{0.9pt}{$\scriptstyle \psi$}}{\pi}{} \, \circ \, \tub[\hspace{-1pt}\raisebox{0.9pt}{$\scriptstyle \psi$}]{\raisebox{0.9pt}{$\scriptstyle \psi$}}{\pi}{} \; &= \; \tub[\hspace{-1pt}\raisebox{0.9pt}{$\scriptstyle \psi$}]{\raisebox{0.9pt}{$\scriptstyle \psi$}}{1}{} \, + \, \tub[\hspace{-1pt}\raisebox{0.9pt}{$\scriptstyle \psi$}]{\raisebox{0.9pt}{$\scriptstyle \psi$}}{\raisebox{0.9pt}{$\scriptstyle \psi$}}{} \, - \, \tub[\hspace{-1pt}\raisebox{0.9pt}{$\scriptstyle \psi$}]{\raisebox{0.9pt}{$\scriptstyle \psi$}}{\pi}{} \; , \\[3pt]
\tub[\hspace{-1pt}\raisebox{0.9pt}{$\scriptstyle \psi$}]{\pi}{\pi}{} \, \circ \, \tub[\hspace{-1pt}\raisebox{0.9pt}{$\scriptstyle \psi$}]{\raisebox{0.9pt}{$\scriptstyle \psi$}}{\pi}{} \; &= \; 0 \; , \\[3pt]
\tub[\pi]{1}{\pi}{} \, \circ \, \tub[\hspace{-1pt}\raisebox{0.9pt}{$\scriptstyle \psi$}]{\pi}{\pi}{} \; &= \; 0 \; , \\[3pt]
\tub[\pi]{\raisebox{0.9pt}{$\scriptstyle \psi$}}{\pi}{} \, \circ \, \tub[\hspace{-1pt}\raisebox{0.9pt}{$\scriptstyle \psi$}]{\pi}{\pi}{} \; &= \; - \tub[\hspace{-1pt}\raisebox{0.9pt}{$\scriptstyle \psi$}]{\raisebox{0.9pt}{$\scriptstyle \psi$}}{1}{} \, + \, \tub[\hspace{-1pt}\raisebox{0.9pt}{$\scriptstyle \psi$}]{\raisebox{0.9pt}{$\scriptstyle \psi$}}{\raisebox{0.9pt}{$\scriptstyle \psi$}}{} \; , \\[3pt]
\tub[\pi]{\pi}{\raisebox{0.9pt}{$\scriptstyle \psi$}}{} \, \circ \, \tub[\hspace{-1pt}\raisebox{0.9pt}{$\scriptstyle \psi$}]{\pi}{\pi}{} \; &= \; \tub[\hspace{-1pt}\raisebox{0.9pt}{$\scriptstyle \psi$}]{\pi}{\pi}{} \; , \\[3pt]
\tub[\pi]{\pi}{\pi}{1} \, \circ \, \tub[\hspace{-1pt}\raisebox{0.9pt}{$\scriptstyle \psi$}]{\pi}{\pi}{} \; &= \; - \tfrac{1}{2} \hspace{1pt}\tub[\hspace{-1pt}\raisebox{0.9pt}{$\scriptstyle \psi$}]{\pi}{\pi}{} \; , \\[3pt]
\tub[\pi]{\pi}{\pi}{\psi} \, \circ \, \tub[\hspace{-1pt}\raisebox{0.9pt}{$\scriptstyle \psi$}]{\pi}{\pi}{} \; &= \; - \tfrac{1}{2} \hspace{1pt}\tub[\hspace{-1pt}\raisebox{0.9pt}{$\scriptstyle \psi$}]{\pi}{\pi}{} \; , \\[3pt]
\tub[\pi]{\pi}{\pi}{\pi} \, \circ \, \tub[\hspace{-1pt}\raisebox{0.9pt}{$\scriptstyle \psi$}]{\pi}{\pi}{} \; &= \; 0 \; , \\[3pt]
\tub[1]{1}{\raisebox{0.9pt}{$\scriptstyle \psi$}}{} \, \circ \, \tub[\pi]{1}{\pi}{} \; &= \; - \tub[\pi]{1}{\pi}{} \; , \\[3pt]
\tub[1]{1}{\pi}{} \, \circ \, \tub[\pi]{1}{\pi}{} \; &= \; 0 \; , \\[3pt]
\tub[1]{\raisebox{0.9pt}{$\scriptstyle \psi$}}{\pi}{}  \, \circ \, \tub[\pi]{1}{\pi}{} \; &= \; 0 \; , \\[3pt]
\tub[1]{\pi}{\pi}{} \, \circ \, \tub[\pi]{1}{\pi}{} \; &= \; \tfrac{1}{2} \hspace{1pt}\tub[\pi]{\pi}{1}{} \, + \, \tfrac{1}{2} \hspace{1pt}\tub[\pi]{\pi}{\raisebox{0.9pt}{$\scriptstyle \psi$}}{} \\[3pt]
&\hphantom{=} \; + \, \tub[\pi]{\pi}{\pi}{1} \, + \, \tub[\pi]{\pi}{\pi}{\psi} \; , \\[3pt]
\tub[\hspace{-1pt}\raisebox{0.9pt}{$\scriptstyle \psi$}]{1}{\pi}{}  \, \circ \, \tub[\pi]{\raisebox{0.9pt}{$\scriptstyle \psi$}}{\pi}{} \; &= \; 0 \; , \\[3pt]
\tub[\hspace{-1pt}\raisebox{0.9pt}{$\scriptstyle \psi$}]{\raisebox{0.9pt}{$\scriptstyle \psi$}}{\raisebox{0.9pt}{$\scriptstyle \psi$}}{}  \, \circ \, \tub[\pi]{\raisebox{0.9pt}{$\scriptstyle \psi$}}{\pi}{} \; &= \; -\tub[\pi]{\raisebox{0.9pt}{$\scriptstyle \psi$}}{\pi}{} \; , \\[3pt]
\tub[\hspace{-1pt}\raisebox{0.9pt}{$\scriptstyle \psi$}]{\raisebox{0.9pt}{$\scriptstyle \psi$}}{\pi}{}  \, \circ \, \tub[\pi]{\raisebox{0.9pt}{$\scriptstyle \psi$}}{\pi}{} \; &= \; 0 \; , \\[3pt]
\tub[\hspace{-1pt}\raisebox{0.9pt}{$\scriptstyle \psi$}]{\pi}{\pi}{}  \, \circ \, \tub[\pi]{\raisebox{0.9pt}{$\scriptstyle \psi$}}{\pi}{} \; &= \; -\tfrac{1}{2} \hspace{1pt}\tub[\pi]{\pi}{1}{} \, - \, \tfrac{1}{2} \hspace{1pt}\tub[\pi]{\pi}{\raisebox{0.9pt}{$\scriptstyle \psi$}}{} \\[3pt]
&\hphantom{=} \hspace{10pt} + \, \tub[\pi]{\pi}{\pi}{1} \, + \, \tub[\pi]{\pi}{\pi}{\psi} \; , \\[3pt]
\tub[\pi]{1}{\pi}{}  \, \circ \, \tub[\pi]{\pi}{\raisebox{0.9pt}{$\scriptstyle \psi$}}{} \; &= \; \tub[\pi]{1}{\pi}{} \; , \\[3pt]
\tub[\pi]{\raisebox{0.9pt}{$\scriptstyle \psi$}}{\pi}{}  \, \circ \, \tub[\pi]{\pi}{\raisebox{0.9pt}{$\scriptstyle \psi$}}{} \; &= \; \tub[\pi]{\raisebox{0.9pt}{$\scriptstyle \psi$}}{\pi}{} \; , \\[3pt]
\tub[\pi]{\pi}{\raisebox{0.9pt}{$\scriptstyle \psi$}}{} \, \circ \, \tub[\pi]{\pi}{\raisebox{0.9pt}{$\scriptstyle \psi$}}{} \; &= \; \tub[\pi]{\pi}{1}{} \; , \\[3pt]
\tub[\pi]{\pi}{\pi}{1} \, \circ \, \tub[\pi]{\pi}{\raisebox{0.9pt}{$\scriptstyle \psi$}}{} \; &= \; \tub[\pi]{\pi}{\pi}{\psi} \; , \\[3pt]
\tub[\pi]{\pi}{\pi}{\psi} \, \circ \, \tub[\pi]{\pi}{\raisebox{0.9pt}{$\scriptstyle \psi$}}{} \; &= \; \tub[\pi]{\pi}{\pi}{1} \; , \\[3pt]
\tub[\pi]{\pi}{\pi}{\pi} \, \circ \, \tub[\pi]{\pi}{\raisebox{0.9pt}{$\scriptstyle \psi$}}{} \; &= \; -\tub[\pi]{\pi}{\pi}{\pi} \; , \\[3pt]
\tub[\pi]{1}{\pi}{} \, \circ \, \tub[\pi]{\pi}{\pi}{1} \; &= \; \tfrac{1}{2} \hspace{1pt} \tub[\pi]{1}{\pi}{} \; , \\[3pt]
\tub[\pi]{\raisebox{0.9pt}{$\scriptstyle \psi$}}{\pi}{} \, \circ \, \tub[\pi]{\pi}{\pi}{1} \; &= \; -\tfrac{1}{2} \hspace{1pt} \tub[\pi]{\raisebox{0.9pt}{$\scriptstyle \psi$}}{\pi}{} \; , \\[3pt]
\tub[\pi]{\pi}{\raisebox{0.9pt}{$\scriptstyle \psi$}}{} \, \circ \, \tub[\pi]{\pi}{\pi}{1} \; &= \; \tub[\pi]{\pi}{\pi}{\psi} \; , \\[3pt]
\tub[\pi]{\pi}{\pi}{1}  \, \circ \, \tub[\pi]{\pi}{\pi}{1} \; &= \; \tfrac{1}{8} \tub[\pi]{\pi}{1}{} \hspace{-1pt} + \tfrac{1}{8} \tub[\pi]{\pi}{\raisebox{0.9pt}{$\scriptstyle \psi$}}{} \hspace{-1pt} + \tfrac{1}{4} \tub[\pi]{\pi}{\pi}{\pi} \hspace{1pt}, \\[3pt]
\tub[\pi]{\pi}{\pi}{\psi}  \, \circ \, \tub[\pi]{\pi}{\pi}{1} \; &= \; \tfrac{1}{8} \tub[\pi]{\pi}{1}{} \hspace{-1pt} + \tfrac{1}{8} \tub[\pi]{\pi}{\raisebox{0.9pt}{$\scriptstyle \psi$}}{} \hspace{-1pt} - \tfrac{1}{4} \tub[\pi]{\pi}{\pi}{\pi} \hspace{1pt}, \\[3pt]
\tub[\pi]{\pi}{\pi}{\pi}  \, \circ \, \tub[\pi]{\pi}{\pi}{1} \; &= \; \tfrac{1}{4} \hspace{1pt}\tub[\pi]{\pi}{1}{} \, - \, \tfrac{1}{4} \hspace{1pt}\tub[\pi]{\pi}{\raisebox{0.9pt}{$\scriptstyle \psi$}}{} \; , \\[3pt]
\tub[\pi]{1}{\pi}{}  \, \circ \, \tub[\pi]{\pi}{\pi}{\psi} \; &= \; \tfrac{1}{2} \hspace{1pt}\tub[\pi]{1}{\pi}{} \; , \\[3pt]
\tub[\pi]{\raisebox{0.9pt}{$\scriptstyle \psi$}}{\pi}{} \, \circ \, \tub[\pi]{\pi}{\pi}{\psi} \; &= \; -\tfrac{1}{2} \hspace{1pt}\tub[\pi]{\raisebox{0.9pt}{$\scriptstyle \psi$}}{\pi}{} \; , \\[3pt]
\tub[\pi]{\pi}{\raisebox{0.9pt}{$\scriptstyle \psi$}}{} \, \circ \, \tub[\pi]{\pi}{\pi}{\psi} \; &= \; \tub[\pi]{\pi}{\pi}{1} \; , \\[3pt]
\tub[\pi]{\pi}{\pi}{1}  \, \circ \, \tub[\pi]{\pi}{\pi}{\psi} \; &= \; \tfrac{1}{8} \tub[\pi]{\pi}{1}{} \hspace{-1pt} + \tfrac{1}{8} \tub[\pi]{\pi}{\raisebox{0.9pt}{$\scriptstyle \psi$}}{} \hspace{-1pt} - \tfrac{1}{4} \tub[\pi]{\pi}{\pi}{\pi} \hspace{1pt}, \\[3pt]
\tub[\pi]{\pi}{\pi}{\psi}  \, \circ \, \tub[\pi]{\pi}{\pi}{\psi} \; &= \; \tfrac{1}{8} \tub[\pi]{\pi}{1}{} \hspace{-1pt} + \tfrac{1}{8} \tub[\pi]{\pi}{\raisebox{0.9pt}{$\scriptstyle \psi$}}{} \hspace{-1pt} + \tfrac{1}{4} \tub[\pi]{\pi}{\pi}{\pi} \hspace{1pt}, \\[3pt]
\tub[\pi]{\pi}{\pi}{\pi}  \, \circ \, \tub[\pi]{\pi}{\pi}{\psi} \; &= \; -\tfrac{1}{4} \tub[\pi]{\pi}{1}{} \, + \, \tfrac{1}{4} \tub[\pi]{\pi}{\raisebox{0.9pt}{$\scriptstyle \psi$}}{} \; , \\[3pt]
\tub[\pi]{1}{\pi}{}  \, \circ \, \tub[\pi]{\pi}{\pi}{\pi} \; &= \; 0 \; , \\[3pt]
\tub[\pi]{\raisebox{0.9pt}{$\scriptstyle \psi$}}{\pi}{}  \, \circ \, \tub[\pi]{\pi}{\pi}{\pi} \; &= \; 0 \; , \\[3pt]
\tub[\pi]{\pi}{\raisebox{0.9pt}{$\scriptstyle \psi$}}{} \, \circ \, \tub[\pi]{\pi}{\pi}{\pi} \; &= \; -\tub[\pi]{\pi}{\pi}{\pi} \; , \\[3pt]
\tub[\pi]{\pi}{\pi}{1}  \, \circ \, \tub[\pi]{\pi}{\pi}{\pi} \; &= \; \tfrac{1}{4} \tub[\pi]{\pi}{1}{} \, - \, \tfrac{1}{4} \tub[\pi]{\pi}{\raisebox{0.9pt}{$\scriptstyle \psi$}}{} \; , \\[3pt]
\tub[\pi]{\pi}{\pi}{\psi}  \, \circ \, \tub[\pi]{\pi}{\pi}{\pi} \; &= \; -\tfrac{1}{4} \tub[\pi]{\pi}{1}{} \, + \, \tfrac{1}{4} \tub[\pi]{\pi}{\raisebox{0.9pt}{$\scriptstyle \psi$}}{} \; , \\[3pt]
\tub[\pi]{\pi}{\pi}{\pi}  \, \circ \, \tub[\pi]{\pi}{\pi}{\pi} \; &= \; \tub[\pi]{\pi}{\pi}{1} \, - \, \tfrac{1}{4} \tub[\pi]{\pi}{\pi}{\psi} \; . 
\end{align*}
The $\dagger$-structure acts on morphisms via
\be
\begin{split}
    \tub[1]{1}{1}{}^{\dagger} \, &= \; \tub[1]{1}{1}{} \; , \;\quad \tub[\hspace{-1pt}\raisebox{0.9pt}{$\scriptstyle \psi$}]{\raisebox{0.9pt}{$\scriptstyle \psi$}}{\raisebox{0.9pt}{$\scriptstyle \psi$}}{}^{\dagger} \; = \; \tub[\hspace{-1pt}\raisebox{0.9pt}{$\scriptstyle \psi$}]{\raisebox{0.9pt}{$\scriptstyle \psi$}}{\raisebox{0.9pt}{$\scriptstyle \psi$}}{} \; , \\[3pt]
\tub[1]{1}{\raisebox{0.9pt}{$\scriptstyle \psi$}}{}^{\dagger} \, &= \; \tub[1]{1}{\raisebox{0.9pt}{$\scriptstyle \psi$}}{} \; , \;\quad \tub[\hspace{-1pt}\raisebox{0.9pt}{$\scriptstyle \psi$}]{\raisebox{0.9pt}{$\scriptstyle \psi$}}{\pi}{}^{\dagger} \; = \; \tub[\hspace{-1pt}\raisebox{0.9pt}{$\scriptstyle \psi$}]{\raisebox{0.9pt}{$\scriptstyle \psi$}}{\pi}{} \; , \\[3pt]
\tub[1]{1}{\pi}{}^{\dagger} \, &= \; \tub[1]{1}{\pi}{} \; , \;\quad \tub[\hspace{-1pt}\raisebox{0.9pt}{$\scriptstyle \psi$}]{\pi}{\pi}{}^{\dagger} \; = \; -\tub[\pi]{\raisebox{0.9pt}{$\scriptstyle \psi$}}{\pi}{} \; , \\[3pt]
\tub[1]{\raisebox{0.9pt}{$\scriptstyle \psi$}}{\pi}{}^{\dagger} \, &= \; \tub[\hspace{-1pt}\raisebox{0.9pt}{$\scriptstyle \psi$}]{1}{\pi}{} \; , \;\quad \tub[\pi]{1}{\pi}{}^{\dagger} \; = \; \tub[1]{\pi}{\pi}{} \; , \\[3pt]
\tub[1]{\pi}{\pi}{}^{\dagger} \; &= \, \tub[\pi]{1}{\pi}{} \; , \;\quad \tub[\pi]{\raisebox{0.9pt}{$\scriptstyle \psi$}}{\pi}{}^{\dagger} \; = \; - \tub[\hspace{-1pt}\raisebox{0.9pt}{$\scriptstyle \psi$}]{\pi}{\pi}{} \; , \\[3pt]
\tub[\hspace{-1pt}\raisebox{0.9pt}{$\scriptstyle \psi$}]{1}{\pi}{}^{\dagger} \; &= \; \tub[1]{\raisebox{0.9pt}{$\scriptstyle \psi$}}{\pi}{} \; , \;\quad \tub[\pi]{\pi}{1}{}^{\dagger} \; = \; \tub[\pi]{\pi}{1}{} \; , \\[3pt]
\tub[\hspace{-1pt}\raisebox{0.9pt}{$\scriptstyle \psi$}]{\raisebox{0.9pt}{$\scriptstyle \psi$}}{1}{}^{\dagger} \; &= \; \tub[\hspace{-1pt}\raisebox{0.9pt}{$\scriptstyle \psi$}]{\raisebox{0.9pt}{$\scriptstyle \psi$}}{1}{} \; , \;\quad \tub[\pi]{\pi}{\raisebox{0.9pt}{$\scriptstyle \psi$}}{}^{\dagger} \; = \; \tub[\pi]{\pi}{\raisebox{0.9pt}{$\scriptstyle \psi$}}{} \; , \\[3pt]
\tub[\pi]{\pi}{\pi}{1}^{\dagger} \; &= \; \tfrac{1}{2} \tub[\pi]{\pi}{\pi}{1} \hspace{-1pt} + \tfrac{1}{2} \tub[\pi]{\pi}{\pi}{\psi} \hspace{-1pt} + \tfrac{1}{2} \tub[\pi]{\pi}{\pi}{\pi} \; , \hspace{10pt} \\[3pt]
\tub[\pi]{\pi}{\pi}{\psi}^{\dagger} \; &= \; \tfrac{1}{2} \tub[\pi]{\pi}{\pi}{1} \hspace{-1pt} + \tfrac{1}{2} \tub[\pi]{\pi}{\pi}{\psi} \hspace{-1pt} - \tfrac{1}{2} \tub[\pi]{\pi}{\pi}{\pi} \; , \\[3pt]
\tub[\pi]{\pi}{\pi}{\pi}^{\dagger} \; &= \; \tub[\pi]{\pi}{\pi}{1} \, - \, \tub[\pi]{\pi}{\pi}{\psi} \; . 
\end{split}
\ee

\vspace{5pt}
\noindent\textbf{Generalised Charges.}
There is a total of eight irreducible generalised charges that can be described as follows:
\begin{itemize}[leftmargin=3ex]
\item There is a one-dimensional charge $U_1$ that acts on the untwisted sector $\mathcal{H}_1 \cong \mathbb{C}$ via
\begin{equation}
\begin{aligned}
U_1\Big(\ntub[1]{1}{\raisebox{0.9pt}{$\scriptstyle \psi$}}{}\Big) \; &= \; 1 \; , \\[3pt]
U_1\Big( \ntub[1]{1}{\pi}{} \Big) \; &= \; 2 \; . \\[2pt]
\end{aligned}
\end{equation}

\item There is a one-dimensional charge $U_{\psi}$ that acts on the twisted sector $\mathcal{H}_{\psi} \cong \mathbb{C}$ via
\begin{equation}
\begin{aligned}
U_{\psi}\Big(\ntub[\hspace{-1pt}\raisebox{0.9pt}{$\scriptstyle \psi$}]{\raisebox{0.9pt}{$\scriptstyle \psi$}}{\raisebox{0.9pt}{$\scriptstyle \psi$}}{}\Big) \; &= \; 1 \; , \\[3pt]
U_{\psi}\Big( \ntub[\hspace{-1pt}\raisebox{0.9pt}{$\scriptstyle \psi$}]{\raisebox{0.9pt}{$\scriptstyle \psi$}}{\pi}{} \Big) \; &= \; - 2 \; .
\end{aligned}
\end{equation}

\item There are three one-dimensional charges $U^{(n)}_{\pi}$ ($n=0,1,2$) that act on the twisted sector $\mathcal{H}_{\pi} \cong \mathbb{C}$ via
\begin{equation}
\begin{aligned}
U^{(n)}_{\pi}\Big(\ntub[\pi]{\pi}{\raisebox{0.9pt}{$\scriptstyle \psi$}}{}\Big) \; &= \; -1 \; , \\[3pt]
U^{(n)}_{\pi}\Big( \ntub[\pi]{\pi}{\pi}{1} \Big) \; &= \; \frac{1}{2} \hspace{1pt} q^n \; , \\[3pt]
U^{(n)}_{\pi}\Big( \ntub[\pi]{\pi}{\pi}{\psi} \Big) \; &= \; - \frac{1}{2} \hspace{1pt} q^n \; , \\[3pt]
U^{(n)}_{\pi}\Big( \ntub[\pi]{\pi}{\pi}{\pi} \Big) \; &= \; q^{-n} \; , \\[3pt]
\end{aligned}
\end{equation}
where $q:= e^{2\pi i / 3}$ denotes a third root of unity.

\item There is a two-dimensional charge $U_{1,\psi}$ that acts on the twisted sectors $\mathcal{H}_{1} \cong \mathcal{H}_{\psi} \cong \mathbb{C}$ via
\begin{equation}
\begin{alignedat}{2}
\;\quad U_{1,\psi}\Big(\ntub[1]{1}{\raisebox{0.9pt}{$\scriptstyle \psi$}}{}\Big) \; &= \; 1 \; , \;\quad &&U_{1,\psi}\Big(\ntub[\hspace{-1pt}\raisebox{0.9pt}{$\scriptstyle \psi$}]{\raisebox{0.9pt}{$\scriptstyle \psi$}}{\raisebox{0.9pt}{$\scriptstyle \psi$}}{}\Big) \; = \; 1 \; , \\[3pt]
\;\quad U_{1,\psi}\Big( \ntub[1]{1}{\pi}{} \Big) \; &= \; - 1 \; , \;\quad &&U_{1,\psi}\Big(\ntub[\hspace{-1pt}\raisebox{0.9pt}{$\scriptstyle \psi$}]{\raisebox{0.9pt}{$\scriptstyle \psi$}}{\pi}{}\Big) \; = \; 1 \; , \\[3pt]
\;\quad U_{1,\psi}\Big( \ntub[1]{\raisebox{0.9pt}{$\scriptstyle \psi$}}{\pi}{} \Big) \; &= \; \sqrt{3} \; , \;\quad &&U_{1,\psi}\Big(\ntub[\hspace{-1pt}\raisebox{0.9pt}{$\scriptstyle \psi$}]{1}{\pi}{}\Big) \; = \; \sqrt{3} \; .
\end{alignedat}
\end{equation}

\item There is a two-dimensional charge $U_{1,\pi}$ that acts on the twisted sectors $\mathcal{H}_{1} \cong \mathcal{H}_{\pi} \cong \mathbb{C}$ via
\begin{equation}
\begin{alignedat}{2}
\;\quad U_{1,\pi}\Big(\ntub[1]{1}{\raisebox{0.9pt}{$\scriptstyle \psi$}}{}\Big) \; &= \; -1 \; , \;\quad &&U_{1,\pi}\Big(\ntub[\pi]{\pi}{\raisebox{0.9pt}{$\scriptstyle \psi$}}{}\Big) \; = \; 1 \; , \\[3pt]
\;\quad U_{1,\pi}\Big( \ntub[1]{1}{\pi}{} \Big) \; &= \; 0 \; , \;\quad &&U_{1,\pi}\Big(\ntub[\pi]{\pi}{\pi}{1}\Big) \; = \; \frac{1}{2} \; , \\[3pt]
\;\quad U_{1,\pi}\Big( \ntub[1]{\pi}{\pi}{} \Big) \; &= \; \sqrt{2} \; , \;\quad &&U_{1,\pi}\Big(\ntub[\pi]{\pi}{\pi}{\psi}\Big) \; = \; \frac{1}{2} \; , \\[3pt]
\;\quad U_{1,\pi}\Big( \ntub[\pi]{1}{\pi}{} \Big) \; &= \; \sqrt{2} \; , \;\quad &&U_{1,\pi}\Big(\ntub[\pi]{\pi}{\pi}{\pi}\Big) \; = \; 0 \; .
\end{alignedat}
\end{equation}

\item There is one two-dimensional charge $U_{\psi,\pi}$ that acts on the twisted sectors $\mathcal{H}_{\psi} \cong \mathcal{H}_{\pi} \cong \mathbb{C}$ via
\begin{equation}
\begin{alignedat}{2}
\;\quad U_{\psi,\pi}\Big( \ntub[\hspace{-1pt}\raisebox{0.9pt}{$\scriptstyle \psi$}]{\raisebox{0.9pt}{$\scriptstyle \psi$}}{\raisebox{0.9pt}{$\scriptstyle \psi$}}{} \Big) \; &= \; -1 \; , \;\quad &&U_{\psi,\pi}\Big(\ntub[\pi]{\pi}{\raisebox{0.9pt}{$\scriptstyle \psi$}}{}\Big) \; = \; 1 \; , \\[3pt]
\;\quad U_{\psi,\pi}\Big( \ntub[\hspace{-1pt}\raisebox{0.9pt}{$\scriptstyle \psi$}]{\raisebox{0.9pt}{$\scriptstyle \psi$}}{\pi}{} \Big) \; &= \; 0 \; , \;\quad &&U_{\psi,\pi}\Big(\ntub[\pi]{\pi}{\pi}{1}\Big) \; = \; -\frac{1}{2} \; , \\[3pt]
\;\quad U_{\psi,\pi}\Big( \ntub[\hspace{-1pt}\raisebox{0.9pt}{$\scriptstyle \psi$}]{\pi}{\pi}{}  \Big) \; &= \; \sqrt{2} \; , \;\quad &&U_{\psi,\pi}\Big(\ntub[\pi]{\pi}{\pi}{\psi}\Big) \; = \; -\frac{1}{2} \; , \\[3pt]
\;\quad U_{\psi,\pi}\Big( \ntub[\pi]{\raisebox{0.9pt}{$\scriptstyle \psi$}}{\pi}{} \Big) \; &= \; - \sqrt{2} \; , \;\quad &&U_{\psi,\pi}\Big(\ntub[\pi]{\pi}{\pi}{\pi}\Big) \; = \; 0 \; .
\end{alignedat}
\end{equation}
\end{itemize}

\vspace{5pt}
\noindent\textbf{Isometry Actions.}
Examples of bases of transition channels for the non-invertible defect $\pi$ that obey Theorem \ref{thm-categorical-wigner} were presented in \eqref{eqn:RepS3basis}. Upon choosing a generalised charge $U$, they induce linear isometries $U(\pi)_X$ associated to the action of $\pi$ on $X$-twisted sector states. For instance, for the generalised charge $U = U_{\psi,\pi}$ supported on the twisted sectors $\cH_\psi \cong \cH_\pi \cong \mathbb{C}$, we have that
    \begin{equation}
    \begin{aligned}
    U_{\psi,\pi}(\pi)_1 \, = \,  \begin{bmatrix} 0 ,\, 1 \end{bmatrix}^\text{T} \!\!: \;\; &\mathcal{H}_{\psi} \, \to \, \mathcal{H}_{\psi} \oplus \mathcal{H}_{\pi} \; , \\[2pt]
    U_{\psi,\pi}(\pi)_{\pi} \, = \,  - \tfrac{1}{2} \begin{bmatrix} \sqrt{2},\, 1 ,\, 1 ,\, 0 \end{bmatrix}^\text{T} \!\!: \;\; &\mathcal{H}_{\pi} \, \to \, \mathcal{H}_{\psi} \oplus \mathcal{H}_{\pi}^{\hspace{1pt}\oplus \hspace{1pt} 3} \; . \\
    \end{aligned}
    \end{equation}
    As a result, we obtain the following transition probabilities for the action of $\pi$ on twisted sector states:
    \begin{equation}
    \quad
    \begin{aligned}
        p\Big( \tfrac{1}{2} \hspace{0.5pt} \tub[\hspace{-1pt}\raisebox{0.9pt}{$\scriptstyle \psi$}]{\raisebox{0.9pt}{$\scriptstyle \psi$}}{\pi}{} \Big) \; &= \; 0 \; , \\
        p\Big( \tfrac{1}{\sqrt{2}} \hspace{0.5pt} \tub[\hspace{-1pt}\raisebox{0.9pt}{$\scriptstyle \psi$}]{\pi}{\pi}{} \Big) \; &= \; 1 \; , 
    \end{aligned}
    \quad
    \begin{aligned}
        p\Big( \tfrac{1}{2} \hspace{0.5pt} \tub[\pi]{\raisebox{0.9pt}{$\scriptstyle \psi$}}{\pi}{} \Big) \; &= \; 1/2 \; , \\
        p\Big( \tub[\pi]{\pi}{\pi}{1} \Big) \; &= \; 1/4 \; , \\
        p\Big( \tub[\pi]{\pi}{\pi}{\psi} \Big) \; &= \; 1/4 \; , \\
        p\Big( \tfrac{1}{\sqrt{2}} \hspace{0.5pt} \tub[\pi]{\pi}{\pi}{\pi} \Big) \; &= \; 0 \; . \\[2pt]
    \end{aligned}
    \end{equation}

\subsection{Fibonacci Symmetry}
\label{app-fib}
\noindent The unitary Fibonacci category $\mathcal{C} = \text{Fib}$ has two simple objects 1 and $\tau$ that fuse according to
\begin{equation}
\label{eq-fib-fusion-rule}
    \tau \otimes \tau \; = \; 1 \oplus \tau \; .
\end{equation}
The quantum dimensions of the non-invertible defect $\tau$ is given by the golden ratio $d_{\tau} = \phi \equiv \tfrac{1}{2}(1 + \sqrt{5})$. The non-trivial component of the associator is
\begin{equation}
\label{eq-fib-associator}
\vspace{-5pt}
\begin{gathered}
\includegraphics[height=1.3cm]{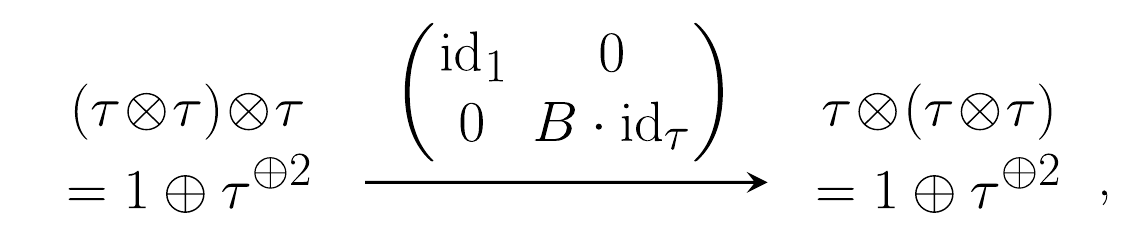}
\end{gathered}
\end{equation}
where the $(2 \!\times \!2)$-matrix $B$ is given by
\begin{equation}
B \; = \;
\begin{pmatrix}
\phi^{-1} \; & \phi^{-\nicefrac{1}{2}} \\
\phi^{-\nicefrac{1}{2}} \; & -\phi^{-1}
\end{pmatrix}\,. 
\end{equation}

\vspace{5pt}
\noindent\textbf{Tube Category.}
The associated tube category was computed in \cite{Bartsch2025} and has morphism spaces
\begin{equation}
\label{eq-fibonacci-tube-morphisms}
\begin{alignedat}{2}
    &\text{T}\mathcal{C}(1,1) \, &&= \; \mathbb{C}\hspace{0.5pt}\raisebox{-1pt}{$\scaleobj{1.5}{[}$} \hspace{0.5pt} \tub[1]{1}{1}{}\hspace{1pt}, \hspace{1pt}\tub[1]{1}{\tau}{} \hspace{0.5pt}\raisebox{-1pt}{$\scaleobj{1.5}{]}$} \; , \\[2pt]
    &\text{T}\mathcal{C}(1,\tau) \, &&= \; \mathbb{C}\hspace{0.5pt}\raisebox{-1pt}{$\scaleobj{1.5}{[}$} \hspace{0.5pt} \tub[1]{\tau}{\tau}{} \hspace{0.5pt}\raisebox{-1pt}{$\scaleobj{1.5}{]}$} \; , \\[2pt]
    &\text{T}\mathcal{C}(\tau,1) \, &&= \; \mathbb{C}\hspace{0.5pt}\raisebox{-1pt}{$\scaleobj{1.5}{[}$} \hspace{0.5pt} \tub[\tau]{1}{\tau}{} \hspace{0.5pt}\raisebox{-1pt}{$\scaleobj{1.5}{]}$} \; , \\[2pt]
    &\text{T}\mathcal{C}(\tau,\tau) \, &&= \; \mathbb{C}\hspace{0.5pt}\raisebox{-1pt}{$\scaleobj{1.5}{[}$} \hspace{0.5pt} \tub[\tau]{\tau}{1}{}\hspace{1pt}, \hspace{1pt}\tub[\tau]{\tau}{\tau}{1}\hspace{1pt}, \hspace{1pt}\tub[\tau]{\tau}{\tau}{\tau} \hspace{0.5pt}\raisebox{-1pt}{$\scaleobj{1.5}{]}$} \; , \\[2pt]
\end{alignedat}
\end{equation}
whose composition rules are given by
\begin{align*}
\tub[1]{1}{\tau}{} \, \circ \, \tub[1]{1}{\tau}{} \; &= \; \tub[1]{1}{1}{} \, + \, \tub[1]{1}{\tau}{} \; , \\[3pt]
\tub[1]{\tau}{\tau}{} \, \circ \, \tub[1]{1}{\tau}{} \; &= \; - \phi^{-1} \cdot \tub[1]{\tau}{\tau}{} \; , \\[3pt]
\tub[\tau]{1}{\tau}{} \, \circ \, \tub[1]{\tau}{\tau}{} \; &= \; \tub[1]{1}{1}{} \, - \, \phi^{-1} \cdot \tub[1]{1}{\tau}{} \; , \\[3pt]
\tub[\tau]{\tau}{\tau}{\raisebox{-1.5pt}{$\scriptstyle 1$}} \, \circ \, \tub[1]{\tau}{\tau}{} \; &= \; \phi^{-1} \cdot \tub[1]{\tau}{\tau}{} \; , \\[3pt]
\tub[\tau]{\tau}{\tau}{\tau} \, \circ \, \tub[1]{\tau}{\tau}{} \; &= \; \phi^{-2} \cdot \tub[1]{\tau}{\tau}{} \; , \\[3pt]
\tub[1]{1}{\tau}{} \, \circ \, \tub[\tau]{1}{\tau}{} \; &= \; - \phi^{-1} \cdot \tub[\tau]{1}{\tau}{} \; , \\[3pt]
\tub[1]{\tau}{\tau}{} \, \circ \, \tub[\tau]{1}{\tau}{} \; &= \; \phi^{-1} \cdot \tub[\tau]{\tau}{1}{} \, + \, \tub[\tau]{\tau}{\tau}{\raisebox{-1.5pt}{$\scriptstyle 1$}} \\[3pt] &\quad\;\, + \, \phi^{-2} \cdot \tub[\tau]{\tau}{\tau}{\tau} \; , \\[3pt]
\tub[\tau]{1}{\tau}{} \, \circ \, \tub[\tau]{\tau}{\tau}{\raisebox{-1.5pt}{$\scriptstyle 1$}} \; &= \; \phi^{-1} \cdot \tub[\tau]{1}{\tau}{} \; , \\[3pt]
\tub[\tau]{1}{\tau}{} \, \circ \, \tub[\tau]{\tau}{\tau}{\tau} \; &= \; \phi^{-2} \cdot \tub[\tau]{1}{\tau}{} \; , \\[3pt]
\tub[\tau]{\tau}{\tau}{\raisebox{-1.5pt}{$\scriptstyle 1$}} \, \circ \, \tub[\tau]{\tau}{\tau}{\raisebox{-1.5pt}{$\scriptstyle 1$}} \; &= \; \phi^{-3} \cdot \tub[\tau]{\tau}{1}{} \, + \, \phi^{-2} \cdot \tub[\tau]{\tau}{\tau}{\tau} \; , \\[3pt]
\tub[\tau]{\tau}{\tau}{\tau} \, \circ \, \tub[\tau]{\tau}{\tau}{\tau} \; &= \; -\phi^{-2} \cdot \tub[\tau]{\tau}{1}{}  \, + \, \tub[\tau]{\tau}{\tau}{\raisebox{-1.5pt}{$\scriptstyle 1$}}  \\[3pt] &\quad\;\, - \, \phi^{-3} \cdot \tub[\tau]{\tau}{\tau}{\tau} \; , \\[3pt]
\tub[\tau]{\tau}{\tau}{\raisebox{-1.5pt}{$\scriptstyle 1$}} \, \circ \, \tub[\tau]{\tau}{\tau}{\tau} \; &= \; \tub[\tau]{\tau}{\tau}{\tau} \, \circ \, \tub[\tau]{\tau}{\tau}{\raisebox{-1.5pt}{$\scriptstyle 1$}} \\[3pt] &= \; \phi^{-2} \cdot \tub[\tau]{\tau}{1}{} \, - \, \phi^{-2} \cdot \tub[\tau]{\tau}{\tau}{\tau} \; .
\end{align*}
The $\dagger$-structure acts on morphisms via
\begin{equation}
\begin{aligned}
\tub[1]{1}{\tau}{}^{\dagger} \; &= \; \tub[1]{1}{\tau}{} \; , \\[3pt]
\tub[1]{\tau}{\tau}{}^{\dagger} \; &= \; \phi \cdot \tub[\tau]{1}{\tau}{} \; , \\[3pt]
\tub[\tau]{1}{\tau}{}^{\dagger} \; &= \; \phi^{-1} \cdot \tub[1]{\tau}{\tau}{} \; , \\[3pt]
\tub[\tau]{\tau}{\tau}{\raisebox{-1.5pt}{$\scriptstyle 1$}}^{\dagger} \; &= \; \phi^{-1} \cdot \tub[\tau]{\tau}{\tau}{\raisebox{-1.5pt}{$\scriptstyle 1$}} \, + \, \phi^{-1} \cdot \tub[\tau]{\tau}{\tau}{\tau} \; , \\[3pt]
\tub[\tau]{\tau}{\tau}{\tau}^{\dagger} \; &= \; \tub[\tau]{\tau}{\tau}{\raisebox{-1.5pt}{$\scriptstyle 1$}} \, - \, \phi^{-1} \cdot \tub[\tau]{\tau}{\tau}{\tau} \; .
\end{aligned}
\end{equation}

\vspace{5pt}
\noindent\textbf{Generalised Charges.}
There is a total of four irreducible generalised charges \cite{Bartsch2025} that can be described as follows:
\begin{itemize}[leftmargin=3ex]
\item There is a one-dimensional charge $U_1$ that acts on the twisted sector $\mathcal{H}_1 \cong \mathbb{C}$ via
\begin{equation}
U_1\Big( \ntub[1]{1}{\tau}{} \Big) \; = \; \phi \; .
\end{equation}

\item There are two one-dimensional charges $U_{\tau}^{\pm}$ that act on the twisted sector $\mathcal{H}_{\tau} \cong \mathbb{C}$ via
\begin{equation}
\begin{aligned}
U_{\tau}^{\pm}\Big( \ntub[\tau]{\tau}{\tau}{\raisebox{-1.5pt}{$\scriptstyle 1$}} \Big) \; &= \; x_{\pm} \; , \\[3pt]
U_{\tau}^{\pm}\Big( \ntub[\tau]{\tau}{\tau}{\tau} \Big) \; &= \; - \phi \cdot \big( 1 + \phi \hspace{1pt} x_{\pm}\big) \; ,
\end{aligned}
\end{equation}
where $x_{\pm}$ are the two solutions of $x^2+x+\phi^{-2}=0\hspace{1pt}$:
\begin{equation}
\label{eqn:xpm}
x_{\pm} \; = \; - \, \frac{1}{2} \, \pm \, i \cdot \sqrt{\frac{3}{4} - \frac{1}{\phi}} \; .
\end{equation}

\item There is one two-dimensional charge $U_{1,\tau}$ that acts on the twisted sectors $\mathcal{H}_1 \cong \mathcal{H}_{\tau} \cong \mathbb{C}$ via
\begin{equation}
\begin{aligned}
U_{1,\tau}\Big( \ntub[1]{1}{\tau}{} \Big) \; &= \; -\phi^{-1} \, , \\
U_{1,\tau}\Big( \ntub[\tau]{\tau}{\tau}{\raisebox{-1.5pt}{$\scriptstyle 1$}} \Big) \; &= \; \phi^{-1}  \, ,\\
U_{1,\tau}\Big( \ntub[\tau]{\tau}{\tau}{\raisebox{-1.5pt}{$\scriptstyle \tau$}} \Big) \; &= \; \phi^{-2} \, ,\\
U_{1,\tau}\Big( \ntub[1]{\tau}{\tau}{} \Big) \; &= \; \phi^{\nicefrac{1}{2}} \cdot \big(1-i\phi^{-1}\big) \, ,\\
U_{1,\tau}\Big( \ntub[\tau]{1}{\tau}{} \Big) \; &= \; \phi^{-\nicefrac{1}{2}} \cdot \big(1+i\phi^{-1}\big) \, .
\end{aligned}
\end{equation}
\end{itemize}

\subsection{Yang-Lee Symmetry}
\label{sec:Yang-Lee}
\noindent 
In order to highlight the importance of requiring the fusion category $\mathcal{C}$ to be \emph{unitary}, we now present an example of a non-unitary fusion category that violates the CPP Theorem~\ref{thm-categorical-wigner}. Concretely, we let $\mathcal{C} = \text{YL}$ be the Yang-Lee category, which has the same simple objects $1$, $\tau$ and fusion rules (\ref{eq-fib-fusion-rule}) as the Fibonacci category, but a different (non-unitary) associator. Concretely, the latter still takes the form (\ref{eq-fib-associator}), but with the $(2\!\times \!2)$-matrix $B$ now replaced by
\begin{equation}
B \; = \;
\begin{pmatrix}
-\phi \; & i \hspace{1pt}\phi^{\nicefrac{1}{2}} \\
i \hspace{1pt}\phi^{\nicefrac{1}{2}} \; & \phi
\end{pmatrix} \; ,
\end{equation}
where $\phi \equiv \tfrac{1}{2}(1 + \sqrt{5})$ is the golden ratio. In particular, the quantum dimension of the non-invertible defect $\tau$ is given by
\begin{equation}
    d_{\tau} \, = \, -\hspace{1pt}\frac{1}{\phi} \; < \; 0 \; .
\end{equation}
 
\vspace{5pt}
\noindent\textbf{Tube Category.}
The associated tube category was computed in \cite{Bartsch2025} and has the same morphism spaces as in (\ref{eq-fibonacci-tube-morphisms}) but different composition rules given by
\begin{align*}
\tub[1]{1}{\tau}{} \, \circ \, \tub[1]{1}{\tau}{} \; &= \; \tub[1]{1}{1}{} \, + \, \tub[1]{1}{\tau}{} \; , \\[3pt]
\tub[1]{\tau}{\tau}{} \, \circ \, \tub[1]{1}{\tau}{} \; &= \; \phi \cdot \tub[1]{\tau}{\tau}{} \; , \\[3pt]
\tub[\tau]{1}{\tau}{} \, \circ \, \tub[1]{\tau}{\tau}{} \; &= \; \tub[1]{1}{1}{} \, + \, \phi \cdot \tub[1]{1}{\tau}{} \; , \\[3pt]
\tub[\tau]{\tau}{\tau}{\raisebox{-1.5pt}{$\scriptstyle 1$}} \, \circ \, \tub[1]{\tau}{\tau}{} \; &= \; -\phi \cdot \tub[1]{\tau}{\tau}{} \; , \\[3pt]
\tub[\tau]{\tau}{\tau}{\tau} \, \circ \, \tub[1]{\tau}{\tau}{} \; &= \; \phi^2 \cdot \tub[1]{\tau}{\tau}{} \; , \\[3pt]
\tub[1]{1}{\tau}{} \, \circ \, \tub[\tau]{1}{\tau}{} \; &= \; \phi \cdot \tub[\tau]{1}{\tau}{} \; , \\[3pt]
\tub[1]{\tau}{\tau}{} \, \circ \, \tub[\tau]{1}{\tau}{} \; &= \; -\phi \cdot \tub[\tau]{\tau}{1}{} \, + \, \tub[\tau]{\tau}{\tau}{\raisebox{-1.5pt}{$\scriptstyle 1$}} \\[3pt] &\quad\;\, + \, \phi^2 \cdot \tub[\tau]{\tau}{\tau}{\tau} \; , \\[3pt]
\tub[\tau]{1}{\tau}{} \, \circ \, \tub[\tau]{\tau}{\tau}{\raisebox{-1.5pt}{$\scriptstyle 1$}} \; &= \; -\phi \cdot \tub[\tau]{1}{\tau}{} \; , \\[3pt]
\tub[\tau]{1}{\tau}{} \, \circ \, \tub[\tau]{\tau}{\tau}{\tau} \; &= \; \phi^2 \cdot \tub[\tau]{1}{\tau}{} \; , \\[3pt]
\tub[\tau]{\tau}{\tau}{\raisebox{-1.5pt}{$\scriptstyle 1$}} \, \circ \, \tub[\tau]{\tau}{\tau}{\raisebox{-1.5pt}{$\scriptstyle 1$}} \; &= \; -\phi^3 \cdot \tub[\tau]{\tau}{1}{} \, + \, \phi^2 \cdot \tub[\tau]{\tau}{\tau}{\tau} \; , \\[3pt]
\tub[\tau]{\tau}{\tau}{\tau} \, \circ \, \tub[\tau]{\tau}{\tau}{\tau} \; &= \; -\phi^2 \cdot \tub[\tau]{\tau}{1}{}  \, + \, \tub[\tau]{\tau}{\tau}{\raisebox{-1.5pt}{$\scriptstyle 1$}}  \\[3pt] &\quad\;\, + \, \phi^3 \cdot \tub[\tau]{\tau}{\tau}{\tau} \; , \\[3pt]
\tub[\tau]{\tau}{\tau}{\raisebox{-1.5pt}{$\scriptstyle 1$}} \, \circ \, \tub[\tau]{\tau}{\tau}{\tau} \; &= \; \tub[\tau]{\tau}{\tau}{\tau} \, \circ \, \tub[\tau]{\tau}{\tau}{\raisebox{-1.5pt}{$\scriptstyle 1$}} \\[3pt] &= \; \phi^2 \cdot \tub[\tau]{\tau}{1}{} \, - \, \phi^2 \cdot \tub[\tau]{\tau}{\tau}{\tau} \; .
\end{align*}

\vspace{5pt}
\noindent\textbf{Generalised Charges.}
There is a total of four irreducible generalised charges \cite{Bartsch2025} that can be described as follows:
\begin{itemize}[leftmargin=3ex]
\item There is a one-dimensional charge $U_1$ that acts on the twisted sector $\mathcal{H}_1 \cong \mathbb{C}$ via
\begin{equation}
U_1\Big( \ntub[1]{1}{\tau}{} \Big) \; = \; - \frac{1}{\phi} \; .
\end{equation}

\item There are two one-dimensional charges $U_{\tau}^{\pm}$ that act on the twisted sector $\mathcal{H}_{\tau} \cong \mathbb{C}$ via
\begin{equation}
\begin{aligned}
U_{\tau}^{\pm}\Big( \ntub[\tau]{\tau}{\tau}{\raisebox{-1.5pt}{$\scriptstyle 1$}} \Big) \; &= \; x_{\pm} \; , \\[3pt]
U_{\tau}^{\pm}\Big( \ntub[\tau]{\tau}{\tau}{\tau} \Big) \; &= \; \frac{1}{\phi} \cdot \Big( 1 - \frac{x_{\pm}}{\phi}\Big) \; ,
\end{aligned}
\end{equation}
where $x_{\pm}$ are the two solutions of $x^2+x+\phi^2=0\hspace{1pt}$:
\begin{equation}
\label{eqn:xpm2}
x_{\pm} \; = \; - \, \frac{1}{2} \, \pm \, i \cdot \sqrt{\phi +\frac{3}{4}} \; .
\end{equation}

\item There is one two-dimensional charge $U_{1,\tau}$ that acts on the twisted sectors $\mathcal{H}_1 \cong \mathcal{H}_{\tau} \cong \mathbb{C}$ via
\begin{equation}
\begin{aligned}
U_{1,\tau}\Big( \ntub[1]{1}{\tau}{} \Big) \; &= \; \phi \, , \\
U_{1,\tau}\Big( \ntub[\tau]{\tau}{\tau}{\raisebox{-1.5pt}{$\scriptstyle 1$}} \Big) \; &= \; -\phi  \, ,\\
U_{1,\tau}\Big( \ntub[\tau]{\tau}{\tau}{\raisebox{-1.5pt}{$\scriptstyle \tau$}} \Big) \; &= \; \phi^2 \, ,\\
U_{1,\tau}\Big( \ntub[1]{\tau}{\tau}{} \Big) \; &= \; \phi^{\nicefrac{1}{2}} \cdot \big(1 + i\phi\big) \, ,\\
U_{1,\tau}\Big( \ntub[\tau]{1}{\tau}{} \Big) \; &= \; \phi^{-\nicefrac{1}{2}} \cdot \big(1-i\phi\big) \, .
\end{aligned}
\end{equation}
\end{itemize}

\vspace{5pt}
\noindent\textbf{Non-Isometry Action.}
In what follows, our goal is to show that $\mathcal{C} = \text{YL}$ violates the CPP Theorem \ref{thm-categorical-wigner}, i.e. does not admit bases of transition channels that assemble into linear isometries for a given defect $A$ within any given generalised charge $U$. To see this, we let $A = \tau$ be the non-invertible defect and make the following ansatz 
\begin{equation}
\label{eq-transition-channel-ansatz}
\begin{split}
    \left\{ a \cdot \tub[1]{1}{\tau}{} \right\} \; & \subset \; \cC^\tau(1,1) \, , \\[3pt] 
    \left\{ b \cdot \tub[1]{\tau}{\tau}{} \right\} \; & \subset \; \cC^\tau(1, \tau) \;,
\end{split}
\end{equation}
for bases of transition channels with trivial incoming twisted sector. In order to determine the parameters $a,b \in \mathbb{C}^{\times}$, we require that (\ref{eq-transition-channel-ansatz}) induces a linear isometry for any given generalised charge $U$. Concretely, by going through the irreducible charges, we obtain the following constraints:
\begin{itemize}[leftmargin=3ex]
    \item For the generalised charge $U = U_1$ supported on the untwisted sector $\cH_1 \cong \mathbb{C}$, we have that
    \begin{equation}
    \qquad U_1(\tau)_1 \, = \, - \hspace{1pt}\frac{a}{\phi} \hspace{1pt} : \;\; \mathcal{H}_1 \; \to \; \mathcal{H}_1 \; .
    \end{equation}
    Demanding this to be an isometry requires
    \begin{equation}
    \label{eq-parameter-condition-1}
        |a| \; = \; \phi \; .
    \end{equation}
    
    \item For the generalised charges $U = U_{1,\tau}$ supported on the twisted sectors $\cH_1 \cong \cH_\tau \cong \mathbb{C}$, we have that
    \begin{equation}
    \begin{aligned}
    \qquad U_{1,\tau}(\tau)_1 \, = \,  \begin{bmatrix} a \cdot \phi \\ b \cdot \phi \cdot (1+i\phi) \end{bmatrix} : \;\; &\mathcal{H}_1 \;\, \to \,\; \mathcal{H}_1 \oplus \mathcal{H}_{\tau} \, , 
    \end{aligned}
    \vspace{4pt}
    \end{equation}
    Demanding this to be an isometry requires
    \begin{equation}
    \label{eq-parameter-condition-2}
        \qquad |a|^2 \hspace{1pt}\phi^2 \, + \, |b|^2 \hspace{1pt} \phi \hspace{2pt} (1+ \phi^2) \; = \; 1 \; .
    \end{equation}
\end{itemize}
As a result, upon plugging (\ref{eq-parameter-condition-1}) into (\ref{eq-parameter-condition-2}) and using the fact that $\phi^4 > 1$, we see that the two conditions are incompatible with each other. In particular, this means that, there exist no bases of transition channels for $\tau$ that obey the CPP Theorem \ref{thm-categorical-wigner}. This highlights the importance of imposing unitarity on the symmetry category $\cC$ in order for the action of symmetry defects to preserve quantum probabilities.

\section{Generalisations}
\label{sec:generalisations}

\noindent
The CPP Theorem \ref{thm-categorical-wigner} presented in the main text applies to generalised symmetries that act on two-dimensional systems quantised on a spatial circle without boundary. It is straightforward to extend the result to more general types of quantum systems. In this section, we present such generalisations for systems with boundary and systems in higher dimensions.

\subsection{Systems with Boundary}
\label{ssec-boundaries}

\noindent
Consider a two-dimensional quantum theory with boundary. We denote by $\mathcal{M}$ the category of boundary conditions, whose objects $M,N \in \mathcal{M}$ label the different boundary conditions of the theory and whose morphisms $\theta \in \mathcal{M}(M,N)$ correspond to topological local operators that can sit at the junction between two boundary conditions:
\begin{equation}
\label{eq-boundary-conditions}
\vspace{-5pt}
\begin{gathered}
\includegraphics[height=0.95cm]{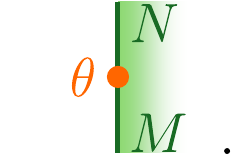}
\end{gathered}
\end{equation}
For simplicity, we will assume that $\mathcal{M}$ is a finite semi-simple linear additive $\dagger$-category in what follows. If the theory admits a fusion category symmetry $\mathcal{C}$, the latter can act on the category $\mathcal{M}$ of boundary conditions by means of a functor $\triangleright: \mathcal{C} \times \mathcal{M} \to \mathcal{M}$, which captures the effect of colliding topological line defects $A \in \mathcal{C}$ in the bulk with boundary conditions $M \in \mathcal{M}$ from the left:
\begin{equation}
\label{eq-module-category}
\vspace{-5pt}
\begin{gathered}
\includegraphics[height=0.95cm]{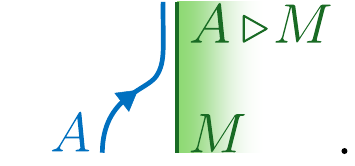}
\end{gathered}
\end{equation}
Mathematically, this turns $\mathcal{M}$ into a (left) module category for the fusion category $\mathcal{C}$ (see e.g. \cite{Etingof2017} for more background on module categories).

Now consider quantising the system on a two-dimensional strip, whose left and right boundaries are labelled by boundary conditions $M,N \in \mathcal{M}$, respectively:
\begin{equation}
\label{eq-strip}
\vspace{-5pt}
\begin{gathered}
\includegraphics[height=1.05cm]{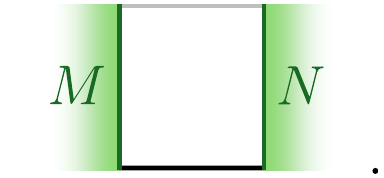}
\end{gathered}
\end{equation}
In this case, there is a space of states associated to the interval with endpoints punctured by boundary conditions $M$ and $N$. We denote the corresponding Hilbert space by $\mathcal{H}_{M,N}$. Symmetry defects $A \in \mathcal{C}$ can act on states $\ket{\Psi} \in \mathcal{H}_{M,N}$ upon being stretched between the two boundaries:
\begin{equation}
\label{eq-strip-action}
\vspace{-5pt}
\begin{gathered}
\includegraphics[height=1.7cm]{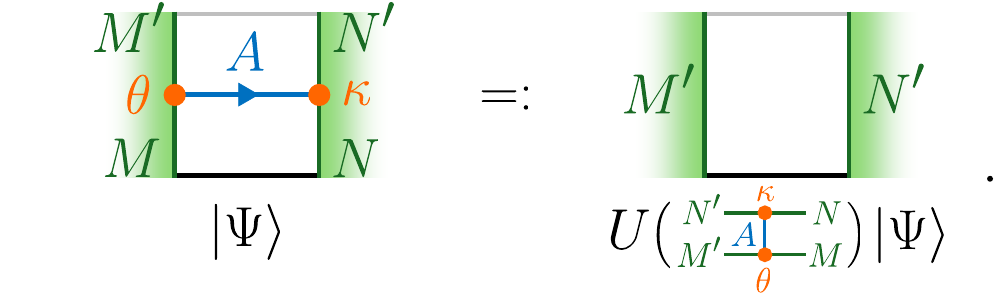}
\end{gathered}
\end{equation}
However, similarly to the case of the cylinder, this requires the choice of two local boundary morphisms
\begin{equation}
\begin{aligned}
    \theta &: \; M' \, \to \, A \triangleright M \; , \\
    \kappa &: \; A \triangleright N \, \to \, N'
\end{aligned}
\end{equation}
that sit at the junctions between $A$, the incoming boundary conditions $M$ and $N$, and the outgoing boundary conditions $M'$ and $N'$. We will refer to pairs of morphisms of this type as \emph{transition channels} between $(M,N)$ and $(M',N')$ and denote the space of all such transition channels for fixed $A$ by 
\begin{equation}
\begin{gathered}
    \quad \mathcal{M}^{A\hspace{-0.5pt}}\big((M,N),(M',N')\big) \; := \\
    \mathcal{M}(M', A \triangleright M) \otimes \mathcal{M}(A \triangleright N, N') \; .
\end{gathered}
\end{equation}
This is naturally a vector space of dimension
\begin{equation}
\label{eqn:multiplicity_strip}
    D^A_{(M,N),(M',N')} \; := \; \text{dim}\big( \hspace{0.5pt}\mathcal{M}^{A\hspace{-0.5pt}}\big((M,N),(M',N')\big) \hspace{0.5pt}\big) \, .
\end{equation}
For a given $\theta \otimes \kappa \in \mathcal{M}^{A\hspace{-0.5pt}}\big((M,N),(M',N')\big)$, we denote by
\begin{equation}
\label{eq-strip-stretching-action}
    U\Big( \hspace{3.5pt} \nstrp[M]{N}{\hspace{-3.5pt}M'}{\hspace{-3.5pt}N'}{A}{\theta}{\kappa} \Big): \; \mathcal{H}_{M,N} \; \to \; \mathcal{H}_{M',N'}
\end{equation}
the corresponding linear map that captures the stretching action of $A$ on states on the interval as illustrated in (\ref{eq-strip-action}). Compatibility with the consecutive actions of two symmetry defects $A,B \in \mathcal{C}$ then requires that
\begin{equation}
\label{eq-strip-composition-rule}
\begin{gathered}
U\Big(\hspace{6pt} \nstrp[M']{N'}{\hspace{-6pt}M''}{\hspace{-6pt}N''}{A}{\theta'}{\kappa'}\hspace{3pt}\Big) \, \circ \, U\Big(\hspace{3.5pt} \nstrp[M]{N}{\hspace{-3.5pt}M'}{\hspace{-3.5pt}N'}{B}{\theta}{\kappa}\Big) \;\, = \,\;  U\Big(\hspace{6pt} \nstrrp[M]{N}{\hspace{-6pt}M''}{\hspace{-6pt}N''}{A\raisebox{0.4pt}{$\scriptstyle\otimes$}\hspace{-0.3pt} B}{\theta' \circ \hspace{1pt} \theta}{\hspace{-0.7pt}\kappa' \circ \hspace{1pt} \kappa }\Big) \; . \\[6pt]
\end{gathered}
\vspace{2pt}
\end{equation}
It is convenient to repackage the above information in terms of the so-called \emph{strip category}\bfootnote{The notion of the strip category $\text{S}_{\mathcal{M}}(\mathcal{C})$ is closely related to that of the \emph{strip algebra} $\text{Strip}_{\mathcal{M}}(\mathcal{C})$ (see e.g. \cite{Cordova2024, Copetti:2024dcz}). Concretely, the latter is given by $\text{Strip}_{\mathcal{M}}(\mathcal{C}) = \text{End}_{\text{S}_{\mathcal{M}}(\mathcal{C})}\big(\raisebox{-.2ex}{\large \text{$\oplus$}}_{P,Q}(P,Q))\big)$, where $P$ and $Q$ run over the simple objects of $\mathcal{M}$. The strip algebra is `Morita equivalent' to the strip category in the sense that there is an equivalence 
\be
\Rep(\text{Strip}_{\cM}(\cC)) \, \cong \, [\text{S}_\cM(\cC), \Hilb]\,,
\ee
where the LHS denotes the category of representations of the strip algebra and the RHS denotes the category of linear functors from $\text{S}_\cM(\cC)$ into the category of Hilbert spaces.
} associated to $\mathcal{C}$ and $\cM$, which is the linear category $\text{S}_{\mathcal{M}}(\mathcal{C})$ whose
\begin{itemize}[label=--,leftmargin=3.5ex]
    \item objects are given by pairs $(M,N) \in \mathcal{M}^{\times 2}$,
    \item morphisms are given by (equivalence classes of) transition channels between $(M,N)$ and $(M',N')$.
\end{itemize}
The stretching action of symmetry defects on states on the interval may then be described as a linear functor
\begin{equation}
\label{eq-generalised-charge-st}
    U: \; \text{S}_{\mathcal{M}}(\mathcal{C}) \; \to \; \text{Hilb} \; ,
\end{equation}
which assigns to each pair $(M,N)$ the Hilbert space $\mathcal{H}_{M,N}$ and to each transition channel a linear map as in (\ref{eq-strip-stretching-action}). Functoriality of $U$ then ensures that the composition rule (\ref{eq-strip-composition-rule}) is satisfied. We furthermore require compatibility with spacetime reflections in the sense that
\begin{equation}
    U\Big( \hspace{3.5pt} \nstrp[M]{N}{\hspace{-3.5pt}M'}{\hspace{-3.5pt}N'}{A}{\theta}{\kappa} \Big)^{\dagger} \; = \; U\Big( \hspace{3pt} \nsstrp[M']{N'}{\hspace{-2pt}M}{\hspace{-2pt}N}{\hspace{-1pt}A^{\!\raisebox{-0.5pt}{$\scriptscriptstyle \vee$}}}{\hspace{-1pt}\raisebox{-0.7pt}{$\scriptstyle\theta^{\dagger}$}}{\hspace{-1pt}\kappa^{\dagger}}  \hspace{3.5pt} \Big) \; ,
\end{equation}
which turns $U$ into a $\dagger$-functor. We will again refer to functors of this type as \emph{generalised charges} in what follows.

The analogue of the CPP Theorem \ref{thm-categorical-wigner} for systems with boundaries can then be formulated as follows:

\begin{theorem}
\label{thm-boundary-wigner}
Let $A \in \mathcal{C}$ be a symmetry defect and $M,N \in \mathcal{M}$ two boundary conditions. Then for all simple $P,Q \in \textup{Irr}(\mathcal{M})$ there exists a tensor product basis
\begin{equation}
    \lbrace e_i^{\raisebox{-2pt}{$\scriptstyle P$}} \otimes f_j^{\raisebox{-3pt}{$\scriptstyle Q$}} \rbrace_{i,j} \; \subset \; \mathcal{M}^{A\hspace{-0.5pt}}\big((M,N),(P,Q)\big)
\end{equation}
    of transition channels such that in any given generalised charge $U: \textup{S}_{\mathcal{M}}(\mathcal{C}) \to \textup{Hilb}$ it holds that
    \vspace{5pt}
    \begin{equation}
    \label{eq-generalised-preservation-st}
        \sum_{\substack{P,Q \\ i,j}} \hspace{1pt} \Big\langle U\Big(  \nstrp[M]{N}{P}{Q}{A}{\hspace{-1pt}\raisebox{-1.5pt}{$\scriptstyle e_i^{\raisebox{-2pt}{$\scriptstyle P$}}$}}{\hspace{-0.5pt}\raisebox{2.5pt}{$\scriptstyle f_j^{\raisebox{-3pt}{$\scriptstyle Q$}}$}} \Big) \hspace{1pt} \Phi \hspace{1pt}, \, U\Big(  \nstrp[M]{N}{P}{Q}{A}{\hspace{-1pt}\raisebox{-1.5pt}{$\scriptstyle e_i^{\raisebox{-2pt}{$\scriptstyle P$}}$}}{\hspace{-0.5pt}\raisebox{2.5pt}{$\scriptstyle f_j^{\raisebox{-3pt}{$\scriptstyle Q$}}$}} \Big) \hspace{1pt} \Psi \Big\rangle \,\; = \,\; \braket{\Phi,\Psi}
    \end{equation}
    for all $\ket{\Phi}, \ket{\Psi} \in \mathcal{H}_{M,N}$. In other words, the operator
    \begin{equation}
        U(A)_{M,N} \; := \; \bigoplus_{\substack{P,Q \\ i,j}} \, U\Big(  \nstrp[M]{N}{P}{Q}{A}{\hspace{-1pt}\raisebox{-1.5pt}{$\scriptstyle e_i^{\raisebox{-2pt}{$\scriptstyle P$}}$}}{\hspace{-0.5pt}\raisebox{2.5pt}{$\scriptstyle f_j^{\raisebox{-3pt}{$\scriptstyle Q$}}$}} \Big)
    \end{equation}
    defines an isometry from the $(M,N)$-twisted Hilbert space $\mathcal{H}_{M,N}$ into the enlarged Hilbert space
    \begin{equation}
        \mathcal{H}_{M,N}^A \; := \; \bigoplus_{P,Q} \, D_{(M,N),(P,Q)}^A \cdot \mathcal{H}_{P,Q} \; ,
    \end{equation}
    where $D_{(M,N),(P,Q)}^A \in \mathbb{N}$ is as in \eqref{eqn:multiplicity_strip}. Moreover, the tensor product basis $\lbrace e_i^{\raisebox{-2pt}{$\scriptstyle P$}} \otimes f_j^{\raisebox{-3pt}{$\scriptstyle Q$}} \rbrace_{i,j}$ is unique up to transformations $\vec{e}^{\hspace{1.5pt}P} \mapsto I \hspace{-2pt}\cdot\hspace{-1pt} \vec{e}^{\hspace{1.5pt}P}$ and $\vec{f}^{\hspace{1.5pt}Q} \mapsto J \hspace{-2pt}\cdot\hspace{-1pt} \vec{f}^{\hspace{1.5pt}Q}$ by unitary matrices $I$ and $J$.
\end{theorem}

\noindent
Similarly to before, the above means that in order for the action of a symmetry defect $A$ on states on the interval to preserve inner products, we have to consider \textit{all} possible outgoing boundary conditions together with \textit{all} possible (linearly independent) transition channels thereto. As in the case of the tube category, the different transition channels then assemble into a quantum channel that captures the full action of symmetry defects on mixed states.

\subsection{Higher Dimensions}
\label{ssec-higher-d}

\noindent
As discussed in the main text of this letter, the CPP Theorem \ref{thm-categorical-wigner} admits direct generalisations to higher spacetime dimensions $d$. To illustrate this, it is convenient to depict the wrapping action of symmetry defects on states on the sphere as a linking action on local operators via a conformal mapping (as shown here for $d=2$):
\begin{equation}
\label{eq-operator-state-map}
\vspace{-5pt}
\begin{gathered}
\includegraphics[height=2.1cm]{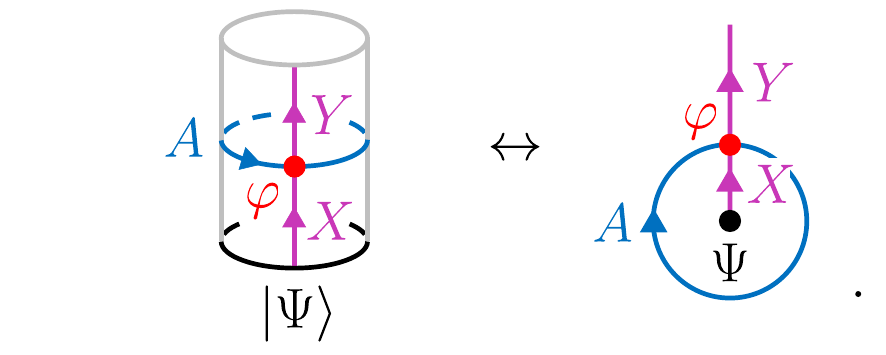}
\end{gathered}
\end{equation}
Now consider a quantum theory in spacetime dimension $d\geq 2$ whose generalised symmetries are described by a unitary\bfootnote{While $n$-categories admit a variety of different $\dagger$-structures parameterised by subgroups $\mathfrak{G} \subset \text{PL}(n)$ \cite{Ferrer:2024vpn}, by a \textit{unitary} higher category we refer to choosing $\mathfrak{G}= \mathbb{Z}_2$, which implements involutory reflections only on the top level of $n$-morphisms.} fusion $(d-1)$-category $\mathcal{C}$. Objects $A,B \in \mathcal{C}$ and morphisms $\varphi, \psi: A \to B$ correspond to codimension-one and -two topological defects and interfaces as before, while 2-morphisms $\Theta: \varphi \Rightarrow \psi$ now correspond to codimension-three topological junctions of interfaces and so on and so forth: 
\begin{equation}
\vspace{-5pt}
\begin{gathered}
\includegraphics[height=1.05cm]{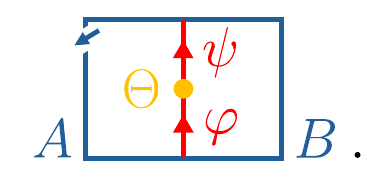}
\end{gathered}
\end{equation}
Similarly to $d=2$, symmetry defects $A \in \mathcal{C}$ can act on elements $\mathcal{O}$ in the Hilbert space $\mathcal{H}_{\mu}$ of local operators that are attached to a genuine line defect\bfootnote{Given a (higher) monoidal category $\mathcal{C}$ with unit object $\mathbf{1} \in \mathcal{C}$, we set $\Omega(\mathcal{C}) := \text{End}_{\hspace{0.5pt}\mathcal{C}}(\mathbf{1})$. Inductively, we set $\Omega^n(\mathcal{C}) := \Omega(\Omega^{n-1}(\mathcal{C}))$.} $\mu \in \Omega^{\hspace{0.5pt}d-2}(\mathcal{C})$ via linking: 
\begin{equation}
\label{eq-higher-d-tube-action}
\vspace{-5pt}
\begin{gathered}
\includegraphics[height=1.83cm]{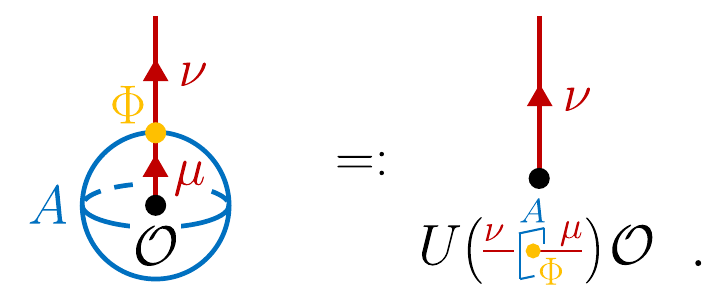}
\end{gathered}
\end{equation}
As before, this requires the choice of a local $(d\!-\!1)$-morphism
\begin{equation}
\label{eq-higher-intersection-morphism}
    \qquad \Phi : \; \text{id}^{d-2}_A \otimes \mu \, \to \, \nu \otimes \text{id}^{d-2}_A
\end{equation}
that sits at the intersection of $A$, the incoming twisted sector line $\mu$, and the outgoing twisted sector line $\nu$. We will refer to $\Phi$ as a \emph{transition channel} between $\mu$ and $\nu$ and denote the space of all transition channels for fixed $A$ by\bfootnote{For an $n$-category $\mathcal{C}$, we denote by $\mathcal{C}^{(p)}(a,b)$ the $(n\!-\!p)$-category of $p$-morphisms between $(p\!-\!1)$-morphisms $a$ and $b$ ($p=1,...,n$).} 
\begin{equation}
\label{eq-higher-transition-channels}
    \mathcal{C}^{\hspace{0.5pt}A\hspace{-0.5pt}}(\mu,\nu) \; := \; \mathcal{C}^{(d-1)}\big(\hspace{1pt}\text{id}^{d-2}_A \otimes \mu \hspace{2pt}, \hspace{1pt} \nu \otimes \text{id}^{d-2}_A\hspace{1pt}\big) \; .
\end{equation}
This is naturally a vector space of dimension
\begin{equation}
\label{eq-higher-transition-channel-dimension}
    D_{\mu\nu}^A \; := \; \text{dim}( \mathcal{C}^{\hspace{0.5pt}A\hspace{-0.5pt}}(\mu,\nu) ) \; .
\end{equation}
For a given choice $\Phi \in \mathcal{C}^{\hspace{0.5pt}A\hspace{-0.5pt}}(\mu,\nu)$, we denote by
\begin{equation}
\label{eq-higher-tube-linking-action}
    U\Big( \nhtub[\mu]{\nu}{A}{\Phi} \Big): \; \mathcal{H}_{\mu} \; \to \; \mathcal{H}_{\nu}
\end{equation}
the corresponding linear map that captures the linking action of $A$ on twisted sector local operators as illustrated in (\ref{eq-higher-d-tube-action}). Compatibility with the consecutive actions of two symmetry defects $A,B \in \mathcal{C}$ then requires that
\begin{equation}
\label{eq-higher-composition-rule}
U\Big(\nhtub[\hspace{-0pt}\nu]{\rho}{A}{\Phi}\Big) \, \circ \, U\Big(\nhtub[\mu]{\nu}{B}{\hspace{1pt}\Psi}\Big) \;\, = \,\;  U\Big(\nhtuub[\mu]{\rho}{\hspace{-2.5pt} A \hspace{0.5pt} \raisebox{0.65pt}{$\scriptstyle \otimes$} B}{\hspace{1pt} \Phi \hspace{0.6pt}\circ\hspace{0.6pt} \Psi}\Big) \; .
\end{equation}
It is again convenient to repackage the above information in terms of the \emph{tube category} of $\mathcal{C}$ as introduced in \cite{Bartsch2023a}, which is the linear category $\text{T}\mathcal{C}$ whose
\begin{itemize}[label=--,leftmargin=3.5ex]
    \item objects are given by genuine lines $\mu \in \Omega^{\hspace{0.5pt}d-2}(\mathcal{C})$,
    \item morphisms are given by (equivalence classes of) transition channels between $\mu$ and $\nu$.
\end{itemize}
The linking action of symmetry defects on twisted sector local operators may then be described as a linear functor
\begin{equation}
\label{eq-higher-generalised-charge}
    U: \; \text{T}\mathcal{C} \; \to \; \text{Hilb} \; ,
\end{equation}
which assigns to each $\mu \in \Omega^{\hspace{0.5pt}d-2}(\mathcal{C})$ the Hilbert space $\mathcal{H}_{\mu}$ and to each transition channel a linear map as in (\ref{eq-higher-tube-linking-action}). Functoriality of $U$ then ensures that the composition rule (\ref{eq-higher-composition-rule}) is satisfied. We furthermore require compatibility with spacetime reflections in the sense that
\begin{equation}
\label{eq-unitarity-condition-4}
    U\Big( \nhtub[\mu]{\nu}{A}{\Phi} \Big)^{\dagger} \; = \; U\Big( \nhtub[\nu]{\mu}{\hspace{0pt}A^{\hspace{-0.4pt}\vee}}{\hspace{-0.7pt}\raisebox{-1.1pt}{\hspace{1.2pt}$\scriptstyle \Phi^{\hspace{-0.5pt}\dagger}$}} \Big) \; ,
\end{equation}
which turns $U$ into a $\dagger$-functor. We will again refer to functors of this type as \emph{generalised charges} in what follows.

The analogue of Theorem \ref{thm-categorical-wigner} for the linking action of $\mathcal{C}$ on twisted sector local operators is then given as follows:

\begin{theorem}
\label{thm-higher_cat_wigner}
    Consider a fixed symmetry defect $A \in \mathcal{C}$ and let $\mu \in \Omega^{d-2}(\mathcal{C})$ a genuine line defect. Then for each simple line $\sigma \in \textup{Irr}(\Omega^{d-2}(\mathcal{C}))$ there exists a basis
    \begin{equation}
        \lbrace e_i^{\raisebox{-2pt}{$\scriptstyle \sigma$}} \rbrace_i \, \subset \, \mathcal{C}^{\hspace{0.5pt}A\hspace{-0.5pt}}(\mu,\sigma)
    \end{equation}
    of transition channels such that in any given generalised charge $U: \textup{T}\mathcal{C} \to \textup{Hilb}$ it holds that
    \begin{equation}
    \label{eq-generalised-preservation-higher-2}
        \sum_{\sigma\hspace{0.5pt},\hspace{1pt}i} \hspace{1pt} \Big\langle U\Big( \nhtub[\mu]{\sigma}{A}{\scriptstyle\hspace{1pt} \raisebox{-0pt}{$\scriptstyle e_i^{\raisebox{-2pt}{$\scriptscriptstyle \sigma$}}$}} \Big) \hspace{1pt} \Phi \hspace{1pt}, \, U\Big( \nhtub[\mu]{\sigma}{A}{\scriptstyle\hspace{1pt} \raisebox{-0pt}{$\scriptstyle e_i^{\raisebox{-2pt}{$\scriptscriptstyle \sigma$}}$}} \Big) \hspace{1pt} \Psi \Big\rangle \,\; = \,\; \braket{\Phi,\Psi}
    \end{equation}
    for all $\ket{\Phi}, \ket{\Psi} \in \mathcal{H}_{\mu}$. In other words, the operator
    \begin{equation}
    \label{eq-higher-isometry}
        U(A)_{\mu} \; := \; \bigoplus_{\sigma\hspace{0.5pt},\hspace{1pt}i} \, U\Big( \nhtub[\mu]{\sigma}{A}{\scriptstyle\hspace{1pt} \raisebox{-0pt}{$\scriptstyle e_i^{\raisebox{-2pt}{$\scriptscriptstyle \sigma$}}$}} \Big)
    \end{equation}
    defines an isometry from the $\mu$-twisted Hilbert space $\mathcal{H}_{\mu}$ into the enlarged Hilbert space
    \begin{equation}
        \mathcal{H}_{\mu}^A \; := \; \bigoplus_{\sigma} \, D_{\mu\sigma}^A \cdot \mathcal{H}_{\sigma} \; ,
    \end{equation}
    where $D_{\mu\sigma}^A \in \mathbb{N}$ is as in (\ref{eq-higher-transition-channel-dimension}). The basis $\lbrace e_i^{\raisebox{-2pt}{$\scriptstyle \sigma$}} \rbrace_i$ is unique up to transformations $\vec{e}^{\hspace{1.5pt}\sigma} \mapsto M \hspace{-2pt}\cdot\hspace{-1pt} \vec{e}^{\hspace{1.5pt}\sigma}$ by unitary matrices $M$.
\end{theorem}

\noindent
Similarly to $d=2$, the above means that in order for the action of a symmetry defect $A$ on $\mu$-twisted sectors to preserve inner products, we have to consider \textit{all} possible outgoing sectors together with \textit{all} possible (linearly independent) transition channels thereinto. For $\mathcal{O} \in \cH_{\mu} \!\setminus\! \{0\}$, we set
\be
\label{eq-transition-probability-higher}
p\Big( \nhtub[\mu]{\sigma}{A}{\scriptstyle\hspace{1pt} \raisebox{-0pt}{$\scriptstyle e_i^{\raisebox{-2pt}{$\scriptscriptstyle \sigma$}}$}} \Big)_{\hspace{-1pt}\mathcal{O}} \; := \; \norm{ \hspace{1pt} U\Big( \nhtub[\mu]{\sigma}{A}{\scriptstyle\hspace{1pt} \raisebox{-0pt}{$\scriptstyle e_i^{\raisebox{-2pt}{$\scriptscriptstyle \sigma$}}$}} \Big) \hspace{1pt} \mathcal{O} \hspace{0.5pt}}^2 \hspace{-1pt}\Big/ \norm{\mathcal{O}}^2 \; ,
\ee
which as a consequence of \eqref{eq-generalised-preservation-higher-2} satisfies
\be
\sum_{\sigma, i} \, p\Big( \nhtub[\mu]{\sigma}{A}{\scriptstyle\hspace{1pt} \raisebox{-0pt}{$\scriptstyle e_i^{\raisebox{-2pt}{$\scriptscriptstyle \sigma$}}$}} \Big)_{\hspace{-1pt}\mathcal{O}} \; = \;\, 1 \;.
\ee
Hence, we can again interpret $p\big( \scaleobj{0.8}{\nhtub[\mu]{\sigma}{A}{\scriptstyle\hspace{1pt} \raisebox{-0pt}{$\scriptstyle e_i^{\raisebox{-2pt}{$\scriptscriptstyle \sigma$}}$}}} \big)_{\mathcal{O}}$ as the probability that the defect $A$ maps a $\mu$-twisted sector operator $\mathcal{O}$ to a $\sigma$-twisted sector operator via the transition channel $e_i^\sigma$.

\subsubsection{Example: 2-Group Symmetry}

\noindent
Consider a quantum theory in spacetime dimension $d=3$ that has a finite 2-group symmetry $\mathcal{G}$ consisting of
\begin{enumerate}[leftmargin=5ex]
    \item a finite 0-form symmetry group $G$,
    \item a finite abelian 1-form symmetry group $A$,
    \item a group action\bfootnote{We will henceforth denote the action of a 0-form element $g \in G$ on a 1-form element $a \in A$ by $g \triangleright a =: {}^ga$.} $\triangleright: G \to \text{Aut}(A)$,
    \item a Postnikov class representative $\alpha \in Z^3_{\,\triangleright}(G,A)$.
\end{enumerate}
We will write $\cG = A[1]\rtimes_{\alpha} G$ for a 2-group specified by the above data in what follows. Let\bfootnote{Here, $\widehat{A} := \text{Hom}(A, U(1))$ denotes the \textit{Pontryagin dual} of $A$.} $\lambda \in Z^2_{\,\triangleright}(G, \widehat{A})$ be a 2-cocycle that captures a mixed anomaly between the 0-form symmetry $G$ and the 1-form symmetry $A$. The corresponding unitary fusion 2-category is denoted by $\mathcal{C} = \text{2Hilb}^{\hspace{1pt}\lambda}_{\hspace{1pt}\cG}$. 

\vspace{5pt}
\noindent\textbf{Tube Category.}
The associated tube category $\text{T}\cC$ was computed in \cite{Bartsch2023a} and has objects $a \in A$ and morphism spaces
\be
\text{T}\cC(a,b) \; = \; \bbC\hspace{0.5pt}\raisebox{-1pt}{$\scaleobj{1.5}{[}$} \hspace{0.5pt} \cchhtub[a]{b}{\raisebox{1pt}{$\scriptstyle g$}} \hspace{1.5pt} \raisebox{-1pt}{$\scaleobj{1.5}{|}$} \hspace{2pt} g \hspace{-1pt}\in\hspace{-1pt} G \;\; \text{s.t.} \;\, {}^{g}a = b
\hspace{1pt}\raisebox{-1pt}{$\scaleobj{1.5}{]}$} \;,
\ee
The composition rule for morphisms is given by
\be
\cchhtub[\hspace{-3pt}{}^{h}\hspace{-0.95pt}a]{\hspace{-6pt}{}^{\mathrlap{\raisebox{-2.5pt}{\hspace{-1pt}\crule[white]{8pt}{8pt}}}g\hspace{-0.5pt}h}\hspace{-0.95pt}a}{\hspace{-0.2pt}\raisebox{1pt}{$\scriptstyle g$}} \, \circ \, \cchhtub[a]{\scriptstyle \hspace{-2.7pt}{}^{\mathrlap{\raisebox{-1.5pt}{\hspace{-5pt}\crule[white]{8pt}{8pt}}}h}\hspace{-0.95pt}a}{\hspace{-0.2pt}h} \; = \; \theta_{a}(\lambda)(g,h) \hspace{1pt} \cdot \hspace{1pt} \cchhtub[a]{\hspace{-6pt}{}^{\mathrlap{\raisebox{-2.5pt}{\hspace{-1pt}\crule[white]{8pt}{8pt}}}g\hspace{-0.5pt}h}\hspace{-0.95pt}a}{\hspace{-2.4pt} \raisebox{1pt}{$\scriptstyle gh$}}  \; ,
\ee
where we defined the multiplicative phases
\be
\theta_{a}(\lambda)(g,h) \;:=\; \langle \lambda(g,h),{}^{gh}a \rangle \, .
\ee
The $\dagger$-structure acts on morphisms via 
\be
\cchhtub[a]{\scriptstyle \hspace{-2.7pt}{}^{\mathrlap{\raisebox{-2.5pt}{\hspace{-5pt}\crule[white]{8pt}{8pt}}}g}\hspace{-0.95pt}a}{\raisebox{1pt}{$\scriptstyle g$}}^{\dagger} \; = \;\, \theta_{a}^*(\lambda)(g^{-1}\hspace{-1pt}, g) \hspace{1pt} \cdot \hspace{1pt} \cchhtub[\hspace{-2.5pt}{}^g\hspace{-0.95pt}a]{a}{\raisebox{1pt}{$\scriptstyle g^{-1}$}}{} \; .
\ee

\vspace{5pt}
\noindent\textbf{Generalised Charges.}
Similarly to the case of pointed fusion categories in $d=2$, the irreducible generalised charges can be labelled by pairs $(a,\rho)$ consisting of
\begin{enumerate}[leftmargin=4ex]
    \item a representative $a\in A$ of a $G$-orbit $[a]\in A/G$,
    \item an irreducible unitary projective representation $\rho$ of the stabiliser $G_a=\{ g\in G \,|\, {}^ga=a \}$ of $a$ on a Hilbert space $\mathcal{V}$ with projective 2-cocycle $\theta_a(\lambda) \in Z^2(G_a,U(1))$.
\end{enumerate}
Concretely, the associated generalised charge $U_{(a,\rho)}$ can be constructed by fixing for each $b\in [a]$ a representative $r_b \in G$ such that ${}^{r_b}b=a$ (with $r_a\equiv 1$) and setting
\be
\label{eqn:hb_stab_2group}
g_{b} \; := \; r_{({}^{h}b)}\cdot g \cdot r_{b}^{-1} \, \in \, G_a
\ee
for $g\in G$ and $b\in [a]$. Then, $U_{(a,\rho)}$ acts on the twisted sectors
\be
\cH_{b} \; = \; \begin{cases}
    \cV & \text{if }b\in [a]\\
    0 & \text{otherwise}
\end{cases}
\ee
via the linear operators
\be
U_{(a,\rho)} \Big( \hspace{1pt}\ncchhtub[b]{\scriptstyle \hspace{-1.5pt}{}^{g}\hspace{-0.95pt}b}{\raisebox{1pt}{$\scriptstyle g$}} \Big) \;:=\; \epsilon_{b}(g) \cdot \rho(g_b) \;,
\ee
where we defined the multiplicative phase
\be
\label{eqn:epsilonphase_2group}
\epsilon_{b}(g) \; := \; \frac{\theta_{b}(\lambda)(r_{({}^gb)},g)}{\theta_{b}(\lambda)(g_b,r_b)} \;.
\ee
This construction depends on the choice of representatives $r_b \in G$ only up to isomorphism.

\vspace{5pt}
\noindent\textbf{Isometry Actions.}
For a given $g \in G$, the following yield bases of transition channels that obey Theorem~\ref{thm-categorical-wigner}:
\begin{equation}
    \Big\{ \cchhtub[b]{\scriptstyle \hspace{-2.7pt}{}^{\mathrlap{\raisebox{-2.5pt}{\hspace{-1pt}\crule[white]{8pt}{8pt}}}g}\hspace{-0.95pt}b}{\hspace{-0pt}\raisebox{1pt}{$\scriptstyle g$}} \Big\} \; \subset \; \cC^{\hspace{0.5pt}g}(b, {}^gb) \; , 
\end{equation}
where $b\in A$. Upon choosing a generalised charge $U = U_{(a,\rho)}$, we then obtain the following linear isometries associated to the action of $g$ on $b$-twisted sectors (which in the present invertible case are in fact unitary as expected):
\be
U_{(a,\rho)}(g)_b \; = \; \delta_{\hspace{0.5pt} b \hspace{0.5pt}\in\hspace{0.5pt} [a]} \cdot \epsilon_{b}(g) \cdot \rho(g_b): \; \cH_{b} \; \rightarrow \; \cH_{({}^{g\hspace{-0.5pt}}b)}\, ,
\ee
In particular, this yields the following transition probability for the action of $g$ on $b$-twisted sectors:
\be
p\Big( \hspace{1pt} \ncchhtub[b]{\scriptstyle \hspace{-1.8pt}{}^{g}\hspace{-0.95pt}b}{\raisebox{1.5pt}{$\scriptstyle g$}} \Big) \; = \; 1\;.
\ee

\subsubsection{Example: $\TwoRep\big((\bbZ_2 \!\times\! \bbZ_2)[1]\rtimes \bbZ_2\big)$}

\noindent
Let $\mathcal{C} = \text{2Rep}(\mathcal{G})$ be the unitary fusion 2-category of finite-dimensional 2-representations of the 2-group
\begin{equation}
    \mathcal{G} \; = \; (\bbZ_2\times \bbZ_2)[1]\rtimes \bbZ_2 \; ,
\end{equation}
where the 0-form part $\mathbb{Z}_2$ acts on the 1-form part $\mathbb{Z}_2 \times \mathbb{Z}_2$ by exchanging the two factors. Following \cite{Bhardwaj:2022maz, Bartsch:2022ytj}, the simple objects of $\mathcal{C}$ together with their categories of 1-morphisms can be depicted as follows:
\be
\label{eq-2rep-d8-category}
\qquad\hspace{5pt}
\begin{tikzpicture}[baseline={(current bounding box.center)}]
    \coordinate[label = {$1$}] (p1) at (-0,0);
    \coordinate[label = {$V$}] (pV) at (2,0);
    \coordinate[label = {$X$}] (p1p) at (-0.8,-1.1);
    \coordinate[label = {$W$}] (pVp) at (2.8,-1.1);
    \coordinate[label = {$D$}] (pD) at (1,-1.3);
    \draw[->] ($(p1)+(0.2,0.3)$) to[out=20,in=90,looseness=6] node[above, outer sep=1pt, xshift=-5pt] {$\Rep(\bbZ_2)$} ($(p1)+(0,0.48)$);
    \draw[->] ($(p1p)+(-0.2,-0.3+0.5)+(-0.05,0)$) to[out=-160,in=-90,looseness=6] node[below, outer sep=1pt, xshift=5pt] {$\Hilb_{\bbZ_2}$} ($(p1p)+(-0.,-0.48+0.5)+(-0.05,0)$);
    \draw[->] ($(pV)+(-0.2,0.3)$) to[out=160,in=90,looseness=6] node[above, outer sep=1pt, xshift=7pt] {$\Rep(\bbZ_2)$} ($(pV)+(0,0.48)$);
    \draw[->] ($(pVp)+(0.2,-0.3+0.5)$) to[out=-20,in=-90,looseness=6] node[below, outer sep=1pt, xshift=0pt] {$\Hilb_{\bbZ_2}$} ($(pVp)+(0,-0.48+0.5)$);
    \draw[->] ($(pD)+(-0.14,0)$) to[out=-125,in=-55,looseness=5] node[below, outer sep=0pt] {$\Hilb$} ($(pD)+(0.14,0)$);
    \draw[->] ($(p1)$) to[out=-90,in=20,looseness=0.8] node[right, outer sep=2pt] {$\Hilb$} ($(p1p)+(0.2,0.3)$);
    \draw[->] ($(p1p)+(0,0.5)$) to[out=90,in=180,looseness=0.7] node[left, outer sep=2pt] {$\Hilb$} ($(p1)+(-0.2,0.2)$);
    \draw[->] ($(pV)$) to[out=-90,in=160,looseness=0.8] node[left, outer sep=2pt] {$\Hilb$} ($(pVp)+(-0.3,0.3)$);
    \draw[->] ($(pVp)+(0,0.5)$) to[out=90,in=0,looseness=0.7] node[right, outer sep=2pt] {$\Hilb$} ($(pV)+(0.2,0.2)$);
\end{tikzpicture}
\hspace{5pt}\raisebox{-37pt}{$.$}
\ee
The fusion rules of simple objects are given by 
\be
\ba
V\otimes V \, & = \, 1 \;, \\[1pt]
V\otimes X \, & = \, X \otimes V \,=\, W \;,\\[1pt]
V\otimes D \, & = \, D \otimes V \,=\, D \;, \\[1pt]
X\otimes X \, & = \, X \oplus X\;,\\[1pt]
X\otimes D \, & = \, D \otimes X \,=\, D \oplus D \;,\\[1pt]
D \otimes D \, & = \, X \oplus W \;.
\ea
\ee
There are three connected components 
\begin{equation}
    \pi_0(\mathcal{C}) \; = \; \big\lbrace \hspace{0.2pt}[1]\hspace{0.5pt}, [V]\hspace{1pt}, [D] \hspace{0.2pt}\big\rbrace
\end{equation}
and the category of genuine line defects is given by
\begin{equation}
    \Omega(\cC) \; = \; \Rep(\bbZ_2) \; .
\end{equation}
We will denote the non-trivial simple genuine line defect by $\gamma \in \Omega(\mathcal{C})$ in what follows. 

\vspace{5pt}
\noindent\textbf{Tube Category.}
The associated tube category $\text{T}\cC$ was studied in \cite{Bartsch:2022mpm, Bartsch:2022ytj, Bartsch2023a} and has morphism spaces
\be
\ba
\text{T}\cC(1,1) \; & = \; \bbC\hspace{0.5pt}\raisebox{-1pt}{$\scaleobj{1.5}{[}$} \hspace{0.5pt} \cchhtub[1]{1}{1}, \cchhtub[1]{1}{\hspace{-0.5pt}V}, \cchhtub[1]{1}{\hspace{-1pt}D} \hspace{0.5pt}\raisebox{-1pt}{$\scaleobj{1.5}{]}$} \,,\\[3pt]
\text{T}\cC(1,\gamma) \; & = \; \bbC\hspace{0.5pt}\raisebox{-1pt}{$\scaleobj{1.5}{[}$} \hspace{0.5pt} \cchhtub[1]{\gamma}{\hspace{-1pt}D} \hspace{0.5pt}\raisebox{-1pt}{$\scaleobj{1.5}{]}$} \,,\\[3pt]
\text{T}\cC(\gamma,1) \; & = \; \bbC\hspace{0.5pt}\raisebox{-1pt}{$\scaleobj{1.5}{[}$} \hspace{0.5pt} \cchhtub[\gamma]{1}{\hspace{-1pt}D} \hspace{0.5pt}\raisebox{-1pt}{$\scaleobj{1.5}{]}$} \,,\\[3pt]
\text{T}\cC(\gamma,\gamma) \; & = \; \bbC\hspace{0.5pt}\raisebox{-1pt}{$\scaleobj{1.5}{[}$} \hspace{0.5pt} \cchhtub[\gamma]{\gamma}{1}, \cchhtub[\gamma]{\gamma}{\hspace{-0.5pt}V}, \cchhtub[\gamma]{\gamma}{\hspace{-1pt}D} \hspace{0.5pt}\raisebox{-1pt}{$\scaleobj{1.5}{]}$}
\ea
\ee
and composition rules described in \cite{Bartsch2023a}. The $\dagger$-structure acts on morphisms via
\be
\begin{alignedat}{2}
\cchhtub[1]{1}{\hspace{-0.5pt}V}^{\dagger} \, &= \; \cchhtub[1]{1}{\hspace{-0.5pt}V} \; , \;\;\quad \cchhtub[\gamma]{\gamma}{\hspace{-0.5pt}V}^{\dagger} \; &&= \; \cchhtub[\gamma]{\gamma}{\hspace{-0.5pt}V} \; , \\[3pt]
\cchhtub[1]{1}{\hspace{-1pt}D}^{\dagger} \, &= \; \cchhtub[1]{1}{\hspace{-1pt}D} \; , \;\;\quad \cchhtub[\gamma]{1}{\hspace{-1pt}D}^{\dagger} \; &&= \; \cchhtub[1]{\gamma}{\hspace{-1pt}D} \; , \\[3pt]
\cchhtub[1]{\gamma}{\hspace{-1pt}D}^{\dagger} \, &= \; \cchhtub[\gamma]{1}{\hspace{-1pt}D} \; , \;\;\quad \cchhtub[\gamma]{\gamma}{\hspace{-1pt}D}^{\dagger} \; &&= \; \cchhtub[\gamma]{\gamma}{\hspace{-1pt}D} \; .
\end{alignedat}
\ee

\vspace{5pt}
\noindent\textbf{Generalised Charges.}
There are five irreducible generalised charges that can be described as follows \cite{Bartsch2023a}:
\begin{itemize}[leftmargin=3ex]
    \item There are two one-dimensional charges $U_1^{\pm}$ that act on the untwisted sector $\cH_1\cong \bbC$ via
    \be
    \begin{aligned}
    U_1^{\pm}\Big( \cchtub[1]{1}{\hspace{-0.5pt}V} \Big) \; &= \; 1 \; , \\[3pt]
    U_1^{\pm}\Big( \cchtub[1]{1}{\hspace{-1pt}D} \Big) \; &= \; \pm \sqrt{2} \; .
    \end{aligned}
    \ee

    \item There are two 1-dimensional charges $U_{\gamma}^{\pm}$ that act on the twisted sector $\cH_{\gamma}\cong \bbC$ via
    \be
    \begin{aligned}
    U_{\gamma}^{\pm}\Big( \cchtub[\gamma]{\gamma}{\hspace{-0.5pt}V} \Big) \; &= \; -1 \; , \\[3pt]
    U_{\gamma}^{\pm}\Big( \cchtub[\gamma]{\gamma}{\hspace{-1pt}D} \Big) \; &= \; \pm \sqrt{2} \; .
    \end{aligned}
    \ee

    \item There is one two-dimensional charge $U_{1, \gamma}$ that acts on the twisted sectors $\cH_1\cong \cH_{\gamma}\cong \bbC$ via
    \be
    \label{eqn:2dimcharge_2rep2gp}
    \begin{alignedat}{2}
    \;\quad U_{1, \gamma}\Big( \cchtub[1]{1}{\hspace{-0.5pt}V} \Big) \; &= \; -1 \; , \;\quad && U_{1, \gamma}\Big( \cchtub[\gamma]{\gamma}{\hspace{-0.5pt}V} \Big) \; = \; -1 \; , \\[3pt]
    \;\quad U_{1, \gamma}\Big( \cchtub[1]{1}{\hspace{-1pt}D} \Big) \; &= \; 0 \; , \;\quad && U_{1, \gamma}\Big( \cchtub[\gamma]{1}{\hspace{-1pt}D} \Big) \; = \; \sqrt{2} \; , \\[3pt]
    \;\quad U_{1, \gamma}\Big( \cchtub[1]{\gamma}{\hspace{-1pt}D} \Big) \; &= \; \sqrt{2} \; , \;\quad && U_{1, \gamma}\Big( \cchtub[\gamma]{\gamma}{\hspace{-1pt}D} \Big) \; = \; 0 \; . \\[10pt]
    \end{alignedat}
    \ee
\end{itemize}

\vspace{5pt}
\noindent\textbf{Isometry Actions.}
The following yield bases of transition channels for the non-invertible defect $D$ that obey Theorem~\ref{thm-higher_cat_wigner}:
\be
\begin{aligned}
    \left\{ \tfrac{1}{\sqrt{2}} \cchhtub[1]{1}{\hspace{-1pt}D} \hspace{1pt} \right\} \, & \subset \, \cC^{\hspace{0.5pt}D}(1,1) \;, & \left\{ \tfrac{1}{\sqrt{2}} \cchhtub[1]{\gamma}{\hspace{-1pt}D} \right\} \, & \subset \, \cC^{\hspace{0.5pt}D}(1,\gamma) \; , \\[3pt]
   \left\{ \tfrac{1}{\sqrt{2}} \cchhtub[\gamma]{1}{\hspace{-1pt}D} \right\} \, & \subset \, \cC^{\hspace{0.5pt}D}(\gamma, 1)\;, & \left\{\tfrac{1}{\sqrt{2}} \cchhtub[\gamma]{\gamma}{\hspace{-1pt}D} \right\} \, & \subset \, \cC^{\hspace{0.5pt}D}(\gamma, \gamma)\; .
\end{aligned}
\vspace{3pt}
\ee
Upon choosing a generalised charge $U$, the above then induce linear isometries $U(D)_{\mu}$ associated to the action of $D$ on $\mu$-twisted sectors. For instance, for $U = U_{1,\gamma}$ supported on the twisted sectors $\cH_1\cong \cH_{\gamma}\cong \bbC$, we have that
\be
\quad
\ba
    U_{1, \gamma}(D)_{1} \, & = \, \begin{bmatrix} 0 \\ 1 \end{bmatrix} : \;\; \cH_{1} \;\, \rightarrow \;\, \cH_{1}\oplus\cH_{\gamma} \,, \\[3pt]
    U_{1, \gamma}(D)_{\gamma} \, & = \, \begin{bmatrix} 1 \\ 0 \end{bmatrix} : \;\; \cH_{\gamma} \;\, \rightarrow \;\, \cH_{1} \oplus \cH_{\gamma} \,.
\ea
\ee
As a result, we obtain the following transition probabilities for the action of $D$ on twisted sectors:
\be
\begin{alignedat}{2}
    p\Big( \tfrac{1}{\sqrt{2}} \cchhtub[1]{1}{\hspace{-1pt}D} \Big) \; = \; 0\;, \qquad p\Big( \tfrac{1}{\sqrt{2}} \cchhtub[\gamma]{1}{\hspace{-1pt}D} \Big) \; = \; 1 \; , \\[2pt]
    p\Big( \tfrac{1}{\sqrt{2}} \cchhtub[1]{\gamma}{\hspace{-1pt}D} \Big) \; = \; 1\;, \qquad p\Big( \tfrac{1}{\sqrt{2}} \cchhtub[\gamma]{\gamma}{\hspace{-1pt}D} \Big) \; = \; 0 \; . \\[2pt]
\end{alignedat}
\ee


\end{document}